# Recovering Purity with Comonads and Capabilities

VIKRAMAN CHOUDHURY, Indiana University, USA and University of Cambridge, UK
NEEL KRISHNASWAMI, University of Cambridge, UK

In this paper, we take a pervasively effectful (in the style of ML) typed lambda calculus, and show how to *extend* it to permit capturing pure expressions with types. Our key observation is that, just as the pure simply-typed lambda calculus can be extended to support effects with a monadic type discipline, an impure typed lambda calculus can be extended to support purity with a *comonadic* type discipline.

We establish the correctness of our type system via a simple denotational model, which we call the *capability space* model. Our model formalises the intuition common to systems programmers that the ability to perform effects should be controlled via access to a permission or capability, and that a program is *capability-safe* if it performs no effects that it does not have a runtime capability for. We then identify the axiomatic categorical structure that the capability space model validates, and use these axioms to give a categorical semantics for our comonadic type system. We then give an equational theory (substitution and the call-by-value $\beta$ and $\eta$ laws) for the imperative lambda calculus, and show its soundness relative to this semantics.

Finally, we give a translation of the pure simply-typed lambda calculus into our comonadic imperative calculus, and show that any two terms which are $\beta\eta$-equal in the STLC are equal in the equational theory of the comonadic calculus, establishing that pure programs can be mapped in an equation-preserving way into our imperative calculus.

## 1 INTRODUCTION

Consider the two following definitions of the familiar `map` functional, which applies a function to each element of a list.

```
map1 : ∀ a b. (a → b) → List a → List b
map1 f []        = []
map1 f (x :: xs) = let zs = map1 f xs in
                   let  z = f x in
                   z :: zs

map2 : ∀ a b. (a → b) → List a → List b
map2 f []        = []
map2 f (x :: xs) = let  z = f x in
                   let zs = map2 f xs in
                   z :: zs
```

In a purely functional language like Haskell, these two definitions are equivalent. But in an *impure* functional language like ML, the difference between these two definitions is *observable*:

Authors' addresses: Vikraman Choudhury, Department of Computer Science, Indiana University, Bloomington, 47408, USA, vikraman@indiana.edu, Department of Computer Science and Technology, University of Cambridge, Cambridge, CB3 0FD, UK, vc378@cl.cam.ac.uk; Neel Krishnaswami, Department of Computer Science and Technology, University of Cambridge, Cambridge, CB3 0FD, UK, nk480@cl.cam.ac.uk.

```
let xs = ["left "; "to "; "right "]

let f s = stdout.print(s); s

let ys = map1 f xs  -- Prints "right to left "
let zs = map2 f xs  -- Prints "left to right "
```

So something as innocuous-seeming as a `print` function can radically change the equational theory of the language: no program transformation that changes the order in which sub-expressions are evaluated is in general sound. This greatly complicates reasoning about programs, as well as hindering many desirable program optimisations such as list fusion and deforestation [Wadler 1990]. Transformations that are unconditionally valid in a pure language must, in an impure language, be gated by complex whole-program analyses tracking the purity of sub-expressions.

*Contributions.* It is received wisdom that much as a drop of ink cannot be removed from a glass of water, once a language supports ambient effects, there is no way to regain the full equational theory of a pure programming language. In this paper, we show that this folk belief is *false*: we extend an ambiently effectful language to support purity. Entertainingly, it turns out that just as monads are a good tool to extend pure languages with effects, **comonads** are a good tool to extend impure languages with purity!

- We take a pervasively effectful lambda calculus in the style of ML and show how to *extend* it with a *comonadic* type discipline modelling the intuitions underpinning the *object-capability model* [Lauer and Needham 1979; Levy 1984; Miller 2006] developed in the systems community. The object-capability model advises that the ability to perform effects should be controlled via access to a permission or capability, and that a program is *capability-safe* precisely when it can only perform effects that it possesses a runtime capability for.
- We show that the typing rules are faithful to the object-capability model by giving our language a denotational semantics, which we call the *capability space* model. Capability spaces are a simple, direct formalisation of the ideas underpinning the object-capability model, which extends the most naive model of the lambda calculus – sets and functions – with just enough structure to model capability-safety. In our model, a type is just a set $X$ (denoting a set of values), together with a relation $w_X$ saying which capabilities each value $x$ may own. Morphisms $f : X \to Y$ are *capability-safe* if the capabilities of $f(x)$ are bounded by the capabilities of $x$.

  It is already known in the systems community that even effectful, untyped lambda-calculi can be made capability-safe by removing features exposing ambient authority. Our model and type system demonstrates that this observation is incomplete – having a comonad witnessing the *denial* of a capability is also very beneficial. In particular, this greatly simplifies the process of *capability taming*, making it possible to make a standard library capability-safe in an incremental fashion.
- We then identify the axiomatic categorical structure the capability space model validates, and use these axioms to give a categorical semantics for our comonadic type system. We then give an equational theory (substitution and the call-by-value $\beta$ and $\eta$ laws) for the imperative lambda calculus, and show its soundness relative to this semantics.
- Finally, we give a translation of the pure simply-typed lambda calculus into our comonadic imperative calculus, and show that any two terms which are $\beta\eta$-equal in the STLC are equal in the equational theory of the comonadic calculus under the translation, establishing that pure programs can be mapped in an equation-preserving way into our imperative calculus.

Detailed proofs of the lemmas and theorems, as well as additional material are given in the supplementary appendices, and we refer to them in the text.

## 2 PURITY FROM CAPABILITIES

The *object-capability* model is a methodology originating in the operating systems community for building secure operating systems and hardware. The idea behind this model is that systems must be able to control permissions to perform potentially dangerous or insecure operations, and that a good way to control access is to tie the right to perform actions to values in a programming language, dubbed *capabilities*. Then, the usual variable-binding and parameter-passing mechanisms of the language can be used to grant rights to perform actions — access to a capability can be prohibited to a client by simply not passing it the capability as an argument. To quote Miller [2006]:

> Our object-capability model is essentially the untyped call-by-value lambda calculus with applicative-order local side effects and a restricted form of **eval** — the model Actors and Scheme are based on. This correspondence of objects, lambda calculus, and capabilities was noticed several times by 1973.

We use this observation to design our language – we begin with the observation that it is possession of the capability to perform effects that distinguishes impure from safe code. In the example in section 1, the operation `f` that distinguished between `map1` and `map2` contained a reference to **`stdout`**, and so had the intrinsic authority to print to the standard output – that is, `f` was not a capability-safe function.

The $c$.print($s$) operation takes the channel $c$ and prints the string $s$ to it. If we did not possess the capability $c$, then we could not invoke the print operation upon this channel. This property is actually fundamental to the object-capability model, which says that the *only* way to access capabilities must be through capability values. Therefore, if we view channels as capabilities, we know that evaluating a piece of code *lacking* any capabilities cannot print at all.

Naturally, there are many data types in a real programming language beyond channels, but each value can access some set of capabilities (eg, a list of files can access any of the channels in the list, or a closure can access any capability it receives as an argument or possesses in its environment). So for each value, we can bound the set of possible effects it enables by the capabilities it owns.

This lets us approximate the notion of a "capability-safe program" in a simple and brutal fashion: we can judge a term to be capability-safe if it can directly access **zero** capabilities. Lacking access to any channels, it has no intrinsic ability to do I/O, and hence must be capability-safe. Furthermore, we introduce **two kinds of variables:** *safe* variables and arbitrary (or *impure*) variables. By restricting the substitution to only permit substituting capability-safe terms for safe variables, the judgement of safety will be stable under substitution. Then, by internalising the safety judgement as a type, we can pass safe values – i.e., values without access to any capabilities – as first-class values.

To understand this, let us begin with a simple call-by-value higher-order functional language extended with types for string constants, channels (or output file handles), and a single effect: outputting a string onto a channel with the expression `chan.print(s)`. There is no monadic or effect typing discipline here; the type of `print` is just as one might see in OCaml or Java.

```
print : Channel → String → Unit
```

For example, here is a simple function to print each element of a pair of strings to a given channel:

```
print_pair : String × String → Channel → Unit
print_pair = fun p chan →
                chan.print(fst p);
                chan.print(snd p)
```

Here, for clarity we use a semicolon for sequencing, and write `print` in method-invocation style *à la* Java (to make it easy to distinguish the file handle from the string argument). [1]

To support capability safety (and thereby obtain purity as a side-effect(!)) we extend the language with a new type constructor `Safe a`, denoting the set of expressions of type `a` which are *capability-safe* – i.e., they own no file handles and so their execution cannot do any printing, unless a capability is passed. We add the introduction form `box(e)` to introduce a value whose type is `Safe a`; the type system accepts this if `e` has type `a` and is recognisably safe, but rejects it otherwise. Here, "recognisably safe" means that the term `e` does not refer to any capability literals, and all of its free variables are safe variables.

To eliminate a value of type `Safe a`, we will use *pattern matching*, writing the elimination form `let box(x) = e1 in e2` to bind the *safe* expression in `e1` to the variable `x`. The only difference from ordinary pattern matching is that the bound variable `x` is marked as a *safe* variable, permitting it to occur inside of *safe* expressions. Intuitively, this makes sense – `e1` evaluates to a *safe* value, and so its result should be allowed to be used by other *safe* expressions.

It turns out that this discipline of tracking whether a variable is *safe* or not is precisely a *comonadic* type discipline, corresponding to the $\Box$ modality in S4 modal logic. Capability-safety is not exactly the same thing as purity, but we will show how to recover *purity from capability-safety* later in this section, and then prove that this encoding works later on. We illustrate the comonadic behaviour of the `Safe` type constructor with the following examples.

If we know that a value is *safe*, we can `extract` it, giving up that information. Also, since `Safe` is only expressing a property of the underlying value, applying it twice achieves nothing, making `duplicate` an isomorphism. This expresses an idempotent comonad, which encodes the property that a value of type `Safe a` is *safe*.

```
extract : ∀ a. Safe a → a
extract box(x) = x

duplicate : ∀ a. Safe a → Safe (Safe a)
duplicate box(x) = box(box(x))
```

Also, observe that we can apply `Safe` functions to `Safe` values to get `Safe` results, thereby making it *almost* an `Applicative` functor, as shown below. Syntactically, `box(f x)` is accepted, since both the variables `f` and `x` are known to be *safe*, and so are permitted to occur inside of a *safe* expression.

```
(⊛) : ∀ a b. Safe (a → b) → Safe a → Safe b
(⊛) box(f) box(x) = box(f x) -- accepted
```

However, arbitrary values are not `Safe` – we cannot mark any value `x` *safe* because it could own capabilities. So this function is rejected.

```
pure : ∀ a. a → Safe a
pure x = box(x) -- REJECTED
```

Nor can we write an `fmap` for `Safe`, which applies an arbitrary function to a *safe* argument, and tries to return a *safe* result.

```
fmap : ∀ a b. (a → b) → Safe a → Safe b
fmap f box(x) = box(f x)  -- REJECTED
```

[1] We are also using a mix of ML and Haskell syntax, which is in line with the theme of this paper.

Semantically, the function `f` may own capabilities, and so it may have side-effects. Syntactically, since `f` is an *impure* variable, it is simply not allowed to occur in the *safe* expression `box(f x)`. Only if we mark both the function and the argument as `Safe` can we apply it, as we saw in (⊛).

However, `Safe` *is* a functor in the semantic sense – the absence of an `fmap` action indicates that this functor lacks *tensorial strength.* [2]

The capability discipline permits typing functions whose behaviour is intermediate between *pure* and effectful. First, suppose we see the following type signature for a print function:

```
safe_print : Safe (Channel → String → Unit)
-- definition not visible
```

Without looking at the definition of `safe_print`, we can make some inferences about its side-effects. Since it is marked `Safe`, we can immediately infer that *if* this function performs a side-effect, it can print *only on the channel* that it binds. In other words, it *cannot* use an ambient capability to perform side-effects.

Similarly, consider the following type declaration:

```
multi_print : Safe (List Channel → String → Unit)
-- definition not visible
```

Again, we do not know anything about the body of the definition (perhaps it prints its string argument to all of the channels it receives, or perhaps not), but due to the typing discipline, we know that `multi_print` is `Safe`, and hence, owns no capabilities of its own. As a result, we can make some inferences about the following two declarations:

```
x : Unit
x = let box(f) = multi_print in
    f [stdout, stderr] "Hello world"
```

```
y : Unit
y = let box(f) = multi_print in
    f [] "Hello world"
```

The definition of `x` passes two channels to `multi_print`, and so it may have an effect (it might use it to print on either of these channels). On the other hand, we *know* that the evaluation of `y` *will not* have an effect – we know that `multi_print` owned no channels, and we did not give it any channels, therefore it can perform *no effects.* The purity of this function *depends on the inputs that were passed to it.* Moreover, we know this without having to see the definition of `multi_print`!

Even though capability-safety is a more primitive notion than purity, it is strong enough to encode purity. Revisiting our `map` example from section 1, we can now rewrite it using the `Safe` type constructor.

```
map : ∀ a b. Safe (Safe a → b) → List (Safe a) → List b
map box(f) []        = []
map box(f) (x :: xs) = let  z = f x in
                       let zs = map box(f) xs in
                       z :: zs
```

Intuitively, a *safe* function can only have an effect if its argument gives it any capabilities, and we can prohibit a function argument from bearing capabilities by giving it a `Safe` type. Hence, we can model the *pure* function space $A \Rightarrow B$ using the impure function space, by giving it the type `Safe(Safe A → B)`.

An additional benefit of the comonadic type discipline is that it dramatically simplifies the process of *capability taming.* A language is capability-safe when programs have no access to

[2]This also means that safety is not definable in Haskell, since all definable functors are strong.

| | | | |
|---|---|---|---|
| TYPES | $A, B$ | $::=$ | $\mathsf{unit} \mid \mathsf{str} \mid \mathsf{cap}$ |
| | | | $\mid A \times B \mid A \Rightarrow B \mid \Box A$ |
| TERMS | $e$ | $::=$ | $() \mid s \mid e_1 . \mathsf{print}(e_2)$ |
| | | | $\mid (e_1, e_2) \mid \mathsf{fst}\ e \mid \mathsf{snd}\ e$ |
| | | | $\mid x \mid \lambda x : A.\ e \mid e_1\ e_2$ |
| | | | $\mid \mathsf{box}\ \boxed{e} \mid \mathsf{let\ box}\ \boxed{x} = e_1\ \mathsf{in}\ e_2$ |
| VALUES | $v$ | $::=$ | $() \mid s \mid (v_1, v_2)$ |
| | | | $\mid x \mid \lambda x : A.\ e \mid \mathsf{box}\ \boxed{e}$ |
| QUALIFIERS | $q, r$ | $::=$ | $\mathbf{s} \mid \mathbf{i}$ |
| CONTEXTS | $\Gamma, \Delta, \Psi$ | $::=$ | $\mid \Gamma, x : A^q$ |
| SUBSTITUTIONS | $\theta, \phi$ | $::=$ | $\langle\rangle \mid \langle \theta, e^q/x \rangle$ |

Fig. 1. Grammar

*ambient authorities*. As a result, capability-safety has historically been understood not just as a property of the language, but also of its standard library. In particular, if the standard library exposes globally-visible channels like `stdout` and `stderr`, any program in the language can refer to them, and thereby have write effects. As a result, a project like Joe-E [Mettler et al. 2010] involves a massive effort to rewrite the whole standard library of Java. In constrast, a language with a safety comonad affords a *gradual approach* – the bindings in the standard library can all be marked *impure* by default, and as the functions are audited, they can gradually be marked *safe*, allowing more and more capability-safe programs to be written. This lets language implementors and programmers gradually opt-in to capability safety, making it easier to migrate language ecosystems, and also illustrates the importance of being able to track the safety of variable bindings.

## 3 TYPING

We give the grammar of our language in figure 1. We have the usual type constructors for unit, products, and functions from the simply-typed lambda calculus. In addition to this, we have the type str for strings, and the type cap representing output channels (used in the imperative $e_1 . \mathsf{print}(e_2)$ statement). Finally, we add the comonadic $\Box$ type constructor which corresponds to the `Safe` type constructor we introduced in section 2.

Despite the fact that there is a *type* cap of channels, and a print operation which uses them, there are no introduction forms for them. This is intentional! The absence of this facility corresponds to the principle of *capability safety* – the only capabilities a program should possess are those that are passed by its caller. So, a complete program will either be a function that receives a capability token as an argument, or have free variables that the system can bind capability tokens to. [3]

The expressions in our language include the usual ones from the simply-typed lambda calculus, constants $s$ for strings, and print. We also have an introduction form box $\boxed{e}$, and a let box elimination form for the $\Box A$ type; we'll explain how these work later. Values are a subset of expressions, but box turns any expression into a value. [4]

We would like a modal type system where we can distinguish between expressions with and without side-effects. Following the style of [Pfenning and Davies 2001] for S4 modal logic, we could build a dual-context calculus. However, such a setup makes it difficult to define substitution; we

[3]Of course, a full system should have the ability to create new private capabilities of its own. We omit this to keep the denotational semantics simple, but we discuss more about it in section 8.

[4]We write sequencing as $e_1 ; e_2$, which is syntactic sugar for $(\lambda x : \mathsf{unit}.\ e_2)\ e_1$.

| | |
|---|---|
| $x : A^q \in \Gamma$ | $x$ is a variable of type $A$ with qualifier $q$ in context $\Gamma$ |
| $\Gamma \vdash e : A$ | $e$ is an expression of type $A$ in context $\Gamma$ |
| $\Gamma \vdash^{\mathsf{s}} e : A$ | $e$ is a *safe* expression of type $A$ in context $\Gamma$ |

(a) Typing Judgements

| | |
|---|---|
| $\Gamma \supseteq \Delta$ | $\Gamma$ is a weakening of context $\Delta$ |
| $\Gamma \vdash \theta : \Delta$ | $\theta$ is a well-formed substitution from context $\Gamma$ to $\Delta$ |

(b) Weakening and Substitution Judgements

| | |
|---|---|
| $\Gamma \vdash e_1 \approx e_2 : A$ | $e_1$ and $e_2$ are equal expressions of type $A$ in context $\Gamma$ |

(c) Equality Judgements

Fig. 2. Judgement forms

$$\frac{}{\Gamma \vdash () : \mathsf{unit}}\ \text{unitI} \qquad \frac{}{\Gamma \vdash s : \mathsf{str}}\ \text{strI} \qquad \frac{\Gamma \vdash e_1 : \mathsf{cap} \qquad \Gamma \vdash e_2 : \mathsf{str}}{\Gamma \vdash e_1\,.\,\mathsf{print}(e_2) : \mathsf{unit}}\ \textsc{Print}$$

$$\frac{\Gamma \vdash e_1 : A \qquad \Gamma \vdash e_2 : B}{\Gamma \vdash (e_1\,,e_2) : A \times B}\ \times\text{I} \qquad \frac{\Gamma \vdash e : A \times B}{\Gamma \vdash \mathsf{fst}\,e : A}\ \times\text{E}_1 \qquad \frac{\Gamma \vdash e : A \times B}{\Gamma \vdash \mathsf{snd}\,e : B}\ \times\text{E}_2$$

$$\frac{x : A^q \in \Gamma}{\Gamma \vdash x : A}\ \textsc{Var} \qquad \frac{\Gamma, x : A^{\mathsf{i}} \vdash e : B}{\Gamma \vdash \lambda x : A.\ e : A \Rightarrow B}\ \Rightarrow\text{I} \qquad \frac{\Gamma \vdash e_1 : A \Rightarrow B \qquad \Gamma \vdash e_2 : A}{\Gamma \vdash e_1\,e_2 : B}\ \Rightarrow\text{E}$$

$$\frac{\Gamma^{\mathsf{s}} \vdash e : A}{\Gamma \vdash^{\mathsf{s}} e : A}\ \textsc{ctx-safe} \qquad \frac{\Gamma \vdash^{\mathsf{s}} e : A}{\Gamma \vdash \mathsf{box}\,\boxed{e} : \Box A}\ \Box\text{I} \qquad \frac{\Gamma \vdash e_1 : \Box A \qquad \Gamma, x : A^{\mathsf{s}} \vdash e_2 : B}{\Gamma \vdash \mathsf{let\ box}\,\boxed{x} = e_1\ \mathsf{in}\ e_2 : B}\ \Box\text{E}$$

Fig. 3. Typing Rules

can avoid dual contexts by tagging terms with qualifiers instead. [5] We use two qualifiers that we can annotate terms with, in the appropriate places. We use $\mathsf{s}$ to tag *safe* terms, and $\mathsf{i}$ to tag *impure* terms. [6]

Next, we define contexts of variables. A well-formed context is either the empty context , or an extended context with a variable $x$ of type $A$ with qualifier $q$. Finally, we give a grammar for substitutions. A substitution is either the empty substitution $\langle\rangle$, or an extended substitution with an expression $e$ substituted for variable $x$ qualified by $q$.

## 3.1 Typing Judgements

In figure 2a we introduce three kinds of judgement forms, and give typing rules in figure 3. We have the usual introduction and elimination rules for constants and products. If a variable is present in the context, we can introduce it, using the Var rule. In the introduction rule for functions $\Rightarrow$ I, we mark the hypothesis as *impure* when forming a $\lambda$-expression, because we do not want to restrict

[5] Since the comonad is idempotent (see subsection 4.5), we could also use the Fitch-style syntax in [Clouston 2018]. However, we follow our syntactic style to stress the similarity with linear logic.

[6] We use different colours to distinguish *safe* and *impure* syntactic objects, and we'll follow this convention henceforth. When we have unknown qualifiers occurring on terms, we *highlight* them in a different colour, and the colour changes to the appropriate one when the qualifier is $\mathsf{s}$ or $\mathsf{i}$.

$$\begin{aligned} ()^{s} &:= \\ (\Gamma, x : A^{s})^{s} &:= \Gamma^{s}, x : A^{s} \\ (\Gamma, x : A^{i})^{s} &:= \Gamma^{s} \end{aligned}$$

(a)

$$\begin{aligned} \langle\rangle^{s} &:= \langle\rangle \\ \langle\theta, e^{s}/x\rangle^{s} &:= \langle\theta^{s}, e^{s}/x\rangle \\ \langle\theta, e^{i}/x\rangle^{s} &:= \theta^{s} \end{aligned}$$

(b)

Fig. 4. Purifying Contexts and Substitutions

$$\frac{}{x : A^{q} \in (\Gamma, x : A^{q})}\ \in\text{-ID} \qquad \frac{x : A^{q} \in \Gamma \qquad (x \neq y)}{x : A^{q} \in (\Gamma, y : B^{r})}\ \in\text{-EX}$$

(a) Context Membership Rules

$$\frac{}{\supseteq}\ \supseteq\text{-ID} \qquad \frac{\Gamma \supseteq \Delta}{\Gamma, x : A^{q} \supseteq \Delta, x : A^{q}}\ \supseteq\text{-CONG} \qquad \frac{\Gamma \supseteq \Delta}{\Gamma, x : A^{q} \supseteq \Delta}\ \supseteq\text{-WK}$$

(b) Weakening Rules

$$\frac{}{\Gamma \vdash \langle\rangle :}\ \text{SUB-ID}$$

$$\frac{\Gamma \vdash \theta : \Delta \qquad \Gamma \vdash^{s} e : A}{\Gamma \vdash \langle\theta, e^{s}/x\rangle : \Delta, x : A^{s}}\ \text{SUB-SAFE} \qquad \frac{\Gamma \vdash \theta : \Delta \qquad \Gamma \vdash v : A}{\Gamma \vdash \langle\theta, v^{i}/x\rangle : \Delta, x : A^{i}}\ \text{SUB-IMPURE}$$

(c) Substitution Rules

Fig. 5. Membership, Weakening and Substitution Rules

function arguments in general. The elimination rule $\Rightarrow$ E, or function application works as usual. The print statement performs side-effects but has the type unit. We need to do more work to add the comonadic type constructor.

We can mark a term as *safe* if it was well-typed in a *safe* context, where every variable has the $s$ annotation. So we define a syntactic *purify* operation, which acts on contexts; applying it drops the terms with the *impure* annotation, as shown in figure 4a. This is expressed by the CTX-SAFE rule, which introduces a *safe* expression using the *safe* judgement form. And then, we can put it in a box using the $\square$I rule, to get a $\square$-typed value.

We give an elimination rule $\square$E using the let box binding form. Given an expression in the $\square$ type, we bind the underlying *safe* expression to the variable $x$. With an extended context that has a free variable $x$ marked *safe*, if we can produce a well-typed expression in the motive type, the elimination is complete.

### 3.2 Weakening and Substitution

Next, we can define syntactic weakening and substitution.

*3.2.1 Membership.* We give the standard rules for the context membership judgement in figure 5a, following Barendregt's variable convention. The only difference is that variables now have an extra safety annotation.

*3.2.2 Weakening.* The context weakening relation follows the usual rules, as shown in figure 5b, with the extra annotation on free variables in contexts. $\Gamma \supseteq \Delta$ indicates that $\Gamma$ has more variables than $\Delta$, and is defined as an inductive relation in figure 5b. We can prove a syntactic weakening lemma.

LEMMA 3.1 (SYNTACTIC WEAKENING). *If* $\Gamma \supseteq \Delta$ *and* $\Delta \vdash e : A$, *then* $\Gamma \vdash e : A$.

*3.2.3 Substitution.* Substitution requires a bit more care. First, we define the judgement $\Gamma \vdash \theta : \Delta$, which says that $\theta$ is a well-formed substitution from context $\Gamma$ to $\Delta$. Since our language is effectful, we restrict the definition of substitutions, in figure 5c to substitute *values* for *impure* variables, while permitting *safe* expressions for *safe* variables.

Furthermore, we define the syntactic substitution function, which applies a substitution on raw terms. This is mostly standard, but when substituting under a binder, we do a renaming of the bound variable by extending the substitution with an appropriately annotated variable. To substitute inside a box-ed expression, we have to *purify* the substitution when using it. We extend the *purify* operation to substitutions as well; it simply drops the *impure* substitutions, as shown in figure 4b.

*Definition 3.2 (Syntactic substitution on raw terms).*

$$
\begin{aligned}
\theta(x) &:= \theta[x] \\
\theta(()) &:= () \\
\theta(s) &:= s \\
\theta((e_1, e_2)) &:= (\theta(e_1), \theta(e_2)) \\
\theta(\mathsf{fst}\ e) &:= \mathsf{fst}\ \theta(e) \\
\theta(\mathsf{snd}\ e) &:= \mathsf{snd}\ \theta(e) \\
\theta(\lambda x.\ e) &:= \lambda y.\ \langle \theta, y^{\mathsf{i}}/x \rangle(e) \\
\theta(e_1\ e_2) &:= \theta(e_1)\ \theta(e_2) \\
\theta(\mathsf{box}\ \boxed{e}) &:= \mathsf{box}\ \boxed{\theta^{\mathsf{s}}(e)} \\
\theta(\mathsf{let\ box}\ \boxed{x} = e_1\ \mathsf{in}\ e_2) &:= \mathsf{let\ box}\ \boxed{y} = \theta(e_1)\ \mathsf{in}\ \langle \theta, y^{\mathsf{s}}/x \rangle(e_2) \\
\theta(e_1.\mathsf{print}(e_2)) &:= \theta(e_1).\mathsf{print}(\theta(e_2))
\end{aligned}
$$

*Definition 3.3 (Syntactic substitution on variables).*

$$
\theta[x] := \begin{cases} \lightning & \theta = \langle\rangle \\ e & \theta = \langle \phi, e^q/x \rangle \\ \phi[x] & \theta = \langle \phi, e^q/y \rangle, x \neq y \end{cases}
$$

Finally, we show the type-correctness of substitution by proving a syntactic substitution theorem.

THEOREM 3.4 (SYNTACTIC SUBSTITUTION). *If* $\Gamma \vdash \theta : \Delta$ *and* $\Delta \vdash e : A$, *then* $\Gamma \vdash \theta(e) : A$.

## 4 SEMANTICS

In this section, we describe a concrete denotational model of capabilities and the abstract categorical structure it models.

### 4.1 Capability Spaces

Let $\mathcal{C}$ be a fixed set of capability names, possibly countably infinite, and with decidable equality. The powerset $\mathfrak{P}(\mathcal{C})$ denotes the set of all subsets of $\mathcal{C}$, and $(\mathfrak{P}(\mathcal{C}); \varnothing, \mathcal{C}, \subseteq, \cap, \cup)$ is the complete lattice ordered by set inclusion.

A capability space $X = (|X|, w_X)$ is a set $|X|$ with a weight relation $w_X : |X| \nrightarrow \mathfrak{P}(\mathcal{C})$ that assigns sets of capabilities to each member in $X$. Intuitively, we think of the set $|X|$ as the set of values of the type $X$, and we think of the weight relation $w_X$ as defining the possible sets of capabilities that each value may own.

We require maps between capability spaces to preserve weights, i.e., a map between the underlying sets $|X|$ and $|Y|$ is a morphism of capability spaces iff for each $x$ in $|X|$, all the weights in $Y$ for $f(x)$ are bounded by the weights in $X$ for $x$. If we think of a function $f : X \longrightarrow Y$ as a term of type $Y$ with a free variable of type $X$, then this condition ensures that the capabilities of the term are limited to at most those of its free variables. In other words, weight-preserving functions are precisely those which are capability-safe; they do not have unauthorised access to arbitrary capabilities, and they *do not have any ambient authority.*

We now formally define the category of capability spaces $\mathbb{C}$, with objects as capability spaces and morphisms as weight-preserving functions.

*Definition 4.1 (Category $\mathbb{C}$ of capability spaces).*

$$
\begin{aligned}
\mathcal{O}bj_{\mathbb{C}} &:= X = (|X| : \mathrm{Set},\ w_X : |X| \nrightarrow \mathfrak{P}(\mathcal{C})) \\
\mathcal{H}om_{\mathbb{C}}(X, Y) &:= \left\{ f \in |X| \to |Y| \;\middle|\; \begin{array}{l} \forall x, C_x, w_X(x, C_x) \Rightarrow \\ \exists C_y \subseteq C_x, w_Y(f(x), C_y) \end{array} \right\}
\end{aligned}
$$

We remark that the definition of this category is inspired by the category of length spaces defined by Hofmann [2003], which again associates intensional information (in his work, memory usage, and in ours, capabilities) to a set-theoretic semantics.

### 4.2 The Direct Semantics

Before describing the categorical structure of capability spaces, we first consider a direct set-theoretic semantics for our language. Since the capability space model is a "structured sets" model, where each object is a set with some additional structure (i.e., the weights), and morphisms are ordinary set-theoretic functions (which are required to preserve this structure), we can interpret an expression $e$ with typing derivation $\Gamma \vdash e : A$, as a function $\Gamma \longrightarrow TA$. This is an ordinary set-theoretic function which takes an element of $\Gamma$ (i.e., a substitution binding each variable to an element of its type) to a monadic computation (using a writer monad, described later in subsection 4.4) producing an element of $A$. To make this clear, we give an interpretation written in the style of a monadic program in Haskell syntax in figure 6.

For example, function application $e_1\, e_2$ exhibits a right-to-left evaluation order: we first evaluate $e_2$ (with environment $\gamma$) to an argument $a$, then evaluate $e_1$ (with environment $\gamma$) to a function $f$, and then apply the argument to the function. The $e_1 . \mathsf{print}(e_2)$ method evaluates $e_1$ to a channel $c$, $e_2$ to a string $s$, and then represents its effect using the writer monad: it returns a map saying that $s$ was printed to the channel $c$. The interpretation of box $\boxed{e}$ is perhaps the most interesting – it interprets $e$ in a context where all capability-bearing bindings are discarded. As a result, even though $e$ is a monadic term, we know that it could not have written to any channels, and so we can then discard (using `fst`) the writer monad's output component without losing any information.

However, while writing the semantics as a naive set-theoretic semantics makes it easy to read, we still have to check that this definition actually does define a genuine weight-preserving morphism

$$\left[\!\left[ \frac{}{\Gamma \vdash () : \mathsf{unit}} \right]\!\right] \gamma := \texttt{return ()} \qquad \left[\!\left[ \frac{}{\Gamma \vdash s : \mathsf{str}} \right]\!\right] \gamma := \texttt{return s}$$

$$\left[\!\left[ \frac{\Gamma \vdash e_1 : A \qquad \Gamma \vdash e_2 : B}{\Gamma \vdash (e_1, e_2) : A \times B} \right]\!\right] \gamma := \begin{array}{l} \texttt{do g} \leftarrow [\![\Gamma \vdash e_2 : B]\!]\gamma \\ \quad \texttt{f} \leftarrow [\![\Gamma \vdash e_1 : A]\!]\gamma \\ \quad \texttt{return (f,g)} \end{array}$$

$$\left[\!\left[ \frac{\Gamma \vdash e : A \times B}{\Gamma \vdash \mathsf{fst}\, e : A} \right]\!\right] \gamma := \begin{array}{l} \texttt{do f} \leftarrow [\![\Gamma \vdash e : A \times B]\!]\gamma \\ \quad \texttt{return (fst f)} \end{array}$$

$$\left[\!\left[ \frac{\Gamma \vdash e : A \times B}{\Gamma \vdash \mathsf{snd}\, e : B} \right]\!\right] \gamma := \begin{array}{l} \texttt{do f} \leftarrow [\![\Gamma \vdash e : A \times B]\!]\gamma \\ \quad \texttt{return (snd f)} \end{array}$$

$$\left[\!\left[ \frac{x : A^q \in \Gamma}{\Gamma \vdash x : A} \right]\!\right] \gamma := \texttt{return } (\gamma\ \texttt{x})$$

$$\left[\!\left[ \frac{\Gamma, x : A^{\mathsf{i}} \vdash e : B}{\Gamma \vdash \lambda x : A.\, e : A \Rightarrow B} \right]\!\right] \gamma := \texttt{return (fun a} \to [\![\Gamma, x : A^{\mathsf{i}} \vdash e : B]\!]\ (\gamma, \texttt{a}))$$

$$\left[\!\left[ \frac{\Gamma \vdash e_1 : A \Rightarrow B \qquad \Gamma \vdash e_2 : A}{\Gamma \vdash e_1\, e_2 : B} \right]\!\right] \gamma := \begin{array}{l} \texttt{do a} \leftarrow [\![\Gamma \vdash e_2 : A]\!]\gamma \\ \quad \texttt{f} \leftarrow [\![\Gamma \vdash e_1 : A \Rightarrow B]\!]\gamma \\ \quad \texttt{f a} \end{array}$$

$$\left[\!\left[ \frac{\Gamma \vdash e_1 : \mathsf{cap} \qquad \Gamma \vdash e_2 : \mathsf{str}}{\Gamma \vdash e_1.\mathsf{print}(e_2) : \mathsf{unit}} \right]\!\right] \gamma := \begin{array}{l} \texttt{do s} \leftarrow [\![\Gamma \vdash e_2 : \mathsf{str}]\!]\gamma \\ \quad \texttt{c} \leftarrow [\![\Gamma \vdash e_1 : \mathsf{cap}]\!]\gamma \\ \quad \texttt{((),fun c'} \to \texttt{if c = c'} \\ \qquad\qquad\qquad \texttt{then s else } \epsilon) \end{array}$$

$$\left[\!\left[ \frac{\Gamma^{\mathsf{s}} \vdash e : A}{\Gamma \vdash \mathsf{box}\, \boxed{e} : \Box A} \right]\!\right] \gamma := \texttt{return (fst } ([\![\Gamma^{\mathsf{s}} \vdash e : A]\!]\gamma^{\mathsf{s}})) \texttt{ -- pure}$$

$$\left[\!\left[ \frac{\Gamma \vdash e_1 : \Box A \qquad \Gamma, x : A^{\mathsf{s}} \vdash e_2 : B}{\Gamma \vdash \mathsf{let\ box}\, \boxed{x} = e_1 \,\mathsf{in}\, e_2 : B} \right]\!\right] \gamma := \begin{array}{l} \texttt{do a} \leftarrow [\![\Gamma \vdash e_1 : \Box A]\!]\gamma \\ \quad [\![\Gamma, x : A^{\mathsf{s}} \vdash e_2 : B]\!]\ (\gamma,\ \texttt{a}) \end{array}$$

Fig. 6. Direct interpretation of expressions

between capability spaces. As the interpretation of box $\boxed{e}$ makes clear, this is not a trivial fact. Indeed, even though this semantics is in fact capability-safe, checking that is an incredibly tedious and error-prone affair – we have to go through every semantic clause and check not just that each and every operation we use is weight-preserving, but that all their compositions are weight-preserving as well.

To manage and organize this work more efficiently, we turn to a categorical semantics. In the categorical semantics, each type is an object, and each type constructor is interpreted as a functor with operators satisfying some universal properties. This way, we can check that the interpretation of each type connective works the way we want in isolation, without having to worry about any interactions with the rest of the calculus. Furthermore, the universal properties make it easy to check that our language satisfies the equational theory that we desire.

Another important benefit is that by formulating the semantics in a categorical style, the semantics and equational theory only depend upon the algebraic structure of the category of capability spaces. That is, we use the *cartesian closed* structure, the *monoidal idempotent comonad*, the *strong* *monad*, and the *cancellation isomorphism* $\Phi$; the proofs of our theorems use the universal property for each categorical construction. Indeed, our semantics is nearly independent of the specific set of effects – we only use the specific definition of the monad in the interpretation of print. Since our theorems depend only upon the algebraic structure, our results will still hold if we switched to another category with this structure. We say more about that in section 8.

We describe this categorical structure of capability spaces in the remainder of this section, and then give the categorical interpretation (which is actually semantically identical to the direct interpretation) in the following section.

### 4.3 Cartesian Closed Structure

We observe that $\mathcal{C}$ inherits the *cartesian closed* structure of Set. The definitions are the same as in the case of sets, but we additionally have to verify that the morphisms are weight-preserving.

*Definition 4.2 (Terminal Object).*

$$\begin{aligned} |1| &:= \{ * \} \\ w_1 &:= \left\{ (*, \varnothing) \right\} \end{aligned}$$

The terminal object 1 is the usual singleton set, and it has no capabilities. For any object $A$, the unique terminal map $! : A \longrightarrow 1$ is given by $!_A(a) = *$, which is evidently weight preserving.

*Definition 4.3 (Product).*

$$\begin{aligned} |A \times B| &:= |A| \times |B| \\ w_{A\times B} &:= \left\{ ((a, b), C_a \cup C_b) \;\middle|\; w_A(a, C_a) \wedge w_B(b, C_b) \right\} \end{aligned}$$

Products are formed by pairing as usual, and the set of capabilities of a pair of values is the union of their capabilities. The projection maps $\pi_i : A_1 \times A_2 \longrightarrow A_i$ are just the projections on the underlying sets, which are weight preserving as well. We verify the universal property in lemma B.1 in the appendix.

*Definition 4.4 (Exponential).*

$$\begin{aligned} |A \longrightarrow B| &:= |A| \longrightarrow |B| \\ w_{A\longrightarrow B} &:= \left\{ (f, C_f) \;\middle|\; \begin{array}{l} \forall a, C_a, w_A(a, C_a) \Rightarrow \\ \exists C_b \subseteq C_f \cup C_a, w_B(f(a), C_b) \end{array} \right\} \end{aligned}$$

Exponentials are given by functions on the underlying sets, but we have to assign capabilities to the closure. We only record those capabilities which are induced by the function, for some value in the domain. That is, for a function closure $f : A \longrightarrow B$, if a given value $a \in A$ has weight assignment $C_a$, and if there is a weight assignment $C_b$ for $f(a)$, then the weight of the closure $f$ is given by all the capabilities it had access to in its environment.

We verify that our definition satisfies the currying isomorphism in lemma B.2 in the appendix, where we name the currying/uncurrying and evaluation maps.

This cartesian closed structure on $\mathcal{C}$ suffices to interpret the simply-typed lambda calculus. To illustrate the semantics, we give some examples of closed terms with their unique capability weightings in figure 7.

| Expression | Type | Weight |
|---|---|---|
| `unit` | `Unit` | $\varnothing$ |
| `stdout` | `Channel` | { `stdout` } |
| `fun c → unit` | `Channel → Unit` | $\varnothing$ |
| `fun c → c` | `Channel → Channel` | $\varnothing$ |
| `fun c → c.print("hello")` | `Channel → Unit` | $\varnothing$ |
| `fun c → stdout.print("hello")` | `Channel → Unit` | { `stdout` } |
| `(c₁,c₂)` | `Channel × Channel` | { $c_1$, $c_2$ } |
| `[stdout,c₁,c₂]` | `List Channel` | { `stdout`, $c_1$, $c_2$ } |

Fig. 7. Expressions and their capability weights

## 4.4 Monad

Our language supports printing strings along a channel, and to model this print effect, we will structure our semantics monadically, in the style of Moggi [1991]. We define a strong monad $T$ on $\mathcal{C}$ as follows.

*Definition 4.5 ($\Sigma^* : \mathcal{C}$).* $\Sigma^*$ is the set of strings, with an empty string $\epsilon : 1 \longrightarrow \Sigma^*$, and a multiplication $\cdot : \Sigma^* \times \Sigma^* \longrightarrow \Sigma^*$ given by concatenation, making it a monoid object. Strings are constants and hence do not have any weights.

*Definition 4.6 ($C : \mathcal{C}$).* $C$ is the object of capabilities in $\mathcal{C}$ such that $w_C = \left\{ (c, \{ c \}) \right\}$ for every $c \in C$. Note that there are no global sections for this object, because maps $1 \longrightarrow C$ *are not weight-preserving*. In other words, we do not have access to arbitrary capabilities, as evident by the lack of an introduction rule for the cap type. This indicates the lack of ambient authority.

*Definition 4.7 ($T : \mathcal{C} \longrightarrow \mathcal{C}$).*

$$\begin{aligned} |T(A)| &:= |A| \times (C \to \Sigma^*) \\ w_{T(A)} &:= \left\{ ((a, o), C_a \cup \left\{ c \;\middle|\; o(c) \neq \epsilon \right\}) \;\middle|\; w_A(a, C_a) \right\} \end{aligned}$$

Using the monoid $(\Sigma^*; \epsilon, \cdot)$, we can define $T$ to be the writer monad which adds an output function that records the output produced in each channel. The weight of a monadic computation is taken to be the weight of the returned value, unioned with all the channels that *anything* was written to. This corresponds to the intuition that a computation which performs I/O on a channel must possess the capability to do so.

*Definition 4.8 ($T$ is a monad).* The unit and multiplication of the monad are defined below. We check that they are morphisms, and state and verify the monad laws in lemma B.3 in the appendix.

$$\begin{array}{rclrcl} \eta_A : A & \longrightarrow & TA & \mu_A : TTA & \longrightarrow & TA \\ a & \mapsto & (a, \lambda c.\epsilon) & ((a, o_1), o_2) & \mapsto & (a, \lambda c.o_2(c) \cdot o_1(c)) \end{array}$$

*Definition 4.9 ($T$ is a strong monad).* $T$ is strong with respect to products, with a natural family of left and right strengthening maps.

$$\begin{array}{rclrcl} \tau_{A,B} : A \times TB & \longrightarrow & T(A \times B) & \sigma_{A,B} : TA \times B & \longrightarrow & T(A \times B) \\ (a, (b, o)) & \mapsto & ((a, b), o) & ((a, o), b) & \mapsto & ((a, b), o) \end{array}$$

We use this to define the natural map $\beta_{A,B}$, which evaluates a pair of effects, as follows. Notice that it evaluates the effect on the right before the one on the left; we expand more on that in lemma B.4 in the appendix, and verify the appropriate coherences.

$$\begin{aligned}\beta_{A,B} &: TA \times TB \to T(A \times B) \\ \beta_{A,B} &:= \tau_{TA,B} \mathbin{;} T\sigma_{A,B} \mathbin{;} \mu_{A\times B}\end{aligned}$$

### 4.5 Comonad

To model the □ type constructor, we define an endofunctor $\square$ on $\mathcal{C}$ below; it keeps values that *do not* possess any capabilities, i.e., values that are *safe*.

*Definition 4.10* ($\square : \mathcal{C} \longrightarrow \mathcal{C}$).

$$\begin{aligned}|\square A| &:= \left\{ a \in |A| \;\middle|\; \forall C_A, w_A(a, C_a) \Rightarrow C_a = \varnothing \right\} \\ w_{\square A} &:= \{ (a, \varnothing) \}\end{aligned}$$

On objects, we simply restrict the set to the subset of values that *only* have the empty set $\varnothing$ of capabilities. $\square$ acts on morphisms by restricting the domain of the function to $|\square A|$. For any weight-preserving function $f$, we see that $\square(f)$ is trivially weight-preserving, as a function between sets with empty capabilities.

This type constructor is especially useful at function type $\square(A \to B)$, since in general the environment can hold capabilities, and the $\square$ constructor lets us rule those out. We further claim that $\square$ is an idempotent strong monoidal comonad.

*Definition 4.11* ($\square$ *is an idempotent comonad*). The counit $\epsilon$ and comultiplication $\delta$ of the comonad are the natural families of maps given by the inclusion and the identity maps on the underlying set. $\delta$ is a natural isomorphism making it idempotent. We state and verify the comonad laws in lemma B.5 in the appendix.

$$\begin{array}{ccc} \epsilon_A : \square A \to A & & \delta_A : \square A \xrightarrow{\sim} \square\square A \\ a \mapsto a & & a \mapsto a \end{array}$$

*Definition 4.12* ($\square$ *is a strong monoidal functor*). The functor is strong monoidal, in that it preserves the monoidal structure of products (and tensors, see the sequel in subsection 4.7). The identity element is preserved, and we have *natural isomorphisms* given by pairing on the underlying sets.

$$\begin{array}{ccc} m^1 : 1 \xrightarrow{\sim} \square 1 & & m^{\times}_{A,B} : (\square A \times \square B) \xrightarrow{\sim} \square(A \times B) \\ * \mapsto * & & (a, b) \mapsto (a, b) \end{array}$$

$$\begin{array}{ccc} m^I : I \xrightarrow{\sim} \square I & & m^{\otimes}_{A,B} : (\square A \otimes \square B) \xrightarrow{\sim} \square(A \otimes B) \\ * \mapsto * & & (a, b) \mapsto (a, b) \end{array}$$

We remark that $\square$ is not a strong comonad, i.e., it does not possess a tensorial strength. This makes it impossible to evaluate an arbitrary function under the comonad, as we saw in section 2. [7]

### 4.6 The Comonad Cancels the Monad

We make the following observation. There is an isomorphism $\Phi_A$, natural in $A$, where the comonad $\square$ cancels the monad $T$. In programming terms, this says that *an effectful computation with no capabilities can perform no effects* — i.e., it is *safe*. Note that this definition works because of the particular definition of the monad $T$ we chose, in which the weight of a computation includes all the channels it printed on. Consequently a computation of weight zero cannot print on any channel, and so must be *safe*! We verify this fact in lemma B.6 in the appendix.

[7] For Haskellers, the $\square$ functor is not a `Functor`!

*Definition 4.13* ($\Phi : \Box T \Rightarrow \Box$).

$$\begin{aligned} \Phi_A : \Box TA &\xrightarrow{\sim} \Box A \\ (a, o) &\mapsto a \end{aligned}$$

This property is crucial and we will exploit it to manage our syntax: we use it to justify treating terms in *safe* contexts as *safe*, without needing a second grammar for *safe* expressions.

## 4.7 Remarks

While the monad and comonad, together with the cartesian closed structure, suffice to interpret our language, it is worth noting that the category $\mathbb{C}$ also admits a *monoidal closed* structure. Particularly, the cartesian closed structure only required a unique assignments of weights for each value, but we chose a weight relation to make the monoidal closed structure work.

### 4.7.1 Monoidal Closed Structure.

*Definition 4.14 (Tensor product).*

$$\begin{aligned} |A \otimes B| &:= |A| \times |B| \\ w_{A \otimes B} &:= \left\{ ((a, b), C_a \cup C_b) \;\middle|\; C_a \sharp C_b \wedge w_A(a, C_a) \wedge w_B(b, C_b) \right\} \\ I &:= 1 \end{aligned}$$

The tensor product is given by pairing, with unit 1, but it only restricts to pairs whose sets of capabilities are disjoint. However, this tensor product also enjoys a right adjoint.

*Definition 4.15 (Linear exponential).*

$$\begin{aligned} |A \multimap B| &:= |A| \to |B| \\ w_{A \multimap B} &:= \left\{ (f, C_f) \;\middle|\; \begin{array}{l} \forall a, C_a, w_A(a, C_a) \wedge C_f \sharp C_a \Rightarrow \\ \exists C_b \subseteq C_f \cup C_a, w_B(f(a), C_b) \end{array} \right\} \end{aligned}$$

The linear exponential works the same way as the exponential, except that we have to restrict it to satisfy the disjointness condition for the tensor product. We verify that this definition satisfies the tensor-hom adjunction in lemma B.7 in the appendix.

This supports an interpretation of a *linear* (actually, affine) type theory. The disjointness conditions in the interpretation of tensor product and linear implication are essentially the same as the disjointness conditions in the definition of the separating conjunction $A * B$ and magic wand $A \mathrel{-\!\!*} B$ in separation logic [Reynolds 2002]. In separation logic, capabilities correspond to ownership of particular memory locations, and in our setting, capabilities correspond to the right to access a channel.

Our model reassuringly suggests that operating systems researchers and program verification researchers both identified the same notion of capability. However, it seems that the fact that these are *exactly* the same idea was overlooked because operating systems researchers focused on the cartesian closed structure, and semanticists focused on the monoidal closed structure!

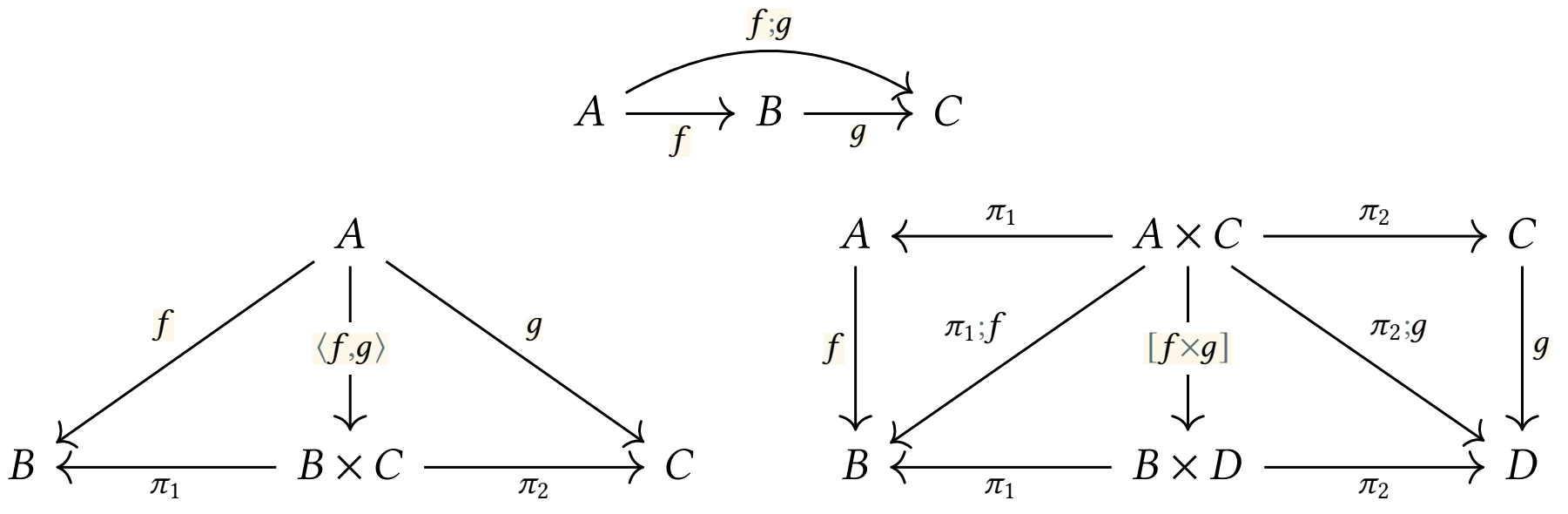

Fig. 8. Composition operations

$$\begin{aligned} [\![\mathsf{unit}]\!] &:= 1 & [\![A \times B]\!] &:= [\![A]\!] \times [\![B]\!] \\ [\![\mathsf{str}]\!] &:= \Sigma^* & [\![A \Rightarrow B]\!] &:= [\![A]\!] \to T[\![B]\!] \\ [\![\mathsf{cap}]\!] &:= \mathcal{C} & [\![\Box A]\!] &:= \Box[\![A]\!] \end{aligned}$$

(a) $[\![A]\!] : \mathcal{O}bj_{\mathcal{C}}$

$$\begin{aligned} [\![\,]\!] &:= 1 \\ [\![\Gamma, x : A^{\mathsf{s}}]\!] &:= [\![\Gamma]\!] \times \Box[\![A]\!] \\ [\![\Gamma, x : A^{\mathsf{i}}]\!] &:= [\![\Gamma]\!] \times [\![A]\!] \end{aligned}$$

(b) $[\![\Gamma]\!] : \mathcal{O}bj_{\mathcal{C}}$

Fig. 9. Interpretation of types and contexts

*4.7.2 Adding Other Effects.* While we used the writer monad for print, we can also define other interesting monads using the capability space model which can be used to interpret a language with other effects. For example, we show how to define an exception monad which allows raising a single exception, and a state monad with a global heap, in appendices B.1 and B.2. For each of these monads, we need to choose a suitable weight assignment, all of which can be cancelled by our *safety* comonad!

## 5 INTERPRETATION

We now interpret the syntax of our language. We adopt some standard notation to work with our categorical combinators. [8] The sequential composition of two arrows, in the diagrammatic order, is $f \mathbin{;} g$. The product of morphisms $f$ and $g$ is $\langle f, g \rangle$ (also called a fork operation in the algebra of programming community [Gibbons 2000]), and $[f \times g]$ is parallel composition with products. We define these using the universal property of products and composition, as shown in figure 8.

### 5.1 Types and Contexts

We interpret types as objects in $\mathbb{C}$, as shown in figure 9a. Note that we use the monad in the interpretation of functions, following the call-by-value computational lambda-calculus interpretation in [Moggi 1989]. We use the comonad to interpret the $\Box$ modality. We use the particular objects $\Sigma^*$ and $\mathcal{C}$ to interpret strings and capabilities respectively.

We interpret contexts as finite products of objects, in figure 9b. The comonad is used to interpret the *safe* variables in the context, while the *impure* variables are just arbitrary objects in $\mathbb{C}$.

[8] We sometimes drop the denotation symbol for brevity, i.e., we write $!_\Gamma$ instead of $!_{[\![\Gamma]\!]}$, or $\delta_{\Gamma^{\mathsf{s}}}$ instead of $\delta_{[\![\Gamma^{\mathsf{s}}]\!]}$.

The judgement $x : A^q \in \Gamma$ is interpreted as a morphism in $\mathcal{H}om_{\mathcal{C}}(\llbracket \Gamma \rrbracket, \llbracket A \rrbracket)$, which we give later in figure 12a. It projects out the appropriately typed and annotated variable from the product in the context. For *safe* variables, we need to use the counit $\epsilon$ to get out of the comonad. [9]

## 5.2 Expressions

We now give an interpretation for expressions $\Gamma \vdash e : A$, and an interpretation for *safe* expressions $\Gamma \vdash^{s} e : A$, in figure 10.

To interpret unitI, we use the terminal map ! to simply get to the terminal object 1, then lift it into the monad using $\eta$, without performing any effects. We do the same for strI, where we use $\ulcorner s \urcorner : \mathcal{H}om_{\mathcal{C}}(1, \Sigma^*)$, which is the global element that picks the literal $s$ in $\Sigma^*$.

For pair introduction $\times$I, we evaluate both components of the pair, and compose, then use the strength of the monad $T$ with the $\beta$ combinator to form the product. [10]

We eliminate products using the $\times E_1$ and $\times E_2$ rules. These are interpreted using the corresponding product projection maps, under the functorial action of $T$.

Variables are introduced using the Var rule, which is interpreted by looking up in the context, for which we use the interpretation of our context membership judgement. This is followed by a trivial lifting into the monad.

To interpret functions using the $\Rightarrow$ I rule, we simply use the currying map, since our context extension is interpreted as a product. Then we lift it into the monad using $\eta$.

To eliminate functions using the $\Rightarrow$ E rule, we evaluate the operator and operand in an application, followed by a use of the monad strength $\beta$ to turn it into a pair. Then we use the evaluation map under the functor $T$ to apply the argument. Since the function is effectful, we have to collapse the effects using a $\mu$.

To interpret the $\square$I rule, we need to interpret the *safe* judgement (defined later), which gives a value of type $\square A$, and then we lift it into the monad.

To eliminate a box-ed value using the $\square$E rule, we first evaluate $f$, which gives a value of type $\square A$, but under the monad $T$. We can use it to introduce a *safe* variable in the context, but we use the strength of the monad to shift the product under the $T$ and get an extended context. We evaluate $g$ under this extended context, and then use a $\mu$ to collapse the effects.

Finally, to interpret the Print rule, we need to perform a non-trivial effect. We define the function $p$ which builds an output function that records the output on channels. Given any channel $c$ and string $s$, it returns a value of type $T1$ containing the trivial value $*$; the output function instantiates a channel $c'$ and tests equality with $c$ – if it equals $c$, we record the string $s$, otherwise we just choose the empty string $\epsilon$. We interpret the arguments of print and apply them to $p$ to evaluate it. The rest of the interpretation is similar to the one for $\Rightarrow$ E, with output type 1.

We used a different interpretation function for *safe* expressions, which we define below.

We first need to interpret the *purify* operation $\mathbf{s}$ on contexts, for which we define the map $\rho(\Gamma)$ in figure 11a. We also need another combinator $\mathcal{M}(\Gamma)$, defined in figure 11b, which uses the monoidal action and the comultiplication of the comonad $\square$ to distribute the $\square$ over the products in $\Gamma$. Note that $\mathcal{M}(\Gamma)$ is an isomorphism because $m$ and $\delta$ are.

Now, the interpretation function for *safe* expressions $\Gamma \vdash^{s} e : A$ uses the ctx-safe rule, and is defined as a morphism in $\mathcal{H}om_{\mathcal{C}}(\llbracket \Gamma \rrbracket, \square\llbracket A \rrbracket)$. We *purify* the context to a *safe* one, so that we can

[9]When interpreting judgements and inference rules, we write $\llbracket \frac{\mathcal{J}_1 \dots \mathcal{J}_n}{\mathcal{J}} \rrbracket$ to mean the interpretation of $\mathcal{J}$, i.e., we recursively define $\llbracket \mathcal{J} \rrbracket$ under the assumption that we have an interpretation for $\mathcal{J}_i$, i.e., $\llbracket \mathcal{J}_1 \rrbracket, \dots, \llbracket \mathcal{J}_n \rrbracket$.

[10]The vigilant reader will have noticed that $\beta$ evaluates the pair from right to left, so the action on the right will be performed first, like OCaml! This is also useful when interpreting function application, because we evaluate the argument first.

$$\left\llbracket \frac{}{\Gamma \vdash () : \mathsf{unit}} \right\rrbracket := !_\Gamma \,;\, \eta_1 \qquad \left\llbracket \frac{}{\Gamma \vdash s : \mathsf{str}} \right\rrbracket := !_\Gamma \,;\, \ulcorner s \urcorner \,;\, \eta_{\Sigma^*}$$

$$\left\llbracket \frac{\Gamma \vdash e_1 : A \qquad \Gamma \vdash e_2 : B}{\Gamma \vdash (e_1\,,e_2) : A \times B} \right\rrbracket := \begin{array}{l} \mathit{let} \begin{cases} f := \llbracket \Gamma \vdash e_1 : A \rrbracket \\ g := \llbracket \Gamma \vdash e_2 : B \rrbracket \end{cases} \\ \mathit{in} \quad \langle f\,,g \rangle \,;\, \beta_{A,B} \end{array}$$

$$\left\llbracket \frac{\Gamma \vdash e : A \times B}{\Gamma \vdash \mathsf{fst}\, e : A} \right\rrbracket := \llbracket \Gamma \vdash e : A \times B \rrbracket \,;\, T\pi_1 \qquad \left\llbracket \frac{\Gamma \vdash e : A \times B}{\Gamma \vdash \mathsf{snd}\, e : B} \right\rrbracket := \llbracket \Gamma \vdash e : A \times B \rrbracket \,;\, T\pi_2$$

$$\left\llbracket \frac{x : A^q \in \Gamma}{\Gamma \vdash x : A} \right\rrbracket := \llbracket x : A^q \in \Gamma \rrbracket \,;\, \eta_A$$

$$\left\llbracket \frac{\Gamma, x : A^{\mathsf{i}} \vdash e : B}{\Gamma \vdash \lambda x : A.\, e : A \Rightarrow B} \right\rrbracket := \mathsf{curry}\,(\llbracket \Gamma, x : A^{\mathsf{i}} \vdash e : B \rrbracket) \,;\, \eta_{A \to TB}$$

$$\left\llbracket \frac{\Gamma \vdash e_1 : A \Rightarrow B \qquad \Gamma \vdash e_2 : A}{\Gamma \vdash e_1\, e_2 : B} \right\rrbracket := \begin{array}{l} \mathit{let} \begin{cases} f := \llbracket \Gamma \vdash e_1 : A \Rightarrow B \rrbracket \\ g := \llbracket \Gamma \vdash e_2 : A \rrbracket \end{cases} \\ \mathit{in} \quad \langle f\,,g \rangle \,;\, \beta_{A \to TB, A} \,;\, T\,\mathsf{ev}_{A,TB} \,;\, \mu_B \end{array}$$

$$\left\llbracket \frac{\Gamma \vdash e_1 : \mathsf{cap} \qquad \Gamma \vdash e_2 : \mathsf{str}}{\Gamma \vdash e_1 . \mathsf{print}(e_2) : \mathsf{unit}} \right\rrbracket := \begin{array}{l} \mathit{let} \begin{cases} f := \llbracket \Gamma \vdash e_1 : \mathsf{cap} \rrbracket \\ g := \llbracket \Gamma \vdash e_2 : \mathsf{str} \rrbracket \\ p : \mathcal{C} \times \Sigma^* \to T1 \\ (c, s) \mapsto \left( *\,,\ \lambda c'. \begin{cases} s & \text{if } c = c' \\ \epsilon & \text{otherwise} \end{cases} \right) \end{cases} \\ \mathit{in} \quad \langle f\,,g \rangle \,;\, \beta_{\mathcal{C},\Sigma^*} \,;\, Tp \,;\, \mu_1 \end{array}$$

$$\left\llbracket \frac{\Gamma \vdash^{\mathsf{s}} e : A}{\Gamma \vdash \mathsf{box}\, \boxed{e} : \Box A} \right\rrbracket := \llbracket \Gamma \vdash^{\mathsf{s}} e : A \rrbracket_p \,;\, \eta_{\Box A}$$

$$\left\llbracket \frac{\Gamma^{\mathsf{s}} \vdash e : A}{\Gamma \vdash^{\mathsf{s}} e : A} \right\rrbracket_p := \rho(\Gamma) \,;\, \mathcal{M}(\Gamma) \,;\, \Box\llbracket \Gamma^{\mathsf{s}} \vdash e : A \rrbracket \,;\, \Phi_A$$

$$\left\llbracket \frac{\Gamma \vdash e_1 : \Box A \qquad \Gamma, x : A^{\mathsf{s}} \vdash e_2 : B}{\Gamma \vdash \mathsf{let\ box}\, \boxed{x} = e_1\, \mathsf{in}\, e_2 : B} \right\rrbracket := \begin{array}{l} \mathit{let} \begin{cases} f := \llbracket \Gamma \vdash e_1 : \Box A \rrbracket \\ g := \llbracket \Gamma, x : A^{\mathsf{s}} \vdash e_2 : B \rrbracket \end{cases} \\ \mathit{in} \quad \langle \mathit{id}_\Gamma\,, f \rangle \,;\, \tau_{\Gamma,\Box A} \,;\, Tg \,;\, \mu_B \end{array}$$

Fig. 10. Interpretation of expressions, $\llbracket \Gamma \vdash e : A \rrbracket : \mathcal{H}om_{\mathcal{C}}\,(\llbracket \Gamma \rrbracket\,, T\llbracket A \rrbracket)$, $\llbracket \Gamma \vdash^{\mathsf{s}} e : A \rrbracket_p : \mathcal{H}om_{\mathcal{C}}\,(\llbracket \Gamma \rrbracket\,, \Box\llbracket A \rrbracket)$

$$\begin{aligned} \rho() &:= \mathit{id}_1 \\ \rho(\Gamma, x : A^{\mathsf{s}}) &:= [\rho(\Gamma) \times \mathit{id}_{\Box A}] \\ \rho(\Gamma, x : A^{\mathsf{i}}) &:= \pi_1 \,;\, \rho(\Gamma) \end{aligned}$$

(a) $\rho(\Gamma) : \mathcal{H}om_{\mathcal{C}}\,(\llbracket \Gamma \rrbracket\,, \llbracket \Gamma^{\mathsf{s}} \rrbracket)$

$$\begin{aligned} \mathcal{M}() &:= \mathit{id}_1 \\ \mathcal{M}(\Gamma, x : A^{\mathsf{s}}) &:= [\mathcal{M}(\Gamma) \times \delta_A] \,;\, m^{\times}_{\Gamma^{\mathsf{s}},\Box A} \\ \mathcal{M}(\Gamma, x : A^{\mathsf{i}}) &:= \mathcal{M}(\Gamma) \end{aligned}$$

(b) $\mathcal{M}(\Gamma) : \mathcal{H}om_{\mathcal{C}}\,(\llbracket \Gamma^{\mathsf{s}} \rrbracket\,, \Box\llbracket \Gamma^{\mathsf{s}} \rrbracket)$

Fig. 11. $\rho(\Gamma)$ and $\mathcal{M}(\Gamma)$

$$\left[\!\left[ \frac{}{x : A^{\mathsf{i}} \in (\Gamma, x : A^{\mathsf{i}})} \right]\!\right] := \pi_2$$

$$\left[\!\left[ \frac{}{x : A^{\mathsf{s}} \in (\Gamma, x : A^{\mathsf{s}})} \right]\!\right] := \pi_2 \,;\, \epsilon_A$$

$$\left[\!\left[ \frac{x : A^{q} \in \Gamma \qquad (x \neq y)}{x : A^{q} \in (\Gamma, y : B^{r})} \right]\!\right] := \pi_1 \,;\, [\![ x : A^{q} \in \Gamma ]\!]$$

(a) $[\![ x : A^{q} \in \Gamma ]\!] : \mathcal{H}om_{\mathcal{C}}([\![\Gamma]\!], [\![A]\!])$

$$\left[\!\left[ \frac{}{\supseteq} \right]\!\right] := id_1$$

$$\left[\!\left[ \frac{\Gamma \supseteq \Delta}{\Gamma, x : A^{q} \supseteq \Delta} \right]\!\right] := \pi_1 \,;\, [\![\Gamma \supseteq \Delta]\!]$$

$$\left[\!\left[ \frac{\Gamma \supseteq \Delta}{\Gamma, x : A^{\mathsf{s}} \supseteq \Delta, x : A^{\mathsf{s}}} \right]\!\right] := [[\![\Gamma \supseteq \Delta]\!] \times id_{\Box A}]$$

$$\left[\!\left[ \frac{\Gamma \supseteq \Delta}{\Gamma, x : A^{\mathsf{i}} \supseteq \Delta, x : A^{\mathsf{i}}} \right]\!\right] := [[\![\Gamma \supseteq \Delta]\!] \times id_{A}]$$

(b) $\mathsf{Wk}(\Gamma \supseteq \Delta) := [\![\Gamma \supseteq \Delta]\!] : \mathcal{H}om_{\mathcal{C}}([\![\Gamma]\!], [\![\Delta]\!])$

Fig. 12. Interpretation of Membership and Weakening

evaluate the expression. However, we need a value in $\Box A$, but the expression interpretation would produce something in $TA$. We can only cancel the monad under the comonad, so we use the $\mathcal{M}(\Gamma)$ map which uses the comultiplication of $\Box$ to do a readjustment. We then evaluate the expression under the $\Box$ in the *safe* context, which gives a monadic value of type $TA$ under the comonad $\Box$. We finally use $\Phi$ to cancel the monad $T$ under the $\Box$.

### 5.3 Weakening and Substitution

We now give semantics for the syntactic weakening and substitution operations.

*5.3.1 Weakening.* For contexts $\Gamma$ and $\Delta$, we interpret the weakening judgement $\Gamma \supseteq \Delta$ as a morphism in $\mathcal{H}om_{\mathcal{C}}([\![\Gamma]\!], [\![\Delta]\!])$, as shown in figure 12b. We also refer to it as the weakening map $\mathsf{Wk}(\Gamma \supseteq \Delta)$. We prove a semantic weakening lemma, analogous to the syntactic weakening lemma 3.1.

Lemma 5.1 (Semantic weakening). *If* $\Gamma \supseteq \Delta$ *and* $\Delta \vdash e : A$*, then*

$$[\![\Gamma \vdash e : A]\!] = \mathsf{Wk}(\Gamma \supseteq \Delta) \,;\, [\![\Delta \vdash e : A]\!].$$

*5.3.2 Substitution.* We now interpret a substitution $\Gamma \vdash \theta : \Delta$ as a morphism in $\mathcal{H}om_{\mathcal{C}}([\![\Gamma]\!], [\![\Delta]\!])$, as shown in figure 13b. However, this is not a trivial iteration of the expression interpretation. The reason is that the interpretation of contexts in figure 9b interprets a variable $x : A^{\mathsf{i}}$ in the context as an element of the type $[\![A]\!]$, and a variable $x : A^{\mathsf{s}}$ as an element of the type $\Box[\![A]\!]$. However, an expression $\Gamma \vdash e : A$ will be interpreted as a morphism in $\mathcal{H}om_{\mathcal{C}}([\![\Gamma]\!], T[\![A]\!])$. Operationally, we resolve this mismatch by only substituting *values* for variables in call-by-value languages, and indeed, our definition of substitutions in figure 5c restricts the definition of substitution to range over values in the rule sub-impure.

Therefore, we mimic this syntactic restriction in the semantics, by giving a separate interpretation only for values, interpreting the judgement $\Gamma \vdash v : A$ as a morphism in $\mathcal{H}om_{\mathcal{C}}([\![\Gamma]\!], [\![A]\!])$, in figure 13a. Note in particular that the value interpretation yields an element of $[\![A]\!]$, as the context interpretation requires, rather than an element of $T[\![A]\!]$. This value interpretation makes use of the expression interpretation in the interpretation of $\lambda$-expressions, but the expression interpretation does not directly refer to the value interpretation. There are alternative presentations such as fine-grain call-by-value [Levy et al. 2003], which have a separate syntactic class of values and value judgements, and hence make the value and expression interpretations mutually recursive. However, we choose not to do that in order to remain close to the usual presentation.

$$\left[\!\left[ \frac{}{\Gamma \vdash () : \mathsf{unit}} \right]\!\right]_v := !_\Gamma$$

$$\left[\!\left[ \frac{\Gamma \vdash v_1 : A \qquad \Gamma \vdash v_2 : B}{\Gamma \vdash (v_1 , v_2) : A \times B} \right]\!\right]_v := \langle [\![\Gamma \vdash v_1 : A]\!]_v , [\![\Gamma \vdash v_2 : B]\!]_v \rangle$$

$$\left[\!\left[ \frac{x : A^q \in \Gamma}{\Gamma \vdash x : A} \right]\!\right]_v := [\![x : A^q \in \Gamma]\!]$$

$$\left[\!\left[ \frac{\Gamma, x : A^{\mathsf{i}} \vdash e : B}{\Gamma \vdash \lambda x : A.\, e : A \Rightarrow B} \right]\!\right]_v := \mathsf{curry}\,([\![\Gamma, x : A^{\mathsf{i}} \vdash e : B]\!])$$

$$\left[\!\left[ \frac{\Gamma \vdash^{\mathsf{s}} e : A}{\Gamma \vdash \mathsf{box}\,\boxed{e} : \Box A} \right]\!\right]_v := [\![\Gamma \vdash^{\mathsf{s}} e : A]\!]_p$$

(a) $[\![\Gamma \vdash v : A]\!]_v : \mathcal{H}om_{\mathcal{C}}\,([\![\Gamma]\!] , [\![A]\!])$

$$\left[\!\left[ \frac{}{\Gamma \vdash \langle\rangle :} \right]\!\right] := !_\Gamma$$

$$\left[\!\left[ \frac{\Gamma \vdash \theta : \Delta \qquad \Gamma \vdash^{\mathsf{s}} e : A}{\Gamma \vdash \langle \theta, e^{\mathsf{s}}/x \rangle : \Delta, x : A^{\mathsf{s}}} \right]\!\right] := \langle [\![\Gamma \vdash \theta : \Delta]\!] , [\![\Gamma \vdash^{\mathsf{s}} e : A]\!]_p \rangle$$

$$\left[\!\left[ \frac{\Gamma \vdash \theta : \Delta \qquad \Gamma \vdash v : A}{\Gamma \vdash \langle \theta, v^{\mathsf{i}}/x \rangle : \Delta, x : A^{\mathsf{i}}} \right]\!\right] := \langle [\![\Gamma \vdash \theta : \Delta]\!] , [\![\Gamma \vdash v : A]\!]_v \rangle$$

(b) $[\![\Gamma \vdash \theta : \Delta]\!] : \mathcal{H}om_{\mathcal{C}}\,([\![\Gamma]\!] , [\![\Delta]\!])$

Fig. 13. Interpretation of values and substitution

Note that box $\boxed{e}$ expressions are also values, and our *safe* interpretation does the right thing for box values, since the interpretation of $\Box A$ uses the comonad, $\Box [\![A]\!]$. With the interpretation of values in hand, we can define the substitution interpretation as follows.

We use the *safe* expression interpretation to interpret the SUB-SAFE rule, and the *impure* value interpretation for the SUB-IMPURE rule.

Finally, we prove the semantic analogue of the syntactic substitution theorem 3.4. We prove two auxiliary lemmas 5.2 and 5.3, characterising the expression interpretation of *safe expressions* and *impure values*. The lemmas show that the interpretation for each ends in a trivial lifting into the monad $T$ using $\eta$. This makes the proof of the semantic substitution theorem 5.4 possible.

LEMMA 5.2 (SAFE INTERPRETATION). *If* $\Gamma \vdash^{\mathsf{s}} e : A$*, then*

$$[\![\Gamma \vdash e : A]\!] = [\![\Gamma \vdash^{\mathsf{s}} e : A]\!]_p \,;\, \epsilon_A \,;\, \eta_A.$$

LEMMA 5.3 (VALUE INTERPRETATION). *If* $\Gamma \vdash v : A$*, then*

$$[\![\Gamma \vdash v : A]\!] = [\![\Gamma \vdash v : A]\!]_v \,;\, \eta_A.$$

THEOREM 5.4 (SEMANTIC SUBSTITUTION). *If* $\Gamma \vdash \theta : \Delta$ *and* $\Delta \vdash e : A$*, then*

$$[\![\Gamma \vdash \theta(e) : A]\!] = [\![\Gamma \vdash \theta : \Delta]\!] \,;\, [\![\Delta \vdash e : A]\!].$$

$$
\begin{aligned}
\mathcal{C} \quad ::= \quad & [\,] \mid e\ \mathcal{C} \mid \mathcal{C}\ e \mid \lambda x : A.\ \mathcal{C} \\
\mid\ & \mathsf{fst}\ \mathcal{C} \mid \mathsf{snd}\ \mathcal{C} \mid (e\,,\mathcal{C}) \mid (\mathcal{C}\,, e) \\
\mid\ & \mathsf{box}\ \boxed{\mathcal{C}} \mid \mathsf{let\ box}\ \boxed{x} = \mathcal{C}\ \mathsf{in}\ e \mid \mathsf{let\ box}\ \boxed{x} = e\ \mathsf{in}\ \mathcal{C} \\
\mathcal{E} \quad ::= \quad & [\,] \mid e\ \mathcal{E} \mid \mathcal{E}\ v \\
\mid\ & \mathsf{fst}\ \mathcal{E} \mid \mathsf{snd}\ \mathcal{E} \mid (e\,,\mathcal{E}) \mid (\mathcal{E}\,, v) \\
\mid\ & \mathsf{let\ box}\ \boxed{x} = \mathcal{E}\ \mathsf{in}\ e \mid \mathsf{let\ box}\ \boxed{x} = v\ \mathsf{in}\ \mathcal{E}
\end{aligned}
$$

Fig. 14. Grammar extended with Evaluation Contexts

## 6 EQUATIONAL THEORY

We have an extension of the call-by-value simply-typed lambda calculus, so we want the usual $\beta\eta$-equations to hold in our theory. However, we also added new expression forms for the $\Box$ type. We want computation and extensionality rules for the box form and the let box binding form. To handle the commuting conversions [Girard et al. 1989], we use evaluation contexts.

We extend our grammar with two kinds of evaluation contexts — a *safe* evaluation context $\mathcal{C}$, and an *impure* evaluation context $\mathcal{E}$, as shown in figure 14. The intuition is that $\mathcal{E}$ allows safe reductions for *impure* expressions, i.e., it picks out the contexts consistent with the evaluation order of the call-by-value simply-typed lambda calculus. The *safe* evaluation context $\mathcal{C}$ allows redexes in every sub-expression; but it is restricted only to *safe* expressions. The hole [] is the empty evaluation context. We use the notation $\mathcal{C}\langle\!\langle e\rangle\!\rangle$ or $\mathcal{E}\langle\!\langle e\rangle\!\rangle$ to indicate that we're replacing the hole in the respective evaluation context with $e$.

We define a judgement form for equality of terms, as shown in figure 2c, and state the rules for the equational theory in figures 15 and 16. We have the usual REFL, SYM, and TRANS rules which give the reflexive, symmetric, and transitive closure, so that the equality relation is an equivalence, and the CONG rules for each term former, which make the relation a congruence closure.

We have the computation rules $\times_1\beta$ and $\times_2\beta$ for pairs; we only allow values for these rules. The $\times\eta$ rule is the extensionality rule for pairs, but again, restricted to values.

The $\Rightarrow\beta$ rule is the usual call-by-value computation rule for an application of a $\lambda$-expression to an argument. [11] Since the calculus has effects, we only allow the operand to be a value. For example, consider the function $f := \lambda x : \mathsf{unit}.\ x \mathbin{;} x$. We can safely $\beta$-reduce $f\ \ ()$ to $() \mathbin{;} ()$, but allowing a $\beta$-reduction for $f\ (c.\mathsf{print}(s))$ would duplicate the effect!

We add $\eta$ rules for functions, but we need to be careful because we have effects. For example, consider the expression $f := c.\mathsf{print}(s) \mathbin{;} \lambda x.\ x$. On $\eta$-expansion, we get $g := \lambda y.\ f\ y$, but now the print operation is suspended in the closure, and doesn't evaluate when we apply $g$. Hence, we add two forms of $\eta$ rules for functions — the $\Rightarrow\eta$-IMPURE rule only allows $\eta$-expansion for values, and the $\Rightarrow\eta$-SAFE rule allows $\eta$-expansion also for expressions that are *safe*.

The computation rule $\Box\beta$ for the $\Box$ type allows computation under the let box binder. If we bind a box-ed expression under the let box binder, we can substitute the underlying expression in the motive. This is safe because $e_1$ is forced to be a *safe* expression.

Finally, we have the $\eta$ expansion rules for the $\Box$ type, which pushes an expression in an evaluation context under a let box binder. The $\Box\eta$–*safe* rule uses the *safe* evaluation context $\mathcal{C}$, while the $\Box\eta$–*impure* rule uses the *impure* evaluation context $\mathcal{E}$. The only difference in the rules is that the $\mathcal{C}$ evaluation context can be plugged with *safe* expressions only.

We prove that our equality rules are sound with respect to our categorical semantics. If two expressions are equal in the equational theory, they have equal interpretations in the semantics.

---

[11] The notation $[v/x]e$ is shorthand for $\langle\langle\Gamma\rangle, v^{\mathsf{i}}/x\rangle(e)$ where $\langle\Gamma\rangle$ is the identity substitution $\Gamma \vdash \langle\Gamma\rangle : \Gamma$.

$$\frac{\Gamma \vdash e : A}{\Gamma \vdash e \approx e : A}\ \textsc{refl} \qquad \frac{\Gamma \vdash e_1 \approx e_2 : A}{\Gamma \vdash e_2 \approx e_1 : A}\ \textsc{sym} \qquad \frac{\Gamma \vdash e_1 \approx e_2 : A \qquad \Gamma \vdash e_2 \approx e_3 : A}{\Gamma \vdash e_1 \approx e_3 : A}\ \textsc{trans}$$

$$\frac{\Gamma \vdash e_1 \approx e_2 : A \times B}{\Gamma \vdash \mathsf{fst}\, e_1 \approx \mathsf{fst}\, e_2 : A}\ \mathsf{fst}\text{-}\textsc{cong} \qquad \frac{\Gamma \vdash e_1 \approx e_2 : A \times B}{\Gamma \vdash \mathsf{snd}\, e_1 \approx \mathsf{snd}\, e_2 : B}\ \mathsf{snd}\text{-}\textsc{cong}$$

$$\frac{\Gamma \vdash e_1 \approx e_2 : A \qquad \Gamma \vdash e_3 \approx e_4 : B}{\Gamma \vdash (e_1 , e_3) \approx (e_2 , e_4) : A \times B}\ \textsc{pair-cong} \qquad \frac{\Gamma, x : A^{\mathsf{i}} \vdash e_1 \approx e_2 : B}{\Gamma \vdash \lambda x : A.\, e_1 \approx \lambda x : A.\, e_2 : A \Rightarrow B}\ \lambda\text{-}\textsc{cong}$$

$$\frac{\Gamma \vdash e_1 \approx e_2 : A \Rightarrow B \qquad \Gamma \vdash e_3 \approx e_4 : A}{\Gamma \vdash e_1\, e_3 \approx e_2\, e_4 : B}\ \textsc{app-cong} \qquad \frac{\Gamma^{\mathsf{s}} \vdash e_1 \approx e_2 : A}{\Gamma \vdash \mathsf{box}\, \boxed{e_1} \approx \mathsf{box}\, \boxed{e_2} : \Box A}\ \mathsf{box}\text{-}\textsc{cong}$$

$$\frac{\Gamma \vdash e_1 \approx e_2 : \Box A \qquad \Gamma, x : A^{\mathsf{s}} \vdash e_3 \approx e_4 : B}{\Gamma \vdash (\mathsf{let\ box}\, \boxed{x} = e_1\, \mathsf{in}\, e_3) \approx (\mathsf{let\ box}\, \boxed{x} = e_2\, \mathsf{in}\, e_4) : B}\ \mathsf{let\ box}\text{-}\textsc{cong}$$

$$\frac{\Gamma \vdash e_1 \approx e_2 : \mathsf{cap} \qquad \Gamma \vdash e_3 \approx e_4 : \mathsf{str}}{\Gamma \vdash e_1 . \mathsf{print}(e_3) \approx e_2 . \mathsf{print}(e_4) : \mathsf{unit}}\ \mathsf{print}\text{-}\textsc{cong}$$

Fig. 15. Equivalence and Congruence rules for the Equational Theory

Theorem 6.1 (Soundness of $\approx$). *If* $\Gamma \vdash e_1 \approx e_2 : A$, *then* $[\![\Gamma \vdash e_1 : A]\!] = [\![\Gamma \vdash e_2 : A]\!]$.

## 7 EMBEDDING

Our language is an extension of the call-by-value simply-typed lambda calculus. But how could we claim that it is really an *extension*? In this section, we show that we can *embed* the simply-typed lambda calculus into our calculus, in an equation preserving way. We state the full simply-typed lambda calculus including its $\beta\eta$-equational theory in figure 17.

We give the grammar and judgements in figures 17a and 17b, typing rules in figure 17c, and the $\beta\eta$-equational theory in figure 17d. Note that we choose to use the base type unit, and we leave out products because their embedding is trivial and uninteresting for our purpose.

We define an embedding function from the simply-typed lambda calculus to our calculus. We use the notation $\underline{X}$ to denote the embedding of a syntactic object $X$ from STLC into our calculus. The syntactic translation of types, contexts, and raw terms is given in figure 18.

To embed the function type, we embed the domain and codomain, but we apply our comonadic type constructor $\Box$ to restrict the domain to a *safe* type. Remarkably, this embedding is quite like the Gödel-McKinsey-Tarski embedding of the intuitionistic propositional calculus into classical S4 modal logic, as outlined in [McKinsey and Tarski 1948], but we do not need to apply the $\Box$ type constructor on the codomain, because our functions are *capability-safe*. We note that this is also similar to the embedding of lax logic into S4 modal logic described in [Pfenning and Davies 2001], as well as the embedding of intuitionistic logic into linear logic [Girard 1987].

When embedding contexts, we mark the variables as *safe* using the $\mathsf{s}$ annotation. To embed functions and applications, we need to use the introduction and elimination forms for $\Box$. When embedding a $\lambda$-expression, the bound variable is embedded as a term of $\Box$ type, so we eliminate

$$\frac{\Gamma \vdash v_1 : A \qquad \Gamma \vdash v_2 : B}{\Gamma \vdash \mathsf{fst}\,(v_1\,,v_2) \approx v_1 : A}\times_1\beta \qquad \frac{\Gamma \vdash v_1 : A \qquad \Gamma \vdash v_2 : B}{\Gamma \vdash \mathsf{snd}\,(v_1\,,v_2) \approx v_2 : B}\times_2\beta$$

$$\frac{\Gamma \vdash v : A \times B}{\Gamma \vdash v \approx (\mathsf{fst}\,v\,,\mathsf{snd}\,v) : A \times B}\times\eta$$

$$\frac{\Gamma, x : A^{\mathsf{i}} \vdash e : B \qquad \Gamma \vdash v : A}{\Gamma \vdash (\lambda x : A.\ e)\ v \approx [v/x]e : B} \Rightarrow\beta$$

$$\frac{\Gamma \vdash v : A \Rightarrow B}{\Gamma \vdash v \approx \lambda x : A.\ v\ x : A \Rightarrow B} \Rightarrow\eta\text{-IMPURE} \qquad \frac{\Gamma \vdash^{\mathsf{s}} e : A \Rightarrow B}{\Gamma \vdash e \approx \lambda x : A.\ e\ x : A \Rightarrow B} \Rightarrow\eta\text{-SAFE}$$

$$\frac{\Gamma^{\mathsf{s}} \vdash e_1 : A \qquad \Gamma, x : A^{\mathsf{s}} \vdash e_2 : B}{\Gamma \vdash \mathsf{let\ box}\ \boxed{x} = \mathsf{box}\ \boxed{e_1}\ \mathsf{in}\ e_2 \approx [e_1/x]e_2 : B}\Box\beta$$

$$\frac{\Gamma \vdash^{\mathsf{s}} e : \Box A \qquad \Gamma \vdash \mathcal{C}\langle\!\langle e \rangle\!\rangle : B \qquad \Gamma \vdash \mathsf{let\ box}\ \boxed{x} = e\ \mathsf{in}\ \mathcal{C}\langle\!\langle \mathsf{box}\ \boxed{x} \rangle\!\rangle : B}{\Gamma \vdash \mathcal{C}\langle\!\langle e \rangle\!\rangle \approx \mathsf{let\ box}\ \boxed{x} = e\ \mathsf{in}\ \mathcal{C}\langle\!\langle \mathsf{box}\ \boxed{x} \rangle\!\rangle : B}\Box\eta\text{-SAFE}$$

$$\frac{\Gamma \vdash e : \Box A \qquad \Gamma \vdash \mathcal{E}\langle\!\langle e \rangle\!\rangle : B \qquad \Gamma \vdash \mathsf{let\ box}\ \boxed{x} = e\ \mathsf{in}\ \mathcal{E}\langle\!\langle \mathsf{box}\ \boxed{x} \rangle\!\rangle : B}{\Gamma \vdash \mathcal{E}\langle\!\langle e \rangle\!\rangle \approx \mathsf{let\ box}\ \boxed{x} = e\ \mathsf{in}\ \mathcal{E}\langle\!\langle \mathsf{box}\ \boxed{x} \rangle\!\rangle : B}\Box\eta\text{-IMPURE}$$

Fig. 16. Equational Theory

the underlying variable using the let box binding form before using it in the body. To embed an application, we simply put the argument in a box.

We first show that this translation preserves typing, i.e., well-typed expressions embed to well-typed expressions. Then, we show that the $\beta\eta$-equational theory of the *pure* call-by-value simply-typed lambda calculus is preserved under the translation. If two expressions are equal in the simply-typed lambda calculus, they *remain equal* after embedding into our imperative calculus.

Theorem 7.1 (Preservation of typing). *If* $\Gamma \vdash_\lambda e : A$, *then* $\underline{\Gamma} \vdash \underline{e} : \underline{A}$.

Theorem 7.2 (Preservation of equality). *If* $\Gamma \vdash_\lambda e_1 \approx e_2 : A$, *then* $\underline{\Gamma} \vdash \underline{e_1} \approx \underline{e_2} : \underline{A}$.

Finally, we show that our imperative calculus is a conservative extension of the simply-typed lambda calculus. To do so, we claim that if two embedded terms are equal in the extended theory, then they must have been equal in the smaller theory. This shows that the equational theory of the imperative calculus does not introduce any extra equations that would destroy the computational properties of the *pure* simply-typed lambda calculus.

Theorem 7.3 (Conservative Extension). *If* $\Gamma \vdash_\lambda e_1 : A, \Gamma \vdash_\lambda e_2 : A$, *and* $\underline{\Gamma} \vdash \underline{e_1} \approx \underline{e_2} : \underline{A}$, *then* $\Gamma \vdash_\lambda e_1 \approx e_2 : A$.

## 8 DISCUSSION AND FUTURE WORK

There has been a vast amount of work on integrating effects into purely functional languages. Ironically though, even the very definition of what a purely functional language is has historically been

| | | | |
|---|---|---|---|
| Types | $A, B$ | $::=$ | $\mathsf{unit} \mid A \Rightarrow B$ |
| Terms | $e$ | $::=$ | $() \mid x \mid \lambda x : A.\ e \mid e_1\ e_2$ |
| Values | $v$ | $::=$ | $() \mid x \mid \lambda x : A.\ e$ |
| Contexts | $\Gamma, \Delta, \Psi$ | $::=$ | $\cdot \mid \Gamma, x : A$ |

(a) Grammar for STLC

| | |
|---|---|
| $x : A \in \Gamma$ | $x$ is a variable of type $A$ in context $\Gamma$ |
| $\Gamma \vdash_\lambda e : A$ | $e$ is an expression of type $A$ in context $\Gamma$ |
| $\Gamma \vdash_\lambda e_1 \approx e_2 : A$ | $e_1$ and $e_2$ are equal expressions of type $A$ in context $\Gamma$ |

(b) Judgements for STLC

$$\frac{}{\Gamma \vdash_\lambda () : \mathsf{unit}}\ \mathsf{unitI} \qquad \frac{x : A \in \Gamma}{\Gamma \vdash_\lambda x : A}\ \textsc{Var}$$

$$\frac{\Gamma, x : A \vdash_\lambda e : B}{\Gamma \vdash_\lambda \lambda x : A.\ e : A \Rightarrow B}\ {\Rightarrow}\mathrm{I} \qquad \frac{\Gamma \vdash_\lambda e_1 : A \Rightarrow B \qquad \Gamma \vdash_\lambda e_2 : A}{\Gamma \vdash_\lambda e_1\ e_2 : B}\ {\Rightarrow}\mathrm{E}$$

(c) Typing rules for STLC

$$\frac{\Gamma \vdash_\lambda e : A}{\Gamma \vdash_\lambda e \approx e : A}\ \textsc{refl} \qquad \frac{\Gamma \vdash_\lambda e_1 \approx e_2 : A}{\Gamma \vdash_\lambda e_2 \approx e_1 : A}\ \textsc{sym} \qquad \frac{\Gamma \vdash_\lambda e_1 \approx e_2 : A \qquad \Gamma \vdash_\lambda e_2 \approx e_3 : A}{\Gamma \vdash_\lambda e_1 \approx e_3 : A}\ \textsc{trans}$$

$$\frac{\Gamma, x : A \vdash_\lambda e_1 \approx e_2 : B}{\Gamma \vdash_\lambda \lambda x : A.\ e_1 \approx \lambda x : A.\ e_2 : A \Rightarrow B}\ \lambda\textsc{-cong}$$

$$\frac{\Gamma \vdash_\lambda e_1 \approx e_2 : A \Rightarrow B \qquad \Gamma \vdash_\lambda e_3 \approx e_4 : A}{\Gamma \vdash_\lambda e_1\ e_3 \approx e_2\ e_4 : B}\ \textsc{app-cong}$$

$$\frac{\Gamma, x : A \vdash_\lambda e_1 : B \qquad \Gamma \vdash_\lambda e_2 : A}{\Gamma \vdash_\lambda (\lambda x : A.\ e_1)\ e_2 \approx [e_2/x]e_1 : B}\ {\Rightarrow}\beta \qquad \frac{\Gamma \vdash_\lambda e : A \Rightarrow B}{\Gamma \vdash_\lambda e \approx \lambda x : A.\ e\ x : A \Rightarrow B}\ {\Rightarrow}\eta$$

(d) Equational Theory for STLC

Fig. 17. The *pure* call-by-value simply-typed lambda calculus

| | | | |
|---|---|---|---|
| Types | $\underline{\mathsf{unit}}$ | $:=$ | $\mathsf{unit}$ |
| | $\underline{A \Rightarrow B}$ | $:=$ | $\square\, \underline{A} \Rightarrow \underline{B}$ |
| Contexts | $\underline{\cdot}$ | $:=$ | $\cdot$ |
| | $\underline{\Gamma, x : A}$ | $:=$ | $\underline{\Gamma}, x : \underline{A}^{\mathsf{s}}$ |
| Terms | $\underline{()}$ | $:=$ | $()$ |
| | $\underline{x}$ | $:=$ | $x$ |
| | $\underline{\lambda x : A.\ e}$ | $:=$ | $\lambda z : \square\, \underline{A}.\ \mathsf{let\ box}\ \boxed{x} = z\ \mathsf{in}\ \underline{e}$ |
| | $\underline{e_1\ e_2}$ | $:=$ | $\underline{e_1}\ \mathsf{box}\ \boxed{\underline{e_2}}$ |

Fig. 18. Embedding STLC

a contested one. Sabry [1998] proposed that a functional language is pure when its behaviour under different evaluation strategies is "morally" the same, in the sense of Danielsson et al. [2006]. That is, if changing the evaluation strategy from call-by-value to (say) call-by-need could only change the divergence/error behaviour of programs in a language, then the language is pure. In contrast, the definition we use in this paper is less sophisticated: we take purity to be the preservation of the $\beta\eta$ equational theory of the simply-typed lambda calculus. However, it lets us prove the correctness of our embedding in an appealingly simple way, by translating derivations of equality. Sabry [1998] also notes that a purely functional language must be a conservative extension of the simply-typed lambda calculus. Using the results of the previous section, our impure calculus also satisfies this requirement, just by extending it with the purity comonad.

The use of substructural type systems to control access to mutable data is also a long-running theme in the development of programming languages. It is so long-running, in fact, that it actually predates linear logic [Girard 1987] by nearly a decade! Reynolds' Syntactic Control of Interference [Reynolds 1978] proposed using a substructural type discipline to prevent aliased access to data structures. The intuition that substructural logic corresponds to ownership of capabilities is also a very old one – O'Hearn [1993] uses it to explain his model of SCI, and Crary et al. [1999] compare their static capabilities to the capabilities in the HYDRA system of Wulf et al. [1974].

However, these comparisons remained informal, due to the fact that semanticists tended to use capabilities in a substructural fashion (e.g., see [Crary et al. 1999; Terauchi and Aiken 2006]), but from the very outset ([Dennis and Horn 1966]) to modern day applications like capability-safe Javascript [Maffeis et al. 2010], systems designers have tended to use capabilities *non-linearly*. In particular, they thought it was desirable for a principal to hand a capability to two different deputies, which is a design principle obviously incompatible with linearity.

The idea that the linear implication and intuitionistic implication could coexist, without one reducing to the other, first arose in the logic of bunched implications [O'Hearn and Pym 1999]. This led to separation logic [Reynolds 2002], which has been very successful at verifying programs with aliasable state. However, even though the semantics of separation logic supports BI, the bulk of the tooling infrastructure for separation logic (such as Smallfoot [Berdine et al. 2006]) have focused on the substructural fragment, often even omitting anything not in the linear fragment.

However, one observation very important to our work did arise from work on separation logic. Dodds et al. [2009] made the critical observation that in addition to being able to assert ownership, it is extremely useful to be able to *deny* the ownership of a capability. Basically, knowing that a client program *lacks* any capabilities can make it safe to invoke it in a secure context.

The comonadic structure behind denial was also known informally: it arises in the work of Morrisett et al. [2005], where the exponential comonad in linear logic is modelled as the *lack* of any heap ownership; and in an intuitionistic context, the work on functional reactive programming [Krishnaswami 2013] used a capability to create temporal values, and a comonad denying ownership of it permitted writing space-leak-free reactive programs. However, both of these papers used operational unary logical relations models, and so did not prove anything about the equational theory.

Equational theories are easier to get with denotational models, and our model derives from the work of Hofmann [2003]. In his work, he developed a denotational model of space-bounded computation, by taking a naive set-theoretic semantics, and then augmenting it with intensional information. His sets were augmented with a *length function* saying how much memory each value used, and in ours, we use a weight function saying how many capabilities each value holds. (In fact, he even notes that his category also forms a model of bunched implications!) We think his approach has a high power-to-weight ratio, and hope we have shown that it has broad applicability as well.

However, this semantics is certainly not the last word: e.g., the semantics in this paper does not model the allocation of new capabilities as a program executes. In the categorical semantics of bunched logics, it is common to use functor categories, such as functors from the *category of* *finite sets and injections* $\mathcal{I}$, to Set, or presheaves over some other monoidal category. The functor category forms a model of BI, inheriting the cartesian closed structure where the limits are computed Kripke-style in Set, and also a monoidal closed structure using the tensor product from the monoidal category and *Day convolution*. In addition, the ability to move to a bigger set permits modelling allocation of new names and channels (e.g., as is done in models of the $\nu$-calculus [Stark 1996]). Our category of capability spaces uses the co-Heyting structure of the powerset lattice, i.e., we use sets weighted in the complete Heyting algebra using $\supseteq$ as implication. This is a subcategory of presheaves on this lattice (seen as a thin category or a poset), and the doubly closed structure is inherited from there. Of course, this category has more structure, which we did not use – for example, it has coproducts and natural numbers, and the comonad commutes with each type constructor, which we can use to extend our calculus to support our initial `map` example. Another natural question is how we might handle recursion, as our explicit description of the category of capability spaces $\mathbb{C}$ in section 4 seems quite tied to Set. By replaying this in a category like CPO rather than Set, we may be able to derive a domain-theoretic analogue of capability spaces.

Another direction for future work lies in the observation that our $\square$ comonad in subsection 4.5 takes away *all* capabilities, yielding a system with a syntax like that of Pfenning and Davies [2001] with an interpretation close to the axiomatic categorical semantics proposed by Alechina et al. [2001] and Kobayashi [1997]. However, we could consider a *graded* or *indexed* version of the same, i.e., $\square_C$, which only takes away a set of capabilities $C \in \mathfrak{P}(\mathcal{C})$ from a value. Our hope would be that this could form a model of systems like bounded linear logic [Dal Lago and Hofmann 2009; Orchard et al. 2019], or other systems of coeffects [Petricek et al. 2014]. This use of qualifiers on contexts to encode linear resource behaviour appeared first in [Terui 2007], and was also used in the quantitative coeffect calculus in [Brunel et al. 2014]. One issue we foresee with indexing is that, while this indexed comonad would still be a strong monoidal functor, it loses the idempotence property, which we used in our interpretation and proofs.

There has also been a great deal of work on using monads and effect systems [Gifford and Lucassen 1986; Moggi 1989; Nielson and Nielson 1999; Wadler 1998] to control the usage of effects. However, the general idea of using a static tag which broadcasts that an effect *may* occur seems somewhat the reverse of the idea of object capabilities, where access to a dynamically-passed value determines whether an effect can occur. The key feature of our system is that the comonad does not say what effects are possible, but rather asserts that effects are *absent*. This manifests in the cancellation law (in subsection 4.6) of the comonad and the monad. Still, the very phrases *"may perform"* and *"does not possess"* hint that some sort of duality ought to exist.

## ACKNOWLEDGMENTS

We would like to acknowledge Marcelo Fiore for stimulating discussions about the ideas in this paper. We are also thankful to the anonymous referees, and the non-anonymous readers of earlier drafts of this paper, for their valuable comments and feedback. The first author is grateful to the Rigbys who provided a welcoming and homely environment during his stay in Cambridge.

## A SUPPLEMENTARY MATERIAL FOR SECTION 3 (TYPING)

LEMMA A.1. *The weakening relation is reflexive.*
PROOF.

(1) $\Gamma$

(2) $\Gamma =$

(3) $\supseteq$ $\supseteq$-ID

(4) $\Gamma = \Gamma', x : A^q$

(5) $\Gamma' \supseteq \Gamma'$ induction hypothesis

(6) $\Gamma', x : A^q \supseteq \Gamma', x : A^q$ $\supseteq$-CONG

(7) $\Gamma \supseteq \Gamma$

□

LEMMA A.2. *The weakening relation is transitive.*
PROOF.

(1) $\Gamma \supseteq \Delta, \Delta \supseteq \Psi$

(2) $\Gamma = , \Delta =$ case $\supseteq$-ID

(3) $\Psi =$ inversion

(4) $\supseteq$ $\supseteq$-ID

(5) $\Gamma = \Gamma', x : A^q, \Delta = \Delta', x : A^q$ case $\supseteq$-CONG

(6) $\Psi = \Psi', x : A^q, \Delta' \supseteq \Psi'$ case $\supseteq$-CONG

(7) $\Gamma' \supseteq \Psi'$ induction hypothesis

(8) $\Gamma', x : A^q \supseteq \Psi', x : A^q$ $\supseteq$-CONG

(9) $\Delta' \supseteq \Psi$ case $\supseteq$-WK

(10) $\Gamma' \supseteq \Psi$ induction hypothesis

(11) $\Gamma', x : A^q \supseteq \Psi$ induction hypothesis

(12) $\Gamma' \supseteq \Delta$ case $\supseteq$-WK

(13) $\Gamma' \supseteq \Psi$ induction hypothesis

(14) $\Gamma', x : A^q \supseteq \Psi$

(15) $\Gamma \supseteq \Psi$

□

LEMMA A.3. *If* $x : A^q \in \Delta$ *and* $\Gamma \supseteq \Delta$, *then* $x : A^q \in \Gamma$.

PROOF. Assuming $\Gamma \supseteq \Delta$, we do induction on $x : A^q \in \Delta$.

◇∈ -ID

| | | |
|---|---|---|
| (1) | $\dfrac{}{x : A^q \in (\Delta', x : A^q)}$ | ∈-ID |
| (2) | $\dfrac{\Gamma' \supseteq \Delta'}{\Gamma', x : A^q \supseteq \Delta', x : A^q}$ | ⊇-CONG |
| (3) | $x : A^q \in (\Gamma', x : A^q)$ | ∈-ID |

◇∈ -EX

| | | |
|---|---|---|
| (1) | $\dfrac{x : A^q \in \Delta' \qquad (x \neq y)}{x : A^q \in (\Delta', y : B^r)}$ | ∈-EX |
| (2) | $\dfrac{\Gamma' \supseteq \Delta'}{\Gamma', y : B^r \supseteq \Delta', y : B^r}$ | ⊇-CONG |
| (3) | $x : A^q \in \Delta'$ | inversion |
| (4) | $\Gamma' \supseteq \Delta'$ | inversion |
| (5) | $x : A^q \in \Gamma'$ | induction hypothesis |
| (6) | $x : A^q \in (\Gamma', y : B^r)$ | ∈-EX |

□

LEMMA A.4. *If* $\Gamma \supseteq \Delta$, *then* $\Gamma^{\mathsf{s}} \supseteq \Delta^{\mathsf{s}}$.

PROOF. We do induction on $\Gamma \supseteq \Delta$.

◇⊇ -ID

| | | |
|---|---|---|
| (1) | $\dfrac{}{\supseteq}$ | ⊇-ID |
| (2) | $\supseteq$ | ⊇-ID |

◇⊇ -CONG

| | | |
|---|---|---|
| (1) | $\dfrac{\Gamma' \supseteq \Delta'}{\Gamma', x : A^q \supseteq \Delta', x : A^q}$ | ⊇-CONG |
| (2) | $\Gamma' \supseteq \Delta'$ | inversion |
| (3) | $\Gamma'^{\mathbf{s}} \supseteq \Delta'^{\mathbf{s}}$ | induction hypothesis |
| (4) | $q = \mathbf{s}$ | |
| (5) | $\Gamma'^{\mathbf{s}}, x : A^{\mathbf{s}} \supseteq \Delta'^{\mathbf{s}}, x : A^{\mathbf{s}}$ | ⊇-CONG (3) |
| (6) | $q = \mathbf{i}$ | |
| (7) | $\Gamma'^{\mathbf{s}} \supseteq \Delta'^{\mathbf{s}}$ | (3) |
| (8) | $(\Gamma', x : A^q)^{\mathbf{s}} \supseteq (\Delta', x : A^q)^{\mathbf{s}}$ | |

◇⊇ -WK

| | | |
|---|---|---|
| (1) | $\dfrac{\Gamma' \supseteq \Delta}{\Gamma', x : A^q \supseteq \Delta}$ | ⊇-WK |
| (2) | $\Gamma' \supseteq \Delta$ | inversion |
| (3) | $\Gamma'^{\mathbf{s}} \supseteq \Delta^{\mathbf{s}}$ | induction hypothesis |
| (4) | $q = \mathbf{s}$ | |
| (5) | $\Gamma'^{\mathbf{s}}, x : A^{\mathbf{s}} \supseteq \Delta^{\mathbf{s}}$ | ⊇-WK (3) |
| (6) | $q = \mathbf{i}$ | |
| (7) | $\Gamma'^{\mathbf{s}} \supseteq \Delta^{\mathbf{s}}$ | (3) |
| (8) | $(\Gamma', x : A^q)^{\mathbf{s}} \supseteq \Delta^{\mathbf{s}}$ | |

□

LEMMA 3.1 (SYNTACTIC WEAKENING). *If* $\Gamma \supseteq \Delta$ *and* $\Delta \vdash e : A$*, then* $\Gamma \vdash e : A$.

PROOF. Assuming $\Gamma \supseteq \Delta$, we do induction on $\Delta \vdash e : A$.

◇VAR

| | | |
|---|---|---|
| (1) | $\dfrac{x : A^q \in \Delta}{\Delta \vdash x : A}$ | VAR |
| (2) | $x : A^q \in \Delta$ | inversion |

(3) $x : A^q \in \Gamma$ lemma A.3

(4) $\Gamma \vdash x : A$ VAR

◇ unitI

(1) $\dfrac{}{\Delta \vdash () : \mathsf{unit}}$ unitI

(2) $\Gamma \vdash () : \mathsf{unit}$ unitI

◇ ×I

(1) $\dfrac{\Delta \vdash e_1 : A \qquad \Delta \vdash e_2 : B}{\Delta \vdash (e_1 , e_2) : A \times B}$ ×I

(2) $\Delta \vdash e_1 : A$ inversion

(3) $\Delta \vdash e_2 : B$ inversion

(4) $\Gamma \vdash e_1 : A$ induction hypothesis

(5) $\Gamma \vdash e_2 : B$ induction hypothesis

(6) $\Gamma \vdash (e_1 , e_2) : A \times B$ ×I

◇ $\times E_i$

(1) $\dfrac{\Delta \vdash e : A \times B}{\Delta \vdash \mathsf{fst}\, e : A}$ $\times E_1$

(2) $\Delta \vdash e : A \times B$ inversion

(3) $\Gamma \vdash e : A \times B$ induction hypothesis

(4) $\Gamma \vdash \mathsf{fst}\, e : A$ $\times E_1$

(1) $\dfrac{\Delta \vdash e : A \times B}{\Delta \vdash \mathsf{snd}\, e : B}$ $\times E_2$

(2) $\Delta \vdash e : A \times B$ inversion

(3) $\Gamma \vdash e : A \times B$ induction hypothesis

(4) $\Gamma \vdash \mathsf{snd}\, e : B$ $\times E_2$

$\diamond\, \Box I$

(1) $\dfrac{\Delta \vdash^{s} e : A}{\Delta \vdash \mathsf{box}\, \boxed{e} : \Box A}$ $\Box I$

(2) $\Delta \vdash^{s} e : A$ inversion

(3) $\Delta^{s} \vdash e : A$ inversion

(4) $\Gamma^{s} \supseteq \Delta^{s}$ lemma A.4

(5) $\Gamma^{s} \vdash e : A$ induction hypothesis

(6) $\Gamma \vdash^{s} e : A$ CTX-SAFE

(7) $\Gamma \vdash \mathsf{box}\, \boxed{e} : \Box A$ $\Box I$

$\diamond\, \Box E$

(1) $\dfrac{\Delta \vdash e_1 : \Box A \qquad \Delta, x : A^{s} \vdash e_2 : B}{\Delta \vdash \mathsf{let\ box}\, \boxed{x} = e_1\ \mathsf{in}\ e_2 : B}$ $\Box E$

(2) $\Delta \vdash e_1 : \Box A$ inversion

(3) $\Delta, x : A^{s} \vdash e_2 : B$ inversion

(4) $\Gamma \vdash e_1 : \Box A$ induction hypothesis (2)

(5) $\Gamma, x : A^{s} \supseteq \Delta, x : A^{s}$ $\supseteq$-CONG

(6) $\Gamma, x : A^{s} \vdash e_2 : B$ induction hypothesis (3) (5)

(7) $\Gamma \vdash \mathsf{let\ box}\, \boxed{x} = e_1\ \mathsf{in}\ e_2 : B$ $\Box E$

$\diamond \Rightarrow I$

(1) $\dfrac{\Delta, x : A^{i} \vdash e : B}{\Delta \vdash \lambda x : A.\ e : A \Rightarrow B}$ $\Rightarrow I$

| | | |
|---|---|---|
| (2) | $\Delta, x : A^{\mathsf{i}} \vdash e : B$ | inversion |
| (3) | $\Gamma, x : A^{\mathsf{i}} \supseteq \Delta, x : A^{\mathsf{i}}$ | $\supseteq$-CONG |
| (4) | $\Gamma, x : A^{\mathsf{i}} \vdash e : B$ | induction hypothesis (3) |
| (5) | $\Gamma \vdash \lambda x.\, e : A \Rightarrow B$ | $\Rightarrow$I |

◇ $\Rightarrow$ E

| | | |
|---|---|---|
| (1) | $\dfrac{\Delta \vdash e_1 : A \Rightarrow B \qquad \Delta \vdash e_2 : A}{\Delta \vdash e_1\, e_2 : B}$ | $\Rightarrow$E |
| (2) | $\Delta \vdash e_1 : A \Rightarrow B$ | inversion |
| (3) | $\Delta \vdash e_2 : A$ | inversion |
| (4) | $\Gamma \vdash e_1 : A \Rightarrow B$ | induction hypothesis (2) |
| (5) | $\Gamma \vdash e_2 : A$ | induction hypothesis (3) |
| (6) | $\Gamma \vdash e_1\, e_2 : B$ | $\Rightarrow$E |

◇ strI

| | | |
|---|---|---|
| (1) | $\dfrac{}{\Delta \vdash s : \mathsf{str}}$ | strI |
| (2) | $\Gamma \vdash s : \mathsf{str}$ | strI |

◇ PRINT

| | | |
|---|---|---|
| (1) | $\dfrac{\Delta \vdash e_1 : \mathsf{cap} \qquad \Delta \vdash e_2 : \mathsf{str}}{\Delta \vdash e_1.\mathsf{print}(e_2) : \mathsf{unit}}$ | PRINT |
| (2) | $\Delta \vdash e_1 : \mathsf{cap}$ | inversion |
| (3) | $\Delta \vdash e_2 : \mathsf{str}$ | inversion |
| (4) | $\Gamma \vdash e_1 : \mathsf{cap}$ | induction hypothesis (2) |
| (5) | $\Gamma \vdash e_2 : \mathsf{str}$ | induction hypothesis (3) |
| (6) | $\Gamma \vdash e_1.\mathsf{print}(e_2) : \mathsf{unit}$ | PRINT |

□

LEMMA A.5. *If* $\Gamma \supseteq \Delta$ *and* $\Delta \vdash \theta : \Psi$, *then* $\Gamma \vdash \theta : \Psi$.

PROOF. Assuming $\Gamma \supseteq \Delta$, we do induction on $\Delta \vdash \theta : \Psi$.

◇ SUB-ID

(1) $\dfrac{}{\Delta \vdash \langle\rangle :}$ SUB-ID

(2) $\Gamma \vdash \langle\rangle :$ SUB-ID

◇ SUB-SAFE

(1) $\dfrac{\Delta \vdash \theta : \Psi' \qquad \Delta \vdash^{s} e : A}{\Delta \vdash \langle \theta, e^{s}/x \rangle : \Psi', x : A^{s}}$ SUB-SAFE

(2) $\Delta \vdash \theta' : \Psi'$ inversion

(3) $\dfrac{\Delta^{s} \vdash e : A}{\Delta \vdash^{s} e : A}$ CTX-SAFE

(4) $\Delta^{s} \vdash e : A$ inversion

(5) $\Gamma \vdash \theta' : \Psi'$ induction hypothesis (2)

(6) $\Gamma^{s} \supseteq \Delta^{s}$ lemma A.4

(7) $\Gamma^{s} \vdash e : A$ syntactic weakening lemma 3.1 (3)

(8) $\Gamma \vdash^{s} e : A$ CTX-SAFE

(9) $\Gamma \vdash \langle \theta', e^{s}/x \rangle : \Psi', x : A^{s}$ SUB-SAFE

◇ SUB-IMPURE

(1) $\dfrac{\Delta \vdash \theta : \Psi' \qquad \Delta \vdash v : A}{\Delta \vdash \langle \theta, v^{i}/x \rangle : \Psi', x : A^{i}}$ SUB-IMPURE

(2) $\Delta \vdash \theta' : \Psi'$ inversion

(3) $\Delta \vdash v : A$ inversion

(4) $\Gamma \vdash \theta' : \Psi'$ induction hypothesis (2)

(5) $\Gamma \vdash v : A$ syntactic weakening lemma 3.1 (3)

(6) $\Gamma \vdash \langle \theta', v^{\mathsf{i}}/x \rangle : \Psi', x : A^{\mathsf{i}}$ SUB-IMPURE

□

LEMMA A.6. *If* $\Gamma \vdash \theta : \Delta$ *then* $\Gamma^{\mathsf{s}} \vdash \theta^{\mathsf{s}} : \Delta^{\mathsf{s}}$.

PROOF. We do induction on $\Gamma \vdash \theta : \Delta$.

(1) $\Gamma \vdash \theta : \Delta$

(2) $\dfrac{}{\Gamma \vdash \langle\rangle :}$ SUB-ID

(3) $\Gamma^{\mathsf{s}} \vdash \langle\rangle :$ SUB-ID

(4) $\dfrac{\Gamma \vdash \theta : \Delta \qquad \Gamma \vdash^{\mathsf{s}} e : A}{\Gamma \vdash \langle \theta, e^{\mathsf{s}}/x \rangle : \Delta, x : A^{\mathsf{s}}}$ SUB-SAFE

(5) $\Gamma \vdash \theta : \Delta$ inversion

(6) $\dfrac{\Gamma^{\mathsf{s}} \vdash e : A}{\Gamma \vdash^{\mathsf{s}} e : A}$ CTX-SAFE

(7) $\Gamma^{\mathsf{s}} \vdash e : A$ inversion

(8) $\Gamma^{\mathsf{s}} \vdash \theta^{\mathsf{s}} : \Delta^{\mathsf{s}}$ induction hypothesis

(9) $(\Gamma^{\mathsf{s}})^{\mathsf{s}} \vdash e : A$ $(\Gamma^{\mathsf{s}})^{\mathsf{s}} = \Gamma^{\mathsf{s}}$

(10) $\Gamma^{\mathsf{s}} \vdash^{\mathsf{s}} e : A$ CTX-SAFE

(11) $\Gamma^{\mathsf{s}} \vdash \langle \theta^{\mathsf{s}}, e^{\mathsf{s}}/x \rangle : \Delta^{\mathsf{s}}, x : A^{\mathsf{s}}$ SUB-SAFE

(12) $\dfrac{\Gamma \vdash \theta : \Delta \qquad \Gamma \vdash v : A}{\Gamma \vdash \langle \theta, v^{\mathsf{i}}/x \rangle : \Delta, x : A^{\mathsf{i}}}$ SUB-IMPURE

(13) $\Gamma \vdash \theta : \Delta$ inversion

(14) $\Gamma^{\mathsf{s}} \vdash \theta^{\mathsf{s}} : \Delta^{\mathsf{s}}$ induction hypothesis

(15) $\Gamma^{\mathsf{s}} \vdash \theta^{\mathsf{s}} : \Delta^{\mathsf{s}}$

□

LEMMA A.7. *For any context* $\Gamma$, *we have* $\Gamma \supseteq \Gamma^{\mathsf{s}}$.

PROOF. We do induction on $\Gamma$.

(1) $\Gamma$

(2) $\Gamma =$

(3) $\supseteq$ $\supseteq$-ID

(4) $\Gamma = \Delta, x : A^{\mathsf{s}}$

(5) $\Delta \supseteq \Delta^{\mathsf{s}}$ induction hypothesis

(6) $\Delta, x : A^{\mathsf{s}} \supseteq \Delta^{\mathsf{s}}, x : A^{\mathsf{s}}$ $\supseteq$-CONG

(7) $\Gamma = \Delta, x : A^{\mathsf{i}}$

(8) $\Delta \supseteq \Delta^{\mathsf{s}}$ induction hypothesis

(9) $\Delta, x : A^{\mathsf{i}} \supseteq \Delta^{\mathsf{s}}$ $\supseteq$-WK

(10) $\Gamma \supseteq \Gamma^{\mathsf{s}}$

□

LEMMA A.8. *If* $\Gamma \vdash \theta : \Delta$ *and* $x : A^q \in \Delta$, *then* $\Gamma \vdash \theta[x] : A$.

PROOF. Assuming $\Gamma \vdash \theta : \Delta$, we do induction on $x : A^q \in \Delta$.
$\diamond \in$ -ID

(1) $\dfrac{}{x : A^q \in (\Delta', x : A^q)}$ $\in$-ID

(2) $q = \mathsf{s}$

(3) $\dfrac{\Gamma \vdash \phi : \Delta' \qquad \Gamma \vdash^{\mathsf{s}} e : A}{\Gamma \vdash \langle \phi, e^{\mathsf{s}}/x \rangle : \Delta', x : A^{\mathsf{s}}}$ SUB-SAFE

(4) $\dfrac{\Gamma^{\mathsf{s}} \vdash e : A}{\Gamma \vdash^{\mathsf{s}} e : A}$ CTX-SAFE

(5) $\Gamma^{\mathsf{s}} \vdash e : A$ inversion

(6) $\Gamma \supseteq \Gamma^{\mathsf{s}}$ lemma A.7

(7) $\Gamma \vdash e : A$ syntactic weakening lemma 3.1

(8) $\Gamma \vdash \langle \phi, e^{\mathsf{s}}/x \rangle[x] : A$ definition

(9) $q = \mathsf{i}$

(10) $\dfrac{\Gamma \vdash \phi : \Delta' \qquad \Gamma \vdash v : A}{\Gamma \vdash \langle \phi, v^{\mathsf{i}}/x \rangle : \Delta', x : A^{\mathsf{i}}}$ SUB-IMPURE

(11) $\Gamma \vdash v : A$ inversion

(12) $\Gamma \vdash \langle \phi, v^{\mathsf{i}}/x \rangle[x] : A$ definition

(13) $\Gamma \vdash \theta[x] : A$

◇∈-EX

(1) $\dfrac{x : A^q \in \Delta' \qquad (x \neq y)}{x : A^q \in (\Delta', y : B^r)}$ ∈-EX

(2) $x : A^q \in \Delta'$ inversion

(3) $q = \mathsf{s}$

(4) $\dfrac{\Gamma \vdash \phi : \Delta' \qquad \Gamma \vdash^{\mathsf{s}} e : B}{\Gamma \vdash \langle \phi, e^{\mathsf{s}}/y \rangle : \Delta', y : B^{\mathsf{s}}}$ SUB-SAFE

(5) $\Gamma \vdash \phi : \Delta'$ inversion

(6) $\Gamma \vdash \phi[x] : A$ induction hypothesis

(7) $\Gamma \vdash \langle \phi, e^{\mathsf{s}}/y \rangle[x] : A$ definition

(8) $q = \mathsf{i}$

(9) $\dfrac{\Gamma \vdash \phi : \Delta' \qquad \Gamma \vdash v : B}{\Gamma \vdash \langle \phi, v^{\mathsf{i}}/y \rangle : \Delta', y : B^{\mathsf{i}}}$ SUB-IMPURE

(10) $\Gamma \vdash \phi : \Delta'$ inversion

(11) $\Gamma \vdash \phi[x] : A$ induction hypothesis

(12) $\Gamma \vdash \langle \phi, v^{\mathsf{i}}/y \rangle[x] : A$ definition

(13) $\Gamma \vdash \theta[x] : A$

□

THEOREM 3.4 (SYNTACTIC SUBSTITUTION). *If* $\Gamma \vdash \theta : \Delta$ *and* $\Delta \vdash e : A$, *then* $\Gamma \vdash \theta(e) : A$.

PROOF. Assuming $\Gamma \vdash \theta : \Delta$, we do induction on $\Delta \vdash e : A$.

◇VAR

(1) $\dfrac{x : A^q \in \Delta}{\Delta \vdash x : A}$ VAR

(2) $x : A^q \in \Delta$ inversion

(3) $\Gamma \vdash \theta[x] : A$ lemma A.8

(4) $\Gamma \vdash \theta(x) : A$ definition

◇ unitI

(1) $\dfrac{}{\Delta \vdash () : \mathsf{unit}}$ unitI

(2) $\Gamma \vdash () : \mathsf{unit}$ unitI

(3) $\Gamma \vdash \theta(()) : \mathsf{unit}$ definition

◇ ×I

(1) $\dfrac{\Delta \vdash e_1 : A \qquad \Delta \vdash e_2 : B}{\Delta \vdash (e_1 , e_2) : A \times B}$ ×I

(2) $\Delta \vdash e_1 : A$ inversion

(3) $\Delta \vdash e_2 : B$ inversion

(4) $\Gamma \vdash \theta(e_1) : A$ induction hypothesis

(5) $\Gamma \vdash \theta(e_2) : B$ induction hypothesis

(6) $\Gamma \vdash (\theta(e_1) , \theta(e_2)) : A \times B$ ×I

(7) $\Gamma \vdash \theta((e_1 , e_2)) : A \times B$ definition

◇ $\times\mathrm{E}_i$

(1) $\dfrac{\Delta \vdash e : A \times B}{\Delta \vdash \mathsf{fst}\, e : A}$ $\times\mathrm{E}_1$

(2) $\Delta \vdash e : A \times B$ inversion

(3) $\Gamma \vdash \theta(e) : A \times B$ induction hypothesis

(4) $\Gamma \vdash \mathsf{fst}\, \theta(e) : B$ $\times\mathrm{E}_1$

(5) $\Gamma \vdash \theta(\mathsf{fst}\, e) : B$ definition

(1) $\dfrac{\Delta \vdash e : A \times B}{\Delta \vdash \mathsf{snd}\, e : B}$ $\times\mathrm{E}_2$

(2) $\Delta \vdash e : A \times B$ inversion

| | | |
|---|---|---|
| (3) | $\Gamma \vdash \theta(e) : A \times B$ | induction hypothesis |
| (4) | $\Gamma \vdash \mathsf{snd}\,\theta(e) : B$ | $\times E_2$ |
| (5) | $\Gamma \vdash \theta(\mathsf{snd}\,e) : B$ | definition |

$\diamond \Rightarrow I$

| | | |
|---|---|---|
| (1) | $\dfrac{\Delta, x : A^{\mathsf{i}} \vdash e : B}{\Delta \vdash \lambda x : A.\ e : A \Rightarrow B}$ | $\Rightarrow$I |
| (2) | $\Delta, x : A^{\mathsf{i}} \vdash e : B$ | inversion |
| (3) | $\Gamma, y : A^{\mathsf{i}} \supseteq \Gamma$ | $\supseteq$-WK |
| (4) | $\Gamma, y : A^{\mathsf{i}} \vdash \theta : \Delta$ | lemma A.5 |
| (5) | $\Gamma, y : A^{\mathsf{i}} \vdash y : A$ | VAR |
| (6) | $\Gamma, y : A^{\mathsf{i}} \vdash \langle \theta, y^{\mathsf{i}}/x \rangle : \Delta, x : A^{\mathsf{i}}$ | SUB-IMPURE (4) (5) |
| (7) | $\Gamma, y : A^{\mathsf{i}} \vdash \langle \theta, y^{\mathsf{i}}/x \rangle(e) : B$ | induction hypothesis (6) (2) |
| (8) | $\Gamma \vdash \lambda y.\ \langle \theta, y^{\mathsf{i}}/x \rangle(e) : A \Rightarrow B$ | $\Rightarrow$I |
| (9) | $\Gamma \vdash \theta(\lambda y.\ e) : A \Rightarrow B$ | definition |

$\diamond \Rightarrow E$

| | | |
|---|---|---|
| (1) | $\dfrac{\Delta \vdash e_1 : A \Rightarrow B \qquad \Delta \vdash e_2 : A}{\Delta \vdash e_1\, e_2 : B}$ | $\Rightarrow$E |
| (2) | $\Delta \vdash e_1 : A \Rightarrow B$ | inversion |
| (3) | $\Delta \vdash e_2 : A$ | inversion |
| (4) | $\Gamma \vdash \theta(e_1) : A \Rightarrow B$ | induction hypothesis (2) |
| (5) | $\Gamma \vdash \theta(e_2) : A$ | induction hypothesis (3) |
| (6) | $\Gamma \vdash \theta(e_1)\, \theta(e_2) : B$ | $\Rightarrow$E |
| (7) | $\Gamma \vdash \theta(e_1\, e_2) : B$ | definition |

$\diamond$ strI

| | | |
|---|---|---|
| (1) | $\dfrac{}{\Delta \vdash s : \mathsf{str}}$ | strI |
| (2) | $\Gamma \vdash s : \mathsf{str}$ | strI |
| (3) | $\Gamma \vdash \theta(s) : \mathsf{str}$ | definition |

◇ Print

| | | |
|---|---|---|
| (1) | $\dfrac{\Delta \vdash e_1 : \mathsf{cap} \qquad \Delta \vdash e_2 : \mathsf{str}}{\Delta \vdash e_1\,.\,\mathsf{print}(e_2) : \mathsf{unit}}$ | Print |
| (2) | $\Delta \vdash e_1 : \mathsf{cap}$ | inversion |
| (3) | $\Delta \vdash e_2 : \mathsf{str}$ | inversion |
| (4) | $\Gamma \vdash \theta(e_1) : \mathsf{cap}$ | induction hypothesis (2) |
| (5) | $\Gamma \vdash \theta(e_2) : \mathsf{str}$ | induction hypothesis (3) |
| (6) | $\Gamma \vdash \theta(e_1)\,.\,\mathsf{print}(\theta(e_2)) : \mathsf{unit}$ | Print |
| (7) | $\Gamma \vdash \theta(e_1\,.\,\mathsf{print}(e_2)) : \mathsf{unit}$ | definition |

◇ □I

| | | |
|---|---|---|
| (1) | $\dfrac{\Delta \vdash^{s} e : A}{\Delta \vdash \mathsf{box}\,\boxed{e} : \Box A}$ | □I |
| (2) | $\dfrac{\Delta^{s} \vdash e : A}{\Delta \vdash^{s} e : A}$ | ctx-safe |
| (3) | $\Delta^{s} \vdash e : A$ | inversion |
| (4) | $\Gamma^{s} \vdash \theta^{s} : \Delta^{s}$ | lemma A.6 |
| (5) | $\Gamma^{s} \vdash \theta^{s}(e) : A$ | induction hypothesis (3) (4) |
| (6) | $\Gamma \vdash^{s} \theta^{s}(e) : A$ | ctx-safe |
| (7) | $\Gamma \vdash \mathsf{box}\,\boxed{\theta^{s}(e)} : \Box A$ | □I |
| (8) | $\Gamma \vdash \theta(\mathsf{box}\,\boxed{e}) : \Box A$ | definition |

◇ □E

| | | |
|---|---|---|
| (1) | $\dfrac{\Delta \vdash e_1 : \Box A \qquad \Delta, x : A^s \vdash e_2 : B}{\Delta \vdash \text{let box}\ \boxed{x} = e_1\ \text{in}\ e_2 : B}$ | $\Box$E |
| (2) | $\Delta \vdash e_1 : \Box A$ | inversion |
| (3) | $\Delta, x : A^s \vdash e_2 : B$ | inversion |
| (4) | $\Gamma, y : A^s \supseteq \Gamma$ | $\supseteq$-WK |
| (5) | $\Gamma, y : A^s \vdash \theta : \Delta$ | lemma A.5 (4) |
| (6) | $y : A^s \in \Gamma^s, y : A^s$ | $\supseteq$-ID |
| (7) | $\Gamma^s, y : A^s \vdash y : A$ | VAR |
| (8) | $\Gamma, y : A^s \vdash \langle \theta, y^s/x \rangle : \Delta, x : A^s$ | SUB-SAFE |
| (9) | $\Gamma, y : A^s \vdash \langle \theta, y^s/x \rangle(e_2) : B$ | induction hypothesis (8) (3) |
| (10) | $\Gamma \vdash \theta(e_1) : \Box A$ | induction hypothesis (2) |
| (11) | $\Gamma \vdash \text{let box}\ \boxed{y} = \theta(e_1)\ \text{in}\ \langle \theta, y^s/x \rangle(e_2) : B$ | $\Box$E (9) (10) |
| (12) | $\Gamma \vdash \theta(\text{let box}\ \boxed{x} = e_1\ \text{in}\ e_2) : B$ | definition |

□

## B SUPPLEMENTARY MATERIAL FOR SECTION 4 (SEMANTICS)

LEMMA B.1.

$$\mathcal{Hom}_{\mathcal{C}}(C, A \times B) \simeq \mathcal{Hom}_{\mathcal{C}}(C, A) \times \mathcal{Hom}_{\mathcal{C}}(C, B)$$

PROOF. Given $f : \mathcal{Hom}_{\mathcal{C}}(C, A)$ and $g : \mathcal{Hom}_{\mathcal{C}}(C, B)$, we define

$$\begin{aligned} \langle f, g \rangle \quad &: \quad \mathcal{Hom}_{\mathcal{C}}(C, A \times B) \\ c \quad &\mapsto \quad (f(c), g(c)) \end{aligned}$$

Assume there exists a $C_c$ such that $w_C(c, C_c)$. Then there exist weights $C_a \subseteq C_c$ and $C_b \subseteq C_c$ such that $w_A(f(c), C_a)$ and $w_B(g(c), C_b)$. Let $C = C_a \cup C_b$, then $C \subseteq C_c$ as well. This gives a weighting for $\langle f, g \rangle$.

Given $h : \mathcal{Hom}_{\mathcal{C}}(C, A \times B)$, we define

$$\begin{aligned} f : \mathcal{Hom}_{\mathcal{C}}(A, C) &:= h \,;\, \pi_1 \\ g : \mathcal{Hom}_{\mathcal{C}}(B, C) &:= h \,;\, \pi_2 \end{aligned}$$

□

LEMMA B.2.

$$\begin{aligned} \mathrm{ev}_{A,B} \;&:\; \mathcal{Hom}_{\mathcal{C}}((A \to B) \times A, B) \\ \mathrm{curry} \;&:\; \mathcal{Hom}_{\mathcal{C}}(C \times A, B) \xrightarrow{\sim} \mathcal{Hom}_{\mathcal{C}}(C, A \to B) \end{aligned}$$

Proof. We define,

$$\begin{aligned} \mathsf{ev}_{A,B} &: \mathcal{H}om_{\mathcal{C}}\left((A \to B) \times A\, , B\right) \\ (f, a) &\mapsto f(a) \end{aligned}$$

Assume there exists a weight $C$ such that $w_{(A\to B)\times A}((f,a),C)$. Then, there exist weights $C_f$ and $C_a$ such that $C = C_f \cup C_a$, $w_{A\to B}(f, C_f)$ and $w_A(a, C_A)$. Hence, there exists a weighting $C_b$ such that $w_B(f(a), C_b)$.

Given $f : \mathcal{H}om_{\mathcal{C}}\left(C \times A\, , B\right)$, we define

$$\begin{aligned} \mathsf{curry}\,(f) &: \mathcal{H}om_{\mathcal{C}}\left(C\, , A \to B\right) \\ c &\mapsto \lambda a.f(c, a) \end{aligned}$$

Assume there exists a $C_c$ such that $w_C(c, C_c)$. We claim that $w_{A\to B}(\mathsf{curry}\,(f), C_c)$. Assume $a$ and $C_a$ such that $w_A(a, C_a)$. Then, $w_B(f(c,a), C_c \cup C_a)$. Choosing, $C_b = C_c \cup C_a$, we have $w_B(f(c,a), C_b)$.

Given $f : \mathcal{H}om_{\mathcal{C}}\left(C\, , A \to B\right)$ we define

$$\begin{aligned} \mathsf{uncurry}\,(f) &: \mathcal{H}om_{\mathcal{C}}\left(C \times A\, , B\right) \\ (c, a) &\mapsto f(c)(a) \end{aligned}$$

Assume there exist weights $C_c$ and $C_a$ such that $w_{C\times A}((c,a), C_c \cup C_a)$, $w_C(c, C_c)$ and $w_A(a, C_a)$. So, there exists $C_f \subseteq C_c$ such that $w_{A\to B}(f(c), C_f)$. Thus, there exists $C_b \subseteq C_f \cup C_a$ such that $w_B(f(c)(a), C_b)$. It follows that $C_b \subseteq C_c \cup C_a$, and $w_B(f(c)(a), C_b)$. □

$$\begin{aligned} \eta_A : A &\to TA \\ a &\mapsto (a, \lambda c.\epsilon) \end{aligned}$$

Assume there exists $C_a$ such that $w_A(a, C_a)$. With $o = \lambda c.\epsilon$, we have that for all $c \in \mathcal{C}$, $o(c) = \epsilon$. Using $C_o = \varnothing$, we have, $w_{T(A)}((a, o), C_a \cup C_o)$.

$$\begin{aligned} \mu_A : TTA &\to TA \\ ((a, o_1), o_2) &\mapsto (a, \lambda c.o_2(c) \cdot o_1(c)) \end{aligned}$$

Let $C_{o_1} = \left\{ c \mid o_1(c) \neq \epsilon \right\}$ and $C_{o_2} = \left\{ c \mid o_2(c) \neq \epsilon \right\}$. Assume there exists $C_a$ such that $w_{TA}((a, o_1), C_a \cup C_{o_2})$, and $w_A(a, C_a \cup C_{o_1} \cup C_{o_2})$. For all $c \in C_{o_1}$, $o_1(c) \neq \epsilon$, and for all $c \in C_{o_2}$, $o_2(c) \neq \epsilon$. So, for all $c \in C_{o_1} \cup C_{o_2}$, $o_2(c) \cdot o_1(c) \neq \epsilon$. Using $C_o = C_{o_1} \cup C_{o_2}$ we have, $w_{T(A)}((a, \lambda c.o_2(c) \cdot o_1(c)), C_a \cup C_o)$.

Lemma B.3. *The following diagrams commute.*

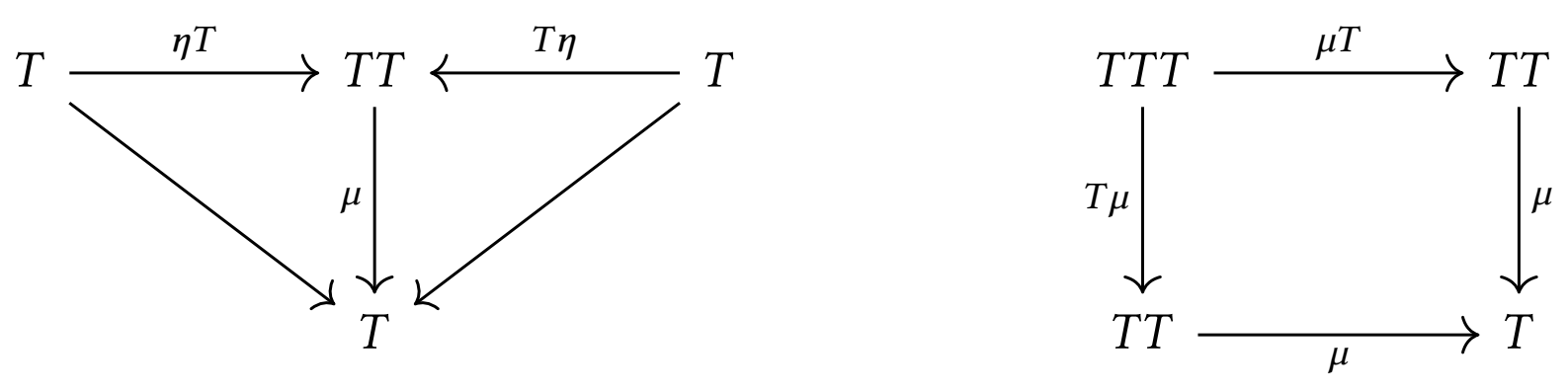

Proof.

$$\begin{array}{rlrl}
 & \mu(\eta T(a,o)) & & \mu(T\eta(a,o)) \\
= & \mu((a,\lambda c.\epsilon),o) & = & \mu((a,o),\lambda c.\epsilon) \\
= & (a,\lambda c.o(c)\cdot\epsilon) & = & (a,\lambda c.\epsilon\cdot o(c)) \\
= & (a,\lambda c.o(c)) & = & (a,\lambda c.o(c)) \\
= & (a,o) & = & (a,o)
\end{array}$$

$$\begin{array}{rlrl}
 & \mu(\mu T(((a,o_1),o_2),o_3)) & & \mu(T\mu(((a,o_1),o_2),o_3)) \\
= & \mu((a,\lambda c.o_2(c)\cdot o_1(c)),o_3) & = & \mu((a,o_1),\lambda c.o_3(c)\cdot o_2(c)) \\
= & (a,\lambda c.o_3(c)\cdot(o_2(c)\cdot o_1(c))) & = & (a,\lambda c.(o_3(c)\cdot o_2(c))\cdot o_1(c)) \\
= & (a,\lambda c.o_3(c)\cdot o_2(c)\cdot o_1(c)) & = & (a,\lambda c.o_3(c)\cdot o_2(c)\cdot o_1(c))
\end{array}$$

□

Lemma B.4. *Strengthening with 1 is irrelevant.*

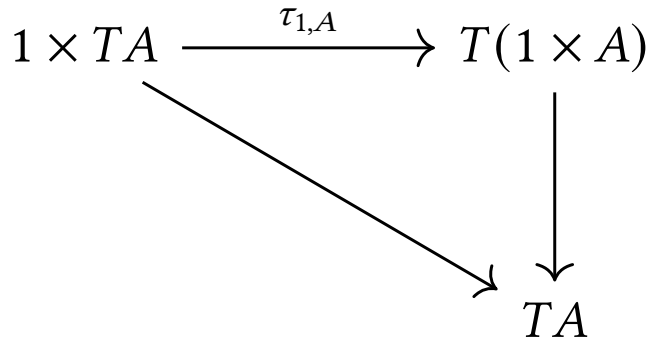

*Consecutive applications of strength commute.*

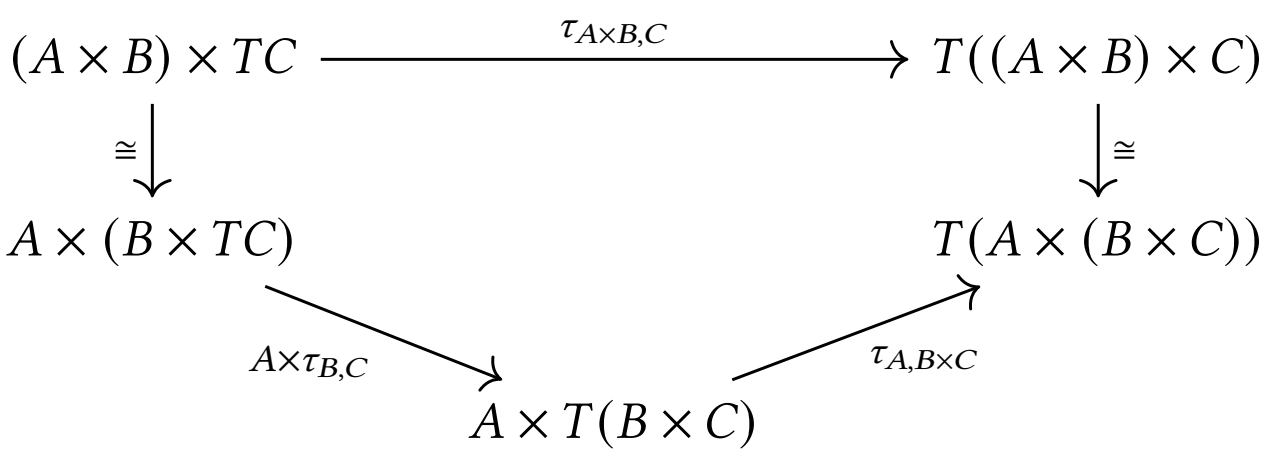

*Strength commutes with monad unit and multiplication.*

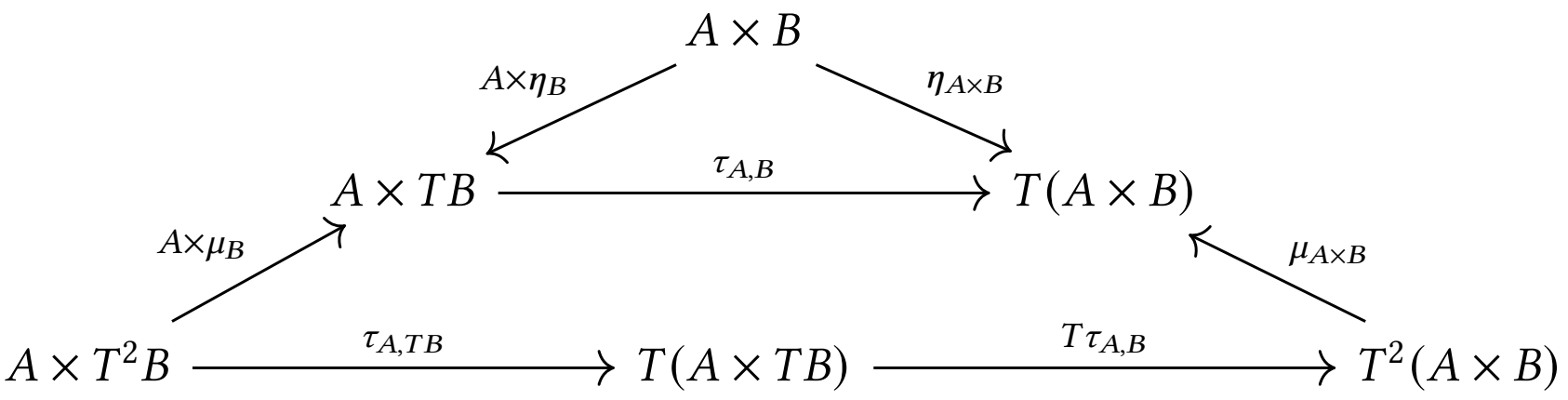

*Left are right strengths are compatible.*

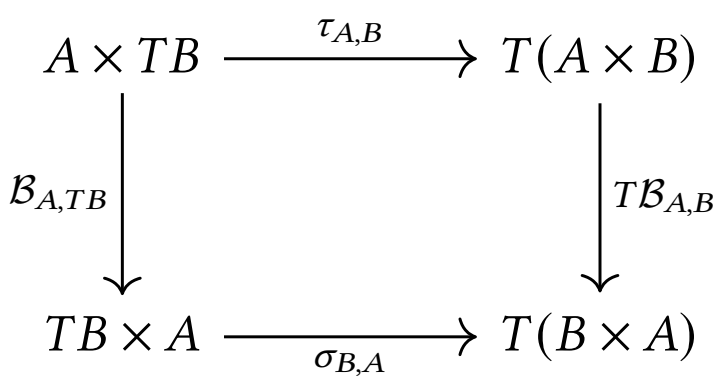

Proof. All monads on Set are strong, and Set is symmetric monoidal for products. Note that, $T$ is *not* a commutative monad, because the following natural transformations are *not* equal.

$$\alpha: \quad TA \times TB \xrightarrow{\sigma_{A,TB}} T(A \times TB) \xrightarrow{T\tau_{A,B}} T^2(A \times B) \xrightarrow{\mu_{A\times B}} T(A \times B)$$

$$\beta: \quad TA \times TB \xrightarrow{\tau_{TA,B}} T(TA \times B) \xrightarrow{T\sigma_{A,B}} T^2(A \times B) \xrightarrow{\mu_{A\times B}} T(A \times B)$$

$$\begin{array}{llll} & \sigma_{A,TB}((a,o_1),(b,o_2)) & & \tau_{TA,B}((a,o_1),(b,o_2)) \\ = & T\tau_{A,B}((a,(b,o_2)),o_1) & = & T\sigma_{A,B}((((a,o_1),b)),o_2) \\ = & \mu_{A\times B}(((a,b),o_2),o_1) & = & \mu_{A\times B}(((a,b),o_1),o_2) \\ = & ((a,b),\lambda c.o_1(c)\cdot o_2(c)) & = & ((a,b),\lambda c.o_2(c)\cdot o_1(c)) \end{array}$$

This means that the order of evaluation matters depending on whether we choose $\alpha$ or $\beta$ for evaluating products. □

Lemma B.5. *The following diagrams commute.*

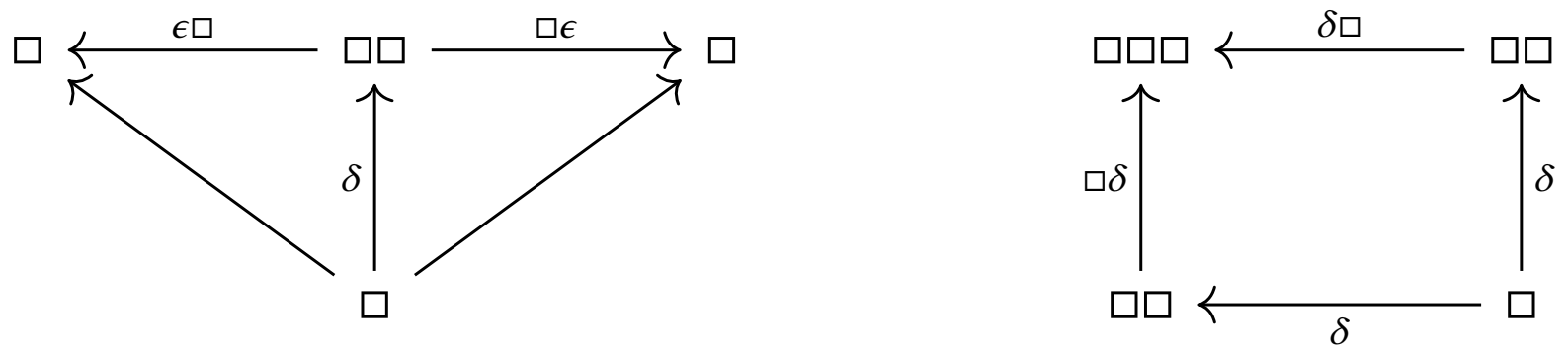

Proof. Since $\delta$ and $\epsilon$ are identities, it follows trivially. Each arrow is weight-preserving because the weight is not altered by $\Box$, $\delta$, or $\epsilon$. □

Lemma B.6.

$$\Box TA \simeq \Box A$$

Proof. Let $a \in |A|$ such that $(a,o) \in |\Box TA|$. Assume $C$, such that, $w_{TA}((a,o),C)$. Then, $C = \varnothing$. Also, there exist $C_a$ and $C_o$ such that $\varnothing = C = C_a \cup C_o$ and $w_A(a,C_a)$. Hence, $w_A(a,\varnothing)$. This gives the map $\Phi_A : \Box TA \longrightarrow \Box A$, which is natural in $A$. We also have $\Box\eta_A : \Box A \longrightarrow \Box TA$ sending $a \in |A|$ to $(a,\lambda c.\epsilon)$. This gives an isomorphism. □

Lemma B.7.

$$\mathcal{H}om_{\mathcal{C}}(\Gamma \otimes A, B) \cong \mathcal{H}om_{\mathcal{C}}(\Gamma, A \multimap B)$$

Proof. We define,

$$\begin{array}{rcl} \mathrm{ev}_{A,B} & : & \mathcal{H}om_{\mathcal{C}}((A \multimap B)\otimes A, B) \\ (f,a) & \mapsto & f(a) \end{array}$$

Assume there exists a weight $C$ such that $w_{(A\multimap B)\otimes A}((f,a),C)$. Then, there exist weights $C_f$ and $C_a$ such that $C_f \sharp C_a$ and $C = C_f \cup C_a$, with $w_{A\multimap B}(f,C_f)$ and $w_A(a,C_A)$. Hence, there exists a weighting $C_b$ such that $w_B(f(a),C_b)$.

Given $f : \mathcal{H}om_{\mathcal{C}}(C \otimes A, B)$, we define

$$\begin{array}{rcl} \mathrm{curry}\,(f) & : & \mathcal{H}om_{\mathcal{C}}(C, A \multimap B) \\ c & \mapsto & \lambda a.f(c,a) \end{array}$$

Assume there exists a $C_c$ such that $w_C(c, C_c)$. We claim that $w_{A \multimap B}(\text{curry}\,(f), C_c)$. Assume $a$ and $C_a$ such that $w_A(a, C_a)$. Then, $C_c \sharp C_a$ and $w_B(f(c, a), C_c \cup C_a)$. Choosing, $C_b = C_c \cup C_a$, we have $w_B(f(c, a), C_b)$.

Given $f : \mathcal{Hom}_{\mathcal{C}}\,(C\,, A \multimap B)$ we define

$$\begin{aligned} \text{uncurry}\,(f) \;&:\; \mathcal{Hom}_{\mathcal{C}}\,(C \otimes A\,, B) \\ (c, a) &\mapsto f(c)(a) \end{aligned}$$

Assume there exist weights $C_c$ and $C_a$ such that $C_c \sharp C_a$ and $w_{C \otimes A}((c, a), C_c \cup C_a)$, with $w_C(c, C_c)$ and $w_A(a, C_a)$. So, there exists $C_f \subseteq C_c$ such that $w_{A \multimap B}(f(c), C_f)$. Since $C_c \sharp C_a$, it is also the case that $C_f \sharp C_a$. Thus, there exists $C_b \subseteq C_f \cup C_a$ such that $w_B(f(c)(a), C_b)$. It follows that $C_b \subseteq C_c \cup C_a$, and $w_B(f(c)(a), C_b)$. □

### B.1 Exception monad

*Definition B.8 ($T : \mathcal{C} \longrightarrow \mathcal{C}$).* Let $E = \left\{ fail \right\}$ be the set of exceptions. We define the monad $T$ as follows.

$$\begin{aligned} |T(A)| &\coloneqq |A| + 1 \\ w_{T(A)} &\coloneqq \left\{ (inl(a), C_a) \;\middle|\; w_A(a, C_a) \right\} \cup \left\{ (inr(tt), E) \right\} \end{aligned}$$

It is not hard to see that the maps are weight preserving.

$$\begin{aligned} \eta_A : A &\to TA & \mu_A : TTA &\to TA \\ a &\mapsto inl(a) & inl(inl(a)) &\mapsto inl(a) \\ & & inl(inr(*)) &\mapsto inr(*) \\ & & inr(*) &\mapsto inr(*) \end{aligned}$$

$\square TA$ restricts the weight to only the *safe* values of $A$, ie, values that cannot throw any exceptions, hence is isomorphic to $\square A$, giving the cancellation law.

### B.2 State monad

*Definition B.9 ($T : \mathcal{C} \longrightarrow \mathcal{C}$).* We use $H = Loc \to Val$ to denote a naive model of a heap, where $Loc$ is a fixed set of global locations. Two heaps are equal if the functions are extensionally equal. We choose the capabilities to be sets in $\mathfrak{P}(Loc)$, and the weight of a computation is given exactly by the heap locations it writes to.

$$\begin{aligned} |T(A)| &\coloneqq H \to |A| \times H \\ w_{T(A)}(f, C) &\Leftrightarrow \begin{cases} \forall h, \exists C' \subseteq C. w_A(a, \pi_1(f(h), C')) \\ \forall h_1, h_2, (\forall l \in C, h_1(l) = h_2(l)) \Rightarrow \pi_1(f(h_1)) = \pi_1(f(h_2)) \\ \forall h, \forall l \notin C, \pi_2(f(h))(l) = h(l) \end{cases} \end{aligned}$$

$$\begin{aligned} \eta_A \;&:\; A \to TA & \mu_A \;&:\; TTA \to TA \\ a &\mapsto \lambda h.(a, h) & f &\mapsto \lambda h. \begin{array}{ll} let & \left\{ (f', h') \coloneqq f(h) \right. \\ in & f'(h') \end{array} \end{aligned}$$

$\square TA$ restricts the only writable locations to the empty set, making the set of values *safe*.

## C SUPPLEMENTARY MATERIAL FOR SECTION 5 (INTERPRETATION)

Lemma C.1. *If* $\Gamma \supseteq \Delta$, *then*

$$\rho(\Gamma) \mathbin{;} \mathcal{M}(\Gamma) \mathbin{;} \Box \mathsf{Wk}(\Gamma^{\mathsf{s}} \supseteq \Delta^{\mathsf{s}}) = \mathsf{Wk}(\Gamma \supseteq \Delta) \mathbin{;} \rho(\Delta) \mathbin{;} \mathcal{M}(\Delta)$$

Proof. We do induction on $\Gamma \supseteq \Delta$.

$\diamond \dfrac{}{\supseteq}$ ⊇-id

$$\rho() \mathbin{;} \mathcal{M}() \mathbin{;} \Box \mathsf{Wk}({}^{\mathsf{s}} \supseteq {}^{\mathsf{s}})$$

=⟨ definition ⟩

$$id_1 \mathbin{;} id_1 \mathbin{;} \Box id_1$$

=⟨ $\Box$ preserves *id* ⟩

$$id_1 \mathbin{;} id_1 \mathbin{;} id_1$$

=⟨ definition ⟩

$$\mathsf{Wk}(\supseteq) \mathbin{;} \rho() \mathbin{;} \mathcal{M}()$$

$\diamond \dfrac{\Gamma \supseteq \Delta}{\Gamma, x : A^q \supseteq \Delta, x : A^q}$ ⊇-cong

When $q = \mathsf{s}$,

$$\rho(\Gamma, x : A^{\mathsf{s}}) \mathbin{;} \mathcal{M}(\Gamma, x : A^{\mathsf{s}}) \mathbin{;} \Box \mathsf{Wk}(\Gamma^{\mathsf{s}}, x : A^{\mathsf{s}} \supseteq \Delta^{\mathsf{s}}, x : A^{\mathsf{s}})$$

=⟨ definition ⟩

$$[\rho(\Gamma) \times id_{\Box A}] \mathbin{;} [\mathcal{M}(\Gamma) \times \delta_A] \mathbin{;} m^{\times}_{\Gamma^{\mathsf{s}},\Box A} \mathbin{;} \Box[\mathsf{Wk}(\Gamma^{\mathsf{s}} \supseteq \Delta^{\mathsf{s}}) \times id_{\Box A}]$$

=⟨ monoidal action of $\Box$ ⟩

$$[\rho(\Gamma) \times id_{\Box A}] \mathbin{;} [\mathcal{M}(\Gamma) \times \delta_A] \mathbin{;} [\Box \mathsf{Wk}(\Gamma^{\mathsf{s}} \supseteq \Delta^{\mathsf{s}}) \times \Box id_{\Box A}]; m^{\times}_{\Delta^{\mathsf{s}},\Box A}$$

=⟨ exchange law ⟩

$$[\rho(\Gamma) \mathbin{;} \mathcal{M}(\Gamma) \mathbin{;} \Box \mathsf{Wk}(\Gamma^{\mathsf{s}} \supseteq \Delta^{\mathsf{s}}) \times id_{\Box A} \mathbin{;} \delta_A \mathbin{;} \Box id_{\Box A}] \mathbin{;} m^{\times}_{\Delta^{\mathsf{s}},\Box A}$$

=⟨ identity law ⟩

$$[\rho(\Gamma) \mathbin{;} \mathcal{M}(\Gamma) \mathbin{;} \Box \mathsf{Wk}(\Gamma^{\mathsf{s}} \supseteq \Delta^{\mathsf{s}}) \times \delta_A] \mathbin{;} m^{\times}_{\Delta^{\mathsf{s}},\Box A}$$

=⟨ induction hypothesis ⟩

$$[\mathsf{Wk}(\Gamma \supseteq \Delta) \mathbin{;} \rho(\Delta) \mathbin{;} \mathcal{M}(\Delta) \times \delta_A] \mathbin{;} m^{\times}_{\Delta^{\mathsf{s}},\Box A}$$

=⟨ identity law ⟩

$$[\mathsf{Wk}(\Gamma \supseteq \Delta) \mathbin{;} \rho(\Delta) \mathbin{;} \mathcal{M}(\Delta) \times id_{\Box A} \mathbin{;} id_{\Box A} \mathbin{;} \delta_A] \mathbin{;} m^{\times}_{\Delta^{\mathsf{s}},\Box A}$$

=⟨ exchange law ⟩

$$[\mathsf{Wk}(\Gamma \supseteq \Delta) \times id_{\Box A}] \mathbin{;} [\rho(\Delta) \times id_{\Box A}] \mathbin{;} [\mathcal{M}(\Delta) \times \delta_A] \mathbin{;} m^{\times}_{\Delta^{\mathsf{s}},\Box A}$$

=⟨ definition ⟩

$$\mathsf{Wk}(\Gamma, x : A^{\mathsf{s}} \supseteq \Delta, x : A^{\mathsf{s}}) \,;\, \rho(\Delta, x : A^{\mathsf{s}}) \,;\, \mathcal{M}(\Delta, x : A^{\mathsf{s}})$$

When $q = \mathbf{i}$,

$$\rho(\Gamma, x : A^{\mathsf{i}}) \,;\, \mathcal{M}(\Gamma, x : A^{\mathsf{i}}) \,;\, \Box\,\mathsf{Wk}((\Gamma, x : A^{\mathsf{i}})^{\mathsf{s}} \supseteq (\Delta, x : A^{\mathsf{i}})^{\mathsf{s}})$$

=⟨ definition ⟩

$$\rho(\Gamma, x : A^{\mathsf{i}}) \,;\, \mathcal{M}(\Gamma, x : A^{\mathsf{i}}) \,;\, \Box\,\mathsf{Wk}(\Gamma^{\mathsf{s}} \supseteq \Delta^{\mathsf{s}})$$

=⟨ definition ⟩

$$\pi_1 \,;\, \rho(\Gamma) \,;\, \mathcal{M}(\Gamma) \,;\, \Box\,\mathsf{Wk}(\Gamma^{\mathsf{s}} \supseteq \Delta^{\mathsf{s}})$$

=⟨ induction hypothesis ⟩

$$\pi_1 \,;\, \mathsf{Wk}(\Gamma \supseteq \Delta) \,;\, \rho(\Delta) \,;\, \mathcal{M}(\Delta)$$

=⟨ definition of $\pi_1$ ⟩

$$\langle \pi_1 \,;\, \mathsf{Wk}(\Gamma \supseteq \Delta) \,,\, \pi_2 \,;\, id_A \rangle \,;\, \pi_1 \,;\, \rho(\Delta) \,;\, \mathcal{M}(\Delta)$$

=⟨ universal property of product ⟩

$$[\mathsf{Wk}(\Gamma \supseteq \Delta) \times id_A] \,;\, \pi_1 \,;\, \rho(\Delta) \,;\, \mathcal{M}(\Delta)$$

=⟨ definition ⟩

$$\mathsf{Wk}(\Gamma, x : A^{\mathsf{i}} \supseteq \Delta, x : A^{\mathsf{i}}) \,;\, \rho(\Delta, x : A^{\mathsf{i}}) \,;\, \mathcal{M}(\Delta, x : A^{\mathsf{i}})$$

$$\diamond \; \frac{\Gamma \supseteq \Delta}{\Gamma, x : A^{q} \supseteq \Delta} \supseteq\text{-WK}$$

When $q = \mathbf{s}$,

$$\rho(\Gamma, x : A^{\mathsf{s}}) \,;\, \mathcal{M}(\Gamma, x : A^{\mathsf{s}}) \,;\, \Box\,\mathsf{Wk}(\Gamma^{\mathsf{s}}, x : A^{\mathsf{s}} \supseteq \Delta^{\mathsf{s}})$$

=⟨ definition ⟩

$$[\rho(\Gamma) \times id_{\Box A}] \,;\, [\mathcal{M}(\Gamma) \times \delta_A] \,;\, m^{\times}_{\Gamma^{\mathsf{s}},\Box A} \,;\, \Box(\pi_1 \,;\, \mathsf{Wk}(\Gamma^{\mathsf{s}} \supseteq \Delta^{\mathsf{s}}))$$

=⟨ $\Box$ preserves composition ⟩

$$[\rho(\Gamma) \times id_{\Box A}] \,;\, [\mathcal{M}(\Gamma) \times \delta_A] \,;\, m^{\times}_{\Gamma^{\mathsf{s}},\Box A} \,;\, \Box\pi_1 \,;\, \Box\,\mathsf{Wk}(\Gamma^{\mathsf{s}} \supseteq \Delta^{\mathsf{s}})$$

=⟨ exchange law ⟩

$$[\rho(\Gamma) \,;\, \mathcal{M}(\Gamma) \times id_{\Box A} \,;\, \delta_A] \,;\, m^{\times}_{\Gamma^{\mathsf{s}},\Box A} \,;\, \Box\pi_1 \,;\, \Box\,\mathsf{Wk}(\Gamma^{\mathsf{s}} \supseteq \Delta^{\mathsf{s}})$$

=⟨ identity law ⟩

$$[\rho(\Gamma) \,;\, \mathcal{M}(\Gamma) \times \delta_A] \,;\, m^{\times}_{\Gamma^{\mathsf{s}},\Box A} \,;\, \Box\pi_1 \,;\, \Box\,\mathsf{Wk}(\Gamma^{\mathsf{s}} \supseteq \Delta^{\mathsf{s}})$$

=⟨ definition of $m^{\times}$ ⟩

$$[\rho(\Gamma) \,;\, \mathcal{M}(\Gamma) \times \delta_A] \,;\, \pi_1 \,;\, \Box\,\mathsf{Wk}(\Gamma^{\mathsf{s}} \supseteq \Delta^{\mathsf{s}})$$

=⟨ universal property of product ⟩

$$\langle \pi_1 \,;\, \rho(\Gamma) \,;\, \mathcal{M}(\Gamma) \,,\, \pi_2 \,;\, \delta_A \rangle \,;\, \pi_1 \,;\, \Box\,\mathsf{Wk}(\Gamma^{\mathsf{s}} \supseteq \Delta^{\mathsf{s}})$$

$=\langle$ definition of $\pi_1$ $\rangle$

$\pi_1 \,;\, \rho(\Gamma) \,;\, \mathcal{M}(\Gamma) \,;\, \Box\mathrm{Wk}(\Gamma^{\mathsf{s}} \supseteq \Delta^{\mathsf{s}})$

$=\langle$ induction hypothesis $\rangle$

$\pi_1 \,;\, \mathrm{Wk}(\Gamma \supseteq \Delta) \,;\, \rho(\Delta) \,;\, \mathcal{M}(\Delta)$

$=\langle$ definition $\rangle$

$\mathrm{Wk}(\Gamma, x : A^{\mathsf{s}} \supseteq \Delta) \,;\, \rho(\Delta) \,;\, \mathcal{M}(\Delta)$

When $q = \mathsf{i}$,

$\rho(\Gamma, x : A^{\mathsf{i}}) \,;\, \mathcal{M}(\Gamma, x : A^{\mathsf{i}}) \,;\, \Box\mathrm{Wk}((\Gamma, x : A^{\mathsf{i}})^{\mathsf{s}} \supseteq \Delta^{\mathsf{s}})$

$=\langle$ definition $\rangle$

$\pi_1 \,;\, \rho(\Gamma) \,;\, \mathcal{M}(\Gamma) \,;\, \Box\mathrm{Wk}(\Gamma^{\mathsf{s}} \supseteq \Delta^{\mathsf{s}})$

$=\langle$ induction hypothesis $\rangle$

$\pi_1 \,;\, \mathrm{Wk}(\Gamma \supseteq \Delta) \,;\, \rho(\Delta) \,;\, \mathcal{M}(\Delta)$

$=\langle$ definition $\rangle$

$\mathrm{Wk}(\Gamma, x : A^{\mathsf{i}} \supseteq \Delta) \,;\, \rho(\Delta) \,;\, \mathcal{M}(\Delta)$

$\square$

Lemma C.2. *If $x : A^q \in \Delta$ and $\Gamma \supseteq \Delta$, then*

$$[\![ x : A^q \in \Gamma ]\!] = \mathrm{Wk}(\Gamma \supseteq \Delta) \,;\, [\![ x : A^q \in \Delta ]\!]$$

Proof. Assume $\Gamma \supseteq \Delta$. We do induction on $x : A^q \in \Delta$ followed by inversion on $\Gamma \supseteq \Delta$.

$$\diamond \quad \frac{}{x : A^q \in (\Gamma, x : A^q)}\ \in\text{-ID}$$

When $q = \mathsf{i}$,

$[\![ x : A^{\mathsf{i}} \in (\Gamma, x : A^{\mathsf{i}}) ]\!]$

$=\langle$ definition $\rangle$

$\pi_2$

$=\langle$ identity law $\rangle$

$\pi_2 \,;\, id_A$

$=\langle$ definition of $\pi_2$ $\rangle$

$\langle \pi_1 \,;\, \mathrm{Wk}(\Gamma \supseteq \Delta) \,,\, \pi_2 \,;\, id_A \rangle \,;\, \pi_2$

$=\langle$ universal property of products $\rangle$

$[\mathrm{Wk}(\Gamma \supseteq \Delta) \times id_A] \,;\, \pi_2$

$=\langle$ definition $\rangle$

$$\mathsf{Wk}(\Gamma, x : A^{\mathbf{i}} \supseteq \Delta, x : A^{\mathbf{i}}) \mathbin{;} [\![ x : A^{\mathbf{i}} \in (\Delta, x : A^{\mathbf{i}}) ]\!]$$

When $q = \mathbf{s}$,

$$[\![ x : A^{\mathbf{s}} \in (\Gamma, x : A^{\mathbf{s}}) ]\!]$$

$=\langle$ definition $\rangle$

$$\pi_2 \mathbin{;} \epsilon_A$$

$=\langle$ identity law $\rangle$

$$\pi_2 \mathbin{;} id_{\Box A} \mathbin{;} \epsilon_A$$

$=\langle$ definition of $\pi_2$ $\rangle$

$$\langle \pi_1 \mathbin{;} \mathsf{Wk}(\Gamma \supseteq \Delta), \pi_2 \mathbin{;} id_{\Box A} \rangle \mathbin{;} \pi_2 \mathbin{;} \epsilon_A$$

$=\langle$ universal property of products $\rangle$

$$[\mathsf{Wk}(\Gamma \supseteq \Delta) \times id_{\Box A}] \mathbin{;} \pi_2 \mathbin{;} \epsilon_A$$

$=\langle$ definition $\rangle$

$$\mathsf{Wk}(\Gamma, x : A^{\mathbf{s}} \supseteq \Delta, x : A^{\mathbf{s}}) \mathbin{;} [\![ x : A^{\mathbf{s}} \in (\Delta, x : A^{\mathbf{s}}) ]\!]$$

$$\diamond \quad \frac{x : A^q \in \Gamma \qquad (x \neq y)}{x : A^q \in (\Gamma, y : B^r)} \in\text{-EX}$$

When $r = \mathbf{i}$,

$$[\![ x : A^q \in (\Gamma, y : B^r) ]\!]$$

$=\langle$ definition $\rangle$

$$\pi_1 \mathbin{;} [\![ x : A^q \in \Gamma ]\!]$$

$=\langle$ induction hypothesis $\rangle$

$$\pi_1 \mathbin{;} \mathsf{Wk}(\Gamma \supseteq \Delta) \mathbin{;} [\![ x : A^q \in \Delta ]\!]$$

$=\langle$ definition of $\pi_2$ $\rangle$

$$\langle \pi_1 \mathbin{;} \mathsf{Wk}(\Gamma \supseteq \Delta), \pi_2 \mathbin{;} id_B \rangle \mathbin{;} \pi_1 \mathbin{;} [\![ x : A^q \in \Delta ]\!]$$

$=\langle$ universal property of products $\rangle$

$$[\mathsf{Wk}(\Gamma \supseteq \Delta) \times id_B] \mathbin{;} \pi_1 \mathbin{;} [\![ x : A^q \in \Delta ]\!]$$

$=\langle$ definition $\rangle$

$$\mathsf{Wk}(\Gamma, y : B^r \supseteq \Delta, y : B^r) \mathbin{;} [\![ x : A^q \in (\Delta, y : B^r) ]\!]$$

When $r = \mathbf{s}$,

$$[\![ x : A^q \in (\Gamma, y : B^r) ]\!]$$

$=\langle$ definition $\rangle$

$\pi_1 \,;\, [\![ x : A^q \in \Gamma ]\!]$

$=\langle$ induction hypothesis $\rangle$

$\pi_1 \,;\, \mathsf{Wk}(\Gamma \supseteq \Delta) \,;\, [\![ x : A^q \in \Delta ]\!]$

$=\langle$ definition of $\pi_2$ $\rangle$

$\langle \pi_1 \,;\, \mathsf{Wk}(\Gamma \supseteq \Delta) \,,\, \pi_2 \,;\, id_{\Box}B \rangle \,;\, \pi_1 \,;\, [\![ x : A^q \in \Delta ]\!]$

$=\langle$ universal property of products $\rangle$

$[\mathsf{Wk}(\Gamma \supseteq \Delta) \times id_{\Box}B] \,;\, \pi_1 \,;\, [\![ x : A^q \in \Delta ]\!]$

$=\langle$ definition $\rangle$

$\mathsf{Wk}(\Gamma, y : B^r \supseteq \Delta, y : B^r) \,;\, [\![ x : A^q \in (\Delta, y : B^r) ]\!]$

$\square$

Lemma 5.1 (Semantic weakening). *If* $\Gamma \supseteq \Delta$ *and* $\Delta \vdash e : A$*, then*

$$[\![ \Gamma \vdash e : A ]\!] = \mathsf{Wk}(\Gamma \supseteq \Delta) \,;\, [\![ \Delta \vdash e : A ]\!].$$

Proof. We proceed by induction on $\Delta \vdash e : A$.

◇ $\dfrac{x : A^q \in \Gamma}{\Gamma \vdash x : A}$ Var

$[\![ \Gamma \vdash x : A ]\!]$

$=\langle$ definition $\rangle$

$[\![ x : A^{\mathsf{i}} \in \Gamma ]\!]$

$=\langle$ lemma C.2 $\rangle$

$\mathsf{Wk}(\Gamma \supseteq \Delta) \,;\, [\![ x : A^{\mathsf{i}} \in \Delta ]\!]$

$=\langle$ definition $\rangle$

$\mathsf{Wk}(\Gamma \supseteq \Delta) \,;\, [\![ \Delta \vdash x : A ]\!]$

◇ $\dfrac{}{\Gamma \vdash () : \mathsf{unit}}$ unitI

$[\![ \Gamma \vdash () : \mathsf{unit} ]\!]$

$=\langle$ definition $\rangle$

$!_\Gamma \,;\, \eta_1$

$=\langle$ universal property of 1 $\rangle$

$\mathsf{Wk}(\Gamma \supseteq \Delta) \,;\, !_\Delta ; \eta_1$

$=\langle$ definition $\rangle$

$\mathsf{Wk}(\Gamma \supseteq \Delta) \,;\, [\![\Delta \vdash () : \mathsf{unit}]\!]$

$$\diamond\ \frac{\Gamma \vdash e_1 : A \qquad \Gamma \vdash e_2 : B}{\Gamma \vdash (e_1\,,e_2) : A \times B}\ \times\mathrm{I}$$

$[\![\Gamma \vdash (e_1\,,e_2) : A \times B]\!]$

$=\langle$ definition $\rangle$

$\langle [\![\Gamma \vdash e_1 : A]\!]\,, [\![\Gamma \vdash e_2 : B]\!] \rangle \,;\, \beta_{A,B}$

$=\langle$ induction hypothesis $\rangle$

$\langle \mathsf{Wk}(\Gamma \supseteq \Delta) \,;\, [\![\Gamma \vdash e_1 : A]\!]\,, \mathsf{Wk}(\Gamma \supseteq \Delta) \,;\, [\![\Gamma \vdash e_2 : B]\!] \rangle \,;\, \beta_{A,B}$

$=\langle$ universal property of products $\rangle$

$\mathsf{Wk}(\Gamma \supseteq \Delta) \,;\, \langle [\![\Delta \vdash e_1 : A]\!]\,, [\![\Delta \vdash e_2 : B]\!] \rangle \,;\, \beta_{A,B}$

$=\langle$ definition $\rangle$

$\mathsf{Wk}(\Gamma \supseteq \Delta) \,;\, [\![\Delta \vdash (e_1\,,e_2) : A \times B]\!]$

$$\diamond\ \frac{\Gamma \vdash e : A \times B}{\Gamma \vdash \mathsf{fst}\,e : A}\ \times\mathrm{E}_1$$

$[\![\Gamma \vdash \mathsf{fst}\,e : A \times B]\!]$

$=\langle$ definition $\rangle$

$[\![\Gamma \vdash e : A \times B]\!] \,;\, T\pi_1$

$=\langle$ induction hypothesis $\rangle$

$\mathsf{Wk}(\Gamma \supseteq \Delta) \,;\, [\![\Delta \vdash e : A \times B]\!] \,;\, T\pi_1$

$=\langle$ definition $\rangle$

$\mathsf{Wk}(\Gamma \supseteq \Delta) \,;\, [\![\Delta \vdash \mathsf{fst}\,e : A]\!]$

$$\diamond\ \frac{\Gamma \vdash e : A \times B}{\Gamma \vdash \mathsf{snd}\,e : B}\ \times\mathrm{E}_2$$

$[\![\Gamma \vdash \mathsf{snd}\,e : A \times B]\!]$

$=\langle$ definition $\rangle$

$[\![\Gamma \vdash e : A \times B]\!] \,;\, T\pi_2$

$=\langle$ induction hypothesis $\rangle$

$\mathsf{Wk}(\Gamma \supseteq \Delta) \,;\, [\![\Delta \vdash e : A \times B]\!] \,;\, T\pi_2$

$=\langle$ definition $\rangle$

$\mathsf{Wk}(\Gamma \supseteq \Delta) \mathbin{;} [\![\Delta \vdash \mathsf{snd}\, e : B]\!]$

$$\diamond \quad \frac{\Gamma, x : A^{\mathsf{i}} \vdash e : B}{\Gamma \vdash \lambda x : A.\ e : A \Rightarrow B} \Rightarrow\mathrm{I}$$

$[\![\Gamma \vdash \lambda x.\ e : A \Rightarrow B]\!]$

$=\langle$ definition $\rangle$

$\mathsf{curry}\,([\![\Gamma, x : A^{\mathsf{i}} \vdash e : B]\!]) \mathbin{;} \eta_{A \to TB}$

$=\langle$ induction hypothesis $\rangle$

$\mathsf{curry}\,(\mathsf{Wk}(\Gamma, x : A^{\mathsf{i}} \supseteq \Delta, x : A^{\mathsf{i}}) \mathbin{;} [\![\Delta, x : A^{\mathsf{i}} \vdash e : B]\!]) \mathbin{;} \eta_{A \to TB}$

$=\langle$ definition $\rangle$

$\mathsf{curry}\,([\mathsf{Wk}(\Gamma \supseteq \Delta) \times id_A] \mathbin{;} [\![\Delta, x : A^{\mathsf{i}} \vdash e : B]\!]) \mathbin{;} \eta_{A \to TB}$

$=\langle$ universal property of exponential $\rangle$

$\mathsf{Wk}(\Gamma \supseteq \Delta) \mathbin{;} \mathsf{curry}\,([\![\Delta, x : A^{\mathsf{i}} \vdash e : B]\!]) \mathbin{;} \eta_{A \to TB}$

$=\langle$ definition $\rangle$

$\mathsf{Wk}(\Gamma \supseteq \Delta) \mathbin{;} [\![\Delta \vdash \lambda x.\ e : A \Rightarrow B]\!]$

$$\diamond \quad \frac{\Gamma \vdash e_1 : A \Rightarrow B \qquad \Gamma \vdash e_2 : A}{\Gamma \vdash e_1\, e_2 : B} \Rightarrow\mathrm{E}$$

$[\![\Gamma \vdash e_1\, e_2 : B]\!]$

$=\langle$ definition $\rangle$

$\langle [\![\Gamma \vdash e_1 : A \Rightarrow B]\!], [\![\Gamma \vdash e_2 : A]\!] \rangle$
$\mathbin{;} \beta_{A \to TB, A} \mathbin{;} T\, \mathsf{ev}_{A, TB} \mathbin{;} \mu_B$

$=\langle$ induction hypothesis $\rangle$

$\langle \mathsf{Wk}(\Gamma \supseteq \Delta) \mathbin{;} [\![\Delta \vdash e_1 : A \Rightarrow B]\!], \mathsf{Wk}(\Gamma \supseteq \Delta) \mathbin{;} [\![\Delta \vdash e_2 : A]\!] \rangle$
$\mathbin{;} \beta_{A \to TB, A} \mathbin{;} T\, \mathsf{ev}_{A, TB} \mathbin{;} \mu_B$

$=\langle$ universal property of products $\rangle$

$\mathsf{Wk}(\Gamma \supseteq \Delta) \mathbin{;} \langle [\![\Delta \vdash e_1 : A \Rightarrow B]\!], [\![\Delta \vdash e_2 : A]\!] \rangle$
$\mathbin{;} \beta_{A \to TB, A} \mathbin{;} T\, \mathsf{ev}_{A, TB} \mathbin{;} \mu_B$

$=\langle$ definition $\rangle$

$\mathsf{Wk}(\Gamma \supseteq \Delta) \mathbin{;} [\![\Delta \vdash e_1\, e_2 : B]\!]$

$$\diamond\ \frac{\Gamma \vdash e_1 : \mathsf{cap} \qquad \Gamma \vdash e_2 : \mathsf{str}}{\Gamma \vdash e_1\,.\,\mathsf{print}(e_2) : \mathsf{unit}}\ \textsc{Print}$$

$$
\begin{aligned}
& [\![\Gamma \vdash e_1\,.\,\mathsf{print}(e_2) : \mathsf{unit}]\!] \\
=\langle\ & \text{definition}\ \rangle \\
& \langle [\![\Gamma \vdash e_1 : \mathsf{cap}]\!], [\![\Gamma \vdash e_2 : \mathsf{str}]\!] \rangle \,;\, \beta_{\mathcal{C},\Sigma^*} \,;\, Tp \,;\, \mu_1 \\
=\langle\ & \text{induction hypothesis}\ \rangle \\
& \langle \mathsf{Wk}(\Gamma \supseteq \Delta) \,;\, [\![\Delta \vdash e_1 : \mathsf{cap}]\!], \mathsf{Wk}(\Gamma \supseteq \Delta) \,;\, [\![\Delta \vdash e_2 : \mathsf{str}]\!] \rangle \,;\, \beta_{\mathcal{C},\Sigma^*} \,;\, Tp \,;\, \mu_1 \\
=\langle\ & \text{universal property of products}\ \rangle \\
& \mathsf{Wk}(\Gamma \supseteq \Delta) \,;\, \langle [\![\Delta \vdash e_1 : \mathsf{cap}]\!], [\![\Delta \vdash e_2 : \mathsf{str}]\!] \rangle \,;\, \beta_{\mathcal{C},\Sigma^*} \,;\, Tp \,;\, \mu_1 \\
=\langle\ & \text{definition}\ \rangle \\
& \mathsf{Wk}(\Gamma \supseteq \Delta) \,;\, [\![\Delta \vdash e_1\,.\,\mathsf{print}(e_2) : \mathsf{unit}]\!]
\end{aligned}
$$

$$\diamond\ \frac{\Gamma \vdash^{s} e : A}{\Gamma \vdash \mathsf{box}\,\boxed{e} : \Box A}\ \Box\mathrm{I}$$

$$
\begin{aligned}
& [\![\Gamma \vdash \mathsf{box}\,\boxed{e} : \Box A]\!] \\
=\langle\ & \text{definition}\ \rangle \\
& [\![\Gamma \vdash^{s} e : A]\!]_p \,;\, \eta_{\Box A} \\
=\langle\ & \text{definition}\ \rangle \\
& \rho(\Gamma) \,;\, \mathcal{M}(\Gamma) \,;\, \Box[\![\Gamma^{s} \vdash e : A]\!] \,;\, \Phi_A \,;\, \eta_{\Box A} \\
=\langle\ & \text{induction hypothesis}\ \rangle \\
& \rho(\Gamma) \,;\, \mathcal{M}(\Gamma) \,;\, \Box(\mathsf{Wk}(\Gamma^{s} \supseteq \Delta^{s}) \,;\, [\![\Delta^{s} \vdash e : A]\!]) \,;\, \Phi_A \,;\, \eta_{\Box A} \\
=\langle\ & \Box \text{ preserves composition}\ \rangle \\
& \rho(\Gamma) \,;\, \mathcal{M}(\Gamma) \,;\, \Box\,\mathsf{Wk}(\Gamma^{s} \supseteq \Delta^{s}) \,;\, \Box[\![\Delta^{s} \vdash e : A]\!] \,;\, \Phi_A \,;\, \eta_{\Box A} \\
=\langle\ & \text{lemma C.1}\ \rangle \\
& \mathsf{Wk}(\Gamma \supseteq \Delta) \,;\, \rho(\Delta) \,;\, \mathcal{M}(\Delta) \,;\, \Box[\![\Delta^{s} \vdash e : A]\!] \,;\, \Phi_A \,;\, \eta_{\Box A} \\
=\langle\ & \text{definition}\ \rangle \\
& \mathsf{Wk}(\Gamma \supseteq \Delta) \,;\, [\![\Delta \vdash^{s} e : A]\!]_p \,;\, \eta_{\Box A} \\
=\langle\ & \text{definition}\ \rangle \\
& \mathsf{Wk}(\Gamma \supseteq \Delta) \,;\, [\![\Delta \vdash \mathsf{box}\,\boxed{e} : \Box A]\!]
\end{aligned}
$$

$$\diamond\ \frac{\Gamma \vdash e_1 : \Box A \qquad \Gamma, x : A^{s} \vdash e_2 : B}{\Gamma \vdash \mathsf{let\ box}\,\boxed{x} = e_1\ \mathsf{in}\ e_2 : B}\ \Box\mathrm{E}$$

$$\llbracket \Gamma \vdash \mathsf{let\ box}\ \boxed{x} = e_1\ \mathsf{in}\ e_2 : B \rrbracket$$

$=\langle$ definition $\rangle$

$$\langle id_\Gamma , \llbracket \Gamma \vdash e_1 : \Box A \rrbracket \rangle \mathbin{;} \tau_{\Gamma,\Box A} \mathbin{;} T\llbracket \Gamma, x : A^s \vdash e_2 : B \rrbracket \mathbin{;} \mu_B$$

$=\langle$ induction hypothesis $\rangle$

$$\langle id_\Gamma , \mathsf{Wk}(\Gamma \supseteq \Delta) \mathbin{;} \llbracket \Delta \vdash e_1 : \Box A \rrbracket \rangle \mathbin{;} \tau_{\Gamma,\Box A} \mathbin{;} T(\mathsf{Wk}(\Gamma, x : A^s \supseteq \Delta, x : A^s) \mathbin{;} \llbracket \Delta, x : A^s \vdash e_2 : B \rrbracket) \mathbin{;} \mu_B$$

$=\langle$ definition $\rangle$

$$\langle id_\Gamma , \mathsf{Wk}(\Gamma \supseteq \Delta) \mathbin{;} \llbracket \Delta \vdash e_1 : \Box A \rrbracket \rangle \mathbin{;} \tau_{\Gamma,\Box A} \mathbin{;} T([\mathsf{Wk}(\Gamma \supseteq \Delta) \times id_{\Box A}] \mathbin{;} \llbracket \Delta, x : A^s \vdash e_2 : B \rrbracket) \mathbin{;} \mu_B$$

$=\langle$ $T$ preserves composition $\rangle$

$$\langle id_\Gamma , \mathsf{Wk}(\Gamma \supseteq \Delta) \mathbin{;} \llbracket \Delta \vdash e_1 : \Box A \rrbracket \rangle \mathbin{;} \tau_{\Gamma,\Box A} \mathbin{;} T[\mathsf{Wk}(\Gamma \supseteq \Delta) \times id_{\Box A}] \mathbin{;} T\llbracket \Delta, x : A^s \vdash e_2 : B \rrbracket \mathbin{;} \mu_B$$

$=\langle$ tensorial strength of $T$ $\rangle$

$$\langle id_\Gamma , \mathsf{Wk}(\Gamma \supseteq \Delta) \mathbin{;} \llbracket \Delta \vdash e_1 : \Box A \rrbracket \rangle \mathbin{;} [\mathsf{Wk}(\Gamma \supseteq \Delta) \times id_{T\Box A}] \mathbin{;} \tau_{\Delta,\Box A} \mathbin{;} T\llbracket \Delta, x : A^s \vdash e_2 : B \rrbracket \mathbin{;} \mu_B$$

$=\langle$ composition of products $\rangle$

$$\langle id_\Gamma \mathbin{;} \mathsf{Wk}(\Gamma \supseteq \Delta) , \mathsf{Wk}(\Gamma \supseteq \Delta) \mathbin{;} \llbracket \Delta \vdash e_1 : \Box A \rrbracket \mathbin{;} id_{T\Box A} \rangle \mathbin{;} \tau_{\Delta,\Box A} \mathbin{;} T\llbracket \Delta, x : A^s \vdash e_2 : B \rrbracket \mathbin{;} \mu_B$$

$=\langle$ identity law $\rangle$

$$\langle \mathsf{Wk}(\Gamma \supseteq \Delta) \mathbin{;} id_\Delta , \mathsf{Wk}(\Gamma \supseteq \Delta) \mathbin{;} \llbracket \Delta \vdash e_1 : \Box A \rrbracket \rangle \mathbin{;} \tau_{\Delta,\Box A} \mathbin{;} T\llbracket \Delta, x : A^s \vdash e_2 : B \rrbracket \mathbin{;} \mu_B$$

$=\langle$ universal property of products $\rangle$

$$\mathsf{Wk}(\Gamma \supseteq \Delta) \mathbin{;} \langle id_\Delta , \llbracket \Delta \vdash e_1 : \Box A \rrbracket \rangle \mathbin{;} \tau_{\Delta,\Box A} \mathbin{;} T\llbracket \Delta, x : A^s \vdash e_2 : B \rrbracket \mathbin{;} \mu_B$$

$=\langle$ definition $\rangle$

$$\mathsf{Wk}(\Gamma \supseteq \Delta) \mathbin{;} \llbracket \Delta \vdash \mathsf{let\ box}\ \boxed{x} = e_1\ \mathsf{in}\ e_2 : B \rrbracket$$

□

Lemma C.3. *If* $\Gamma \supseteq \Delta$ *and* $\Delta \vdash^s e : A$*, then*

$$\llbracket \Gamma \vdash^s e : A \rrbracket_p = \mathsf{Wk}(\Gamma \supseteq \Delta) \mathbin{;} \llbracket \Delta \vdash^s e : A \rrbracket_p.$$

Proof.

$$\llbracket \Gamma \vdash^s e : A \rrbracket_p$$

$=\langle$ definition $\rangle$

$$\rho(\Gamma) \,;\, \mathcal{M}(\Gamma) \,;\, \Box[\![\Gamma^{\mathsf{s}} \vdash e : A]\!] \,;\, \Phi_A$$

$=\langle$ semantic weakening lemma 5.1 $\rangle$

$$\rho(\Gamma) \,;\, \mathcal{M}(\Gamma) \,;\, \Box(\mathsf{Wk}(\Gamma^{\mathsf{s}} \supseteq \Delta^{\mathsf{s}}) \,;\, [\![\Delta^{\mathsf{s}} \vdash e : A]\!]) \,;\, \Phi_A$$

$=\langle$ $\Box$ preserves composition $\rangle$

$$\rho(\Gamma) \,;\, \mathcal{M}(\Gamma) \,;\, \Box\,\mathsf{Wk}(\Gamma^{\mathsf{s}} \supseteq \Delta^{\mathsf{s}}) \,;\, \Box[\![\Delta^{\mathsf{s}} \vdash e : A]\!] \,;\, \Phi_A$$

$=\langle$ lemma C.1 $\rangle$

$$\mathsf{Wk}(\Gamma \supseteq \Delta) \,;\, \rho(\Delta) \,;\, \mathcal{M}(\Delta) \,;\, [\![\Delta^{\mathsf{s}} \vdash e : A]\!] \,;\, \Phi_A$$

$=\langle$ definition $\rangle$

$$\mathsf{Wk}(\Gamma \supseteq \Delta) \,;\, [\![\Delta \vdash^{\mathsf{s}} e : A]\!]_p$$

□

Lemma C.4. *If* $\Gamma \supseteq \Delta$ *and* $\Delta \vdash v : A$, *then*

$$[\![\Gamma \vdash v : A]\!]_v = \mathsf{Wk}(\Gamma \supseteq \Delta) \,;\, [\![\Delta \vdash v : A]\!]_v.$$

Proof. Assuming $\Gamma \supseteq \Delta$, we do induction on $\Delta \vdash v : A$.

◇ $$\frac{x : A^q \in \Gamma}{\Gamma \vdash x : A}\ \textsc{Var}$$

$$[\![\Gamma \vdash v : A]\!]_v$$

$=\langle$ definition $\rangle$

$$[\![x : A^q \in \Gamma]\!]$$

$=\langle$ lemma C.2 $\rangle$

$$\mathsf{Wk}(\Gamma \supseteq \Delta) \,;\, [\![x : A^q \in \Delta]\!]$$

$=\langle$ definition $\rangle$

$$\mathsf{Wk}(\Gamma \supseteq \Delta) \,;\, [\![\Delta \vdash x : A]\!]_v$$

◇ $$\frac{}{\Gamma \vdash () : \mathsf{unit}}\ \mathsf{unitI}$$

$$[\![\Gamma \vdash () : \mathsf{unit}]\!]_v$$

$=\langle$ definition $\rangle$

$$!_\Gamma$$

$=\langle$ universal property of 1 $\rangle$

$$\mathsf{Wk}(\Gamma \supseteq \Delta) \,;\, !_\Delta$$

$=\langle$ definition $\rangle$

$$\mathsf{Wk}(\Gamma \supseteq \Delta) \,;\, [\![\Gamma \vdash () : \mathsf{unit}]\!]_v$$

$$\diamond \; \frac{\Gamma \vdash e_1 : A \qquad \Gamma \vdash e_2 : B}{\Gamma \vdash (e_1 , e_2) : A \times B} \times\mathrm{I}$$

$$\begin{aligned}
& [\![\Gamma \vdash (v_1 , v_2) : A \times B]\!]_v \\
=\langle & \text{ definition } \rangle \\
& \langle [\![\Gamma \vdash v_1 : A]\!]_v , [\![\Gamma \vdash v_2 : B]\!]_v \rangle \\
=\langle & \text{ induction hypothesis } \rangle \\
& \langle \mathsf{Wk}(\Gamma \supseteq \Delta) \,;\, [\![\Delta \vdash v_1 : A]\!]_v , \mathsf{Wk}(\Gamma \supseteq \Delta) \,;\, [\![\Delta \vdash v_2 : B]\!]_v \rangle \\
=\langle & \text{ universal property of products } \rangle \\
& \mathsf{Wk}(\Gamma \supseteq \Delta) \,;\, \langle [\![\Delta \vdash v_1 : A]\!]_v , [\![\Delta \vdash v_2 : B]\!]_v \rangle \\
=\langle & \text{ definition } \rangle \\
& \mathsf{Wk}(\Gamma \supseteq \Delta) \,;\, [\![\Delta \vdash (v_1 , v_2) : A \times B]\!]_v
\end{aligned}$$

$$\diamond \; \frac{\Gamma, x : A^{\mathsf{i}} \vdash e : B}{\Gamma \vdash \lambda x : A.\ e : A \Rightarrow B} \Rightarrow\mathrm{I}$$

$$\begin{aligned}
& [\![\Gamma \vdash \lambda x.\ e : A \Rightarrow B]\!]_v \\
=\langle & \text{ definition } \rangle \\
& \mathsf{curry}\,([\![\Gamma, x : A^{\mathsf{i}} \vdash e : B]\!]) \\
=\langle & \text{ semantic weakening lemma 5.1 } \rangle \\
& \mathsf{curry}\,(\mathsf{Wk}(\Gamma, x : A^{\mathsf{i}} \supseteq \Delta, x : A^{\mathsf{i}}) \,;\, [\![\Delta, x : A^{\mathsf{i}} \vdash e : B]\!]) \\
=\langle & \text{ definition } \rangle \\
& \mathsf{curry}\,([\mathsf{Wk}(\Gamma \supseteq \Delta) \times id_A]; [\![\Delta, x : A^{\mathsf{i}} \vdash e : B]\!]) \\
=\langle & \text{ universal property of exponential } \rangle \\
& \mathsf{Wk}(\Gamma \supseteq \Delta) \,;\, \mathsf{curry}\,([\![\Delta, x : A^{\mathsf{i}} \vdash e : B]\!]) \\
=\langle & \text{ definition } \rangle \\
& \mathsf{Wk}(\Gamma \supseteq \Delta) \,;\, [\![\Delta \vdash \lambda x.\ e : A \Rightarrow B]\!]_v
\end{aligned}$$

$$\diamond \; \frac{\Gamma \vdash^{\mathsf{s}} e : A}{\Gamma \vdash \mathsf{box}\, \boxed{e} : \square A} \square\mathrm{I}$$

$$\begin{aligned}
& [\![\Gamma \vdash \mathsf{box}\, \boxed{e} : \square A]\!]_v \\
=\langle & \text{ definition } \rangle \\
& [\![\Gamma \vdash^{\mathsf{s}} e : A]\!]_p
\end{aligned}$$

$=\langle$ lemma C.3 $\rangle$

$\mathsf{Wk}(\Gamma \supseteq \Delta) \,;\, [\![\Delta \vdash^{s} e : A]\!]_p$

$=\langle$ definition $\rangle$

$\mathsf{Wk}(\Gamma \supseteq \Delta) \,;\, [\![\Delta \vdash \mathsf{box}\,\boxed{e} : \Box A]\!]_v$

$\square$

Lemma C.5. *If* $\Gamma \supseteq \Delta$ *and* $\Delta \vdash \theta : \Psi$, *then*

$$[\![\Gamma \vdash \theta : \Psi]\!] = \mathsf{Wk}(\Gamma \supseteq \Delta) \,;\, [\![\Delta \vdash \theta : \Psi]\!]$$

Proof. Assume $\Gamma \supseteq \Delta$. We proceed by induction on $\Delta \vdash \theta : \Psi$.

$$\diamond \ \frac{}{\Gamma \vdash \langle\rangle :} \ \text{SUB-ID}$$

$[\![\Gamma \vdash \langle\rangle : ]\!]$

$=\langle$ definition $\rangle$

$!_\Gamma$

$=\langle$ universal property of 1 $\rangle$

$\mathsf{Wk}(\Gamma \supseteq \Delta) \,;\, !_\Delta$

$=\langle$ definition $\rangle$

$\mathsf{Wk}(\Gamma \supseteq \Delta) \,;\, [\![\Delta \vdash \langle\rangle : ]\!]$

$$\diamond \ \frac{\Gamma \vdash \theta : \Delta \qquad \Gamma \vdash^{s} e : A}{\Gamma \vdash \langle \theta, e^{s}/x \rangle : \Delta, x : A^{s}} \ \text{SUB-SAFE}$$

$[\![\Gamma \vdash \langle \theta, e^{s}/x \rangle : \Psi, x : A^{s}]\!]$

$=\langle$ definition $\rangle$

$\langle [\![\Gamma \vdash \theta : \psi]\!] \,,\, [\![\Gamma \vdash^{s} e : A]\!]_p \rangle$

$=\langle$ induction hypothesis $\rangle$

$\langle \mathsf{Wk}(\Gamma \supseteq \Delta) \,;\, [\![\Delta \vdash \theta : \psi]\!] \,,\, [\![\Gamma \vdash^{s} e : A]\!]_p \rangle$

$=\langle$ lemma C.3 $\rangle$

$\langle \mathsf{Wk}(\Gamma \supseteq \Delta) \,;\, [\![\Delta \vdash \theta : \psi]\!] \,,\, \mathsf{Wk}(\Gamma \supseteq \Delta) \,;\, [\![\Delta \vdash^{s} e : A]\!]_p \rangle$

$=\langle$ universal property of products $\rangle$

$\mathsf{Wk}(\Gamma \supseteq \Delta) \,;\, \langle [\![\Delta \vdash \theta : \psi]\!] \,,\, [\![\Delta \vdash^{s} e : A]\!]_p \rangle$

$=\langle$ definition $\rangle$

$$\mathrm{Wk}(\Gamma \supseteq \Delta) \mathbin{;} [\![ \Delta \vdash \langle \theta, e^{\mathsf{s}}/x \rangle : \Psi, x : A^{\mathsf{s}} ]\!]$$

$$\diamond \frac{\Gamma \vdash \theta : \Delta \qquad \Gamma \vdash v : A}{\Gamma \vdash \langle \theta, v^{\mathsf{i}}/x \rangle : \Delta, x : A^{\mathsf{i}}} \text{ SUB-IMPURE}$$

$$
\begin{aligned}
& [\![ \Gamma \vdash \langle \theta, v^{\mathsf{i}}/x \rangle : \Psi, x : A^{\mathsf{i}} ]\!] \\
=\langle & \text{ definition } \rangle \\
& \langle [\![ \Gamma \vdash \theta : \Psi ]\!], [\![ \Gamma \vdash v : A ]\!]_v \rangle \\
=\langle & \text{ induction hypothesis } \rangle \\
& \langle \mathrm{Wk}(\Gamma \supseteq \Delta) \mathbin{;} [\![ \Delta \vdash \theta : \Psi ]\!], [\![ \Gamma \vdash v : A ]\!]_v \rangle \\
=\langle & \text{ lemma C.4 } \rangle \\
& \langle \mathrm{Wk}(\Gamma \supseteq \Delta) \mathbin{;} [\![ \Delta \vdash \theta : \Psi ]\!], \mathrm{Wk}(\Gamma \supseteq \Delta) \mathbin{;} [\![ \Delta \vdash v : A ]\!]_v \rangle \\
=\langle & \text{ universal property of products } \rangle \\
& \mathrm{Wk}(\Gamma \supseteq \Delta) \mathbin{;} \langle [\![ \Delta \vdash \theta : \Psi ]\!], [\![ \Delta \vdash v : A ]\!]_v \rangle \\
=\langle & \text{ definition } \rangle \\
& \mathrm{Wk}(\Gamma \supseteq \Delta) \mathbin{;} [\![ \Delta \vdash \langle \theta, v^{\mathsf{i}}/x \rangle : \Psi, x : A^{\mathsf{i}} ]\!]
\end{aligned}
$$

□

Lemma C.6. *If* $\Gamma^{\mathsf{s}} \vdash e : A^{\mathsf{s}}$, *then*

$$\rho(\Gamma) \mathbin{;} \mathcal{M}(\Gamma) \mathbin{;} \Box [\![ \Gamma^{\mathsf{s}} \vdash^{\mathsf{s}} e : A ]\!]_p = [\![ \Gamma \vdash^{\mathsf{s}} e : A ]\!]_p \mathbin{;} \delta_A$$

Proof.

$$
\begin{aligned}
& \rho(\Gamma) \mathbin{;} \mathcal{M}(\Gamma) \mathbin{;} \Box [\![ \Gamma^{\mathsf{s}} \vdash^{\mathsf{s}} e : A ]\!]_p \\
=\langle & \text{ definition } \rangle \\
& \rho(\Gamma) \mathbin{;} \mathcal{M}(\Gamma) \mathbin{;} \Box(\rho(\Gamma^{\mathsf{s}}) \mathbin{;} \mathcal{M}(\Gamma^{\mathsf{s}}) \mathbin{;} \Box [\![ \Gamma^{\mathsf{s}} \vdash e : A ]\!] \mathbin{;} \Phi_A) \\
=\langle & \ \Box \text{ preserves composition } \rangle \\
& \rho(\Gamma) \mathbin{;} \mathcal{M}(\Gamma) \mathbin{;} \Box\, \rho(\Gamma^{\mathsf{s}}) \mathbin{;} \Box\, \mathcal{M}(\Gamma^{\mathsf{s}}) \mathbin{;} \Box\Box [\![ \Gamma^{\mathsf{s}} \vdash e : A ]\!] \mathbin{;} \Box\, \Phi_A \\
=\langle & \text{ definition } \rangle \\
& \rho(\Gamma) \mathbin{;} \mathcal{M}(\Gamma) \mathbin{;} \Box id_{\Gamma^{\mathsf{s}}} \mathbin{;} \delta_{\Gamma^{\mathsf{s}}} \mathbin{;} \delta^{-1}_{\Gamma^{\mathsf{s}}} \mathbin{;} \Box [\![ \Gamma^{\mathsf{s}} \vdash e : A ]\!] \mathbin{;} \Phi_A \mathbin{;} \delta_A \\
=\langle & \text{ simplification } \rangle \\
& \rho(\Gamma) \mathbin{;} \mathcal{M}(\Gamma) \mathbin{;} \Box [\![ \Gamma^{\mathsf{s}} \vdash e : A ]\!] \mathbin{;} \Phi_A \mathbin{;} \delta_A \\
=\langle & \text{ definition } \rangle \\
& [\![ \Gamma \vdash^{\mathsf{s}} e : A ]\!]_p \mathbin{;} \delta_A
\end{aligned}
$$

□

**Lemma C.7.** *If* $\Gamma \vdash \theta : \Delta$, *then*

$$\rho(\Gamma) ; \mathcal{M}(\Gamma) ; \Box[\![\Gamma^s \vdash \theta^s : \Delta^s]\!] = [\![\Gamma \vdash \theta : \Delta]\!] ; \rho(\Delta) ; \mathcal{M}(\Delta)$$

**Proof.** We do induction on $\Gamma \vdash \theta : \Delta$.

$$\diamond \quad \frac{}{\Gamma \vdash \langle\rangle :} \text{ SUB-ID}$$

$$\rho(\Gamma) ; \mathcal{M}(\Gamma) ; \Box[\![\Gamma^s \vdash \langle\rangle : ]\!]$$

$=\langle$ definition $\rangle$

$$\rho(\Gamma) ; \mathcal{M}(\Gamma) ; \Box !_{\Gamma^s}$$

$=\langle$ definition $\rangle$

$$\rho(\Gamma) ; \mathcal{M}(\Gamma) ; !_{\Box\Gamma^s}$$

$=\langle$ universal property of 1 $\rangle$

$$!_\Gamma$$

$=\langle$ identity law $\rangle$

$$!_\Gamma ; id_1 ; id_1$$

$=\langle$ definition $\rangle$

$$[\![\Gamma \vdash \langle\rangle : ]\!] ; \rho() ; \mathcal{M}()$$

$$\diamond \quad \frac{\Gamma \vdash \theta : \Delta \qquad \Gamma \vdash^s e : A}{\Gamma \vdash \langle \theta, e^s/x \rangle : \Delta, x : A^s} \text{ SUB-SAFE}$$

$$\rho(\Gamma) ; \mathcal{M}(\Gamma) ; \Box[\![\Gamma^s \vdash \langle \theta^s, e^s/x \rangle : \Delta^s, x : A^s]\!]$$

$=\langle$ definition $\rangle$

$$\rho(\Gamma) ; \mathcal{M}(\Gamma) ; \Box\langle [\![\Gamma^s \vdash \theta^s : \Delta^s]\!] , [\![\Gamma^s \vdash^s e : A]\!]_p \rangle$$

$=\langle$ monoidal action of $\Box$ $\rangle$

$$\rho(\Gamma) ; \mathcal{M}(\Gamma) ; \langle \Box[\![\Gamma^s \vdash \theta^s : \Delta^s]\!] , \Box[\![\Gamma^s \vdash^s e : A]\!]_p \rangle ; m^\times_{\Delta^s, \Box A}$$

$=\langle$ universal property of products $\rangle$

$$\langle \rho(\Gamma) ; \mathcal{M}(\Gamma) ; \Box[\![\Gamma^s \vdash \theta^s : \Delta^s]\!] , \rho(\Gamma) ; \mathcal{M}(\Gamma) ; \Box[\![\Gamma^s \vdash^s e : A]\!]_p \rangle ; m^\times_{\Delta^s, \Box A}$$

$=\langle$ induction hypothesis $\rangle$

$$\langle [\![\Gamma \vdash \theta : \Delta]\!] ; \rho(\Delta) ; \mathcal{M}(\Delta) , \rho(\Gamma) ; \mathcal{M}(\Gamma) ; \Box[\![\Gamma^s \vdash^s e : A]\!]_p \rangle ; m^\times_{\Delta^s, \Box A}$$

$=\langle$ lemma C.6 $\rangle$

$$\langle [\![\Gamma \vdash \theta : \Delta]\!] ; \rho(\Delta) ; \mathcal{M}(\Delta) , [\![\Gamma \vdash^s e : A]\!]_p ; \delta_A \rangle ; m^\times_{\Delta^s, \Box A}$$

$=\langle$ identity law $\rangle$

$$\langle [\![\Gamma \vdash \theta : \Delta]\!] ; \rho(\Delta) ; \mathcal{M}(\Delta) , [\![\Gamma \vdash^{s} e : A]\!]_p ; id_{\Box A} ; \delta_A \rangle ; m^{\times}_{\Delta^s, \Box A}$$

$=\langle$ universal property of products $\rangle$

$$\langle [\![\Gamma \vdash \theta : \Delta]\!] , [\![\Gamma \vdash^{s} e : A]\!]_p \rangle ; [\rho(\Delta) ; \mathcal{M}(\Delta) \times id_{\Box A} ; \delta_A] ; m^{\times}_{\Delta^s, \Box A}$$

$=\langle$ exchange law $\rangle$

$$\langle [\![\Gamma \vdash \theta : \Delta]\!] , [\![\Gamma \vdash^{s} e : A]\!]_p \rangle ; [\rho(\Delta) \times id_{\Box A}] ; [\mathcal{M}(\Delta) \times \delta_A] ; m^{\times}_{\Delta^s, \Box A}$$

$=\langle$ definition $\rangle$

$$[\![\Gamma \vdash \langle \theta, e^{s}/x \rangle : \Delta, x : A^{s}]\!] ; \rho(\Delta, x : A^{s}) ; \mathcal{M}(\Delta, x : A^{s})$$

$$\diamond \frac{\Gamma \vdash \theta : \Delta \qquad \Gamma \vdash v : A}{\Gamma \vdash \langle \theta, v^{i}/x \rangle : \Delta, x : A^{i}} \text{ SUB-IMPURE}$$

$$\rho(\Gamma) ; \mathcal{M}(\Gamma) ; \Box [\![\Gamma^{s} \vdash \langle \theta, v^{i}/x \rangle^{s} : (\Delta, x : A^{i})^{s}]\!]$$

$=\langle$ definition $\rangle$

$$\rho(\Gamma) ; \mathcal{M}(\Gamma) ; \Box [\![\Gamma^{s} \vdash \theta^{s} : \Delta^{s}]\!]$$

$=\langle$ induction hypothesis $\rangle$

$$[\![\Gamma \vdash \theta : \Delta]\!] ; \rho(\Delta) ; \mathcal{M}(\Delta)$$

$=\langle$ definition of $\pi_1$ $\rangle$

$$\langle [\![\Gamma \vdash \theta : \Delta]\!] , [\![\Gamma \vdash v : A]\!]_v \rangle ; \pi_1 ; \rho(\Delta) ; \mathcal{M}(\Delta)$$

$=\langle$ definition $\rangle$

$$[\![\Gamma \vdash \langle \theta, v^{i}/x \rangle : \Delta, x : A^{i}]\!] ; \rho(\Delta, x : A^{i}) ; \mathcal{M}(\Delta, x : A^{i})$$

□

Lemma C.8. *For any context* $\Gamma$,

$$\mathrm{Wk}(\Gamma \supseteq \Gamma^{s}) = \rho(\Gamma)$$

Proof. We do induction on $\Gamma$.

$\diamond\, \Gamma =$

$$\mathrm{Wk}( \supseteq {}^{s})$$

$=\langle$ definition $\rangle$

$$\mathrm{Wk}( \supseteq )$$

$=\langle$ definition $\rangle$

$$id_1$$

$=\langle$ definition $\rangle$

$$\rho()$$

$\diamond\, \Gamma = \Delta, x : A^q$
When $q = \mathsf{s}$,

$$\begin{aligned}
& \mathrm{Wk}(\Delta, x : A^{\mathsf{s}} \supseteq \Delta^{\mathsf{s}}, x : A^{\mathsf{s}}) \\
=& \langle \text{ definition } \rangle \\
& [\mathrm{Wk}(\Delta \supseteq \Delta^{\mathsf{s}}) \times id_{\square A}] \\
=& \langle \text{ induction hypothesis } \rangle \\
& [\rho(\Delta) \times id_{\square A}] \\
=& \langle \text{ definition } \rangle \\
& \rho(\Delta, x : A^{\mathsf{s}})
\end{aligned}$$

When $q = \mathsf{i}$,

$$\begin{aligned}
& \mathrm{Wk}(\Delta, x : A^{\mathsf{i}} \supseteq \Delta^{\mathsf{s}}) \\
=& \langle \text{ definition } \rangle \\
& \pi_1 \mathbin{;} \mathrm{Wk}(\Delta \supseteq \Delta^{\mathsf{s}}) \\
=& \langle \text{ induction hypothesis } \rangle \\
& \pi_1 \mathbin{;} \rho(\Delta) \\
=& \langle \text{ definition } \rangle \\
& \rho(\Delta, x : A^{\mathsf{i}})
\end{aligned}$$

□

Lemma 5.2 (Safe interpretation). *If* $\Gamma \vdash^{\mathsf{s}} e : A$, *then*

$$[\![\Gamma \vdash e : A]\!] = [\![\Gamma \vdash^{\mathsf{s}} e : A]\!]_p \mathbin{;} \epsilon_A \mathbin{;} \eta_A.$$

Proof. Assume $\Gamma \vdash^{\mathsf{s}} e : A$. By inversion, we have $\Gamma^{\mathsf{s}} \vdash e : A$.

$$\begin{aligned}
& [\![\Gamma \vdash e : A]\!] \\
=& \langle \text{ semantic weakening lemma 5.1 } \rangle \\
& \mathrm{Wk}(\Gamma \supseteq \Gamma^{\mathsf{s}}) \mathbin{;} [\![\Gamma^{\mathsf{s}} \vdash e : A]\!] \\
=& \langle \text{ lemma C.8 } \rangle \\
& \rho(\Gamma) \mathbin{;} [\![\Gamma^{\mathsf{s}} \vdash e : A]\!] \\
=& \langle \text{ definition } \rangle \\
& \rho(\Gamma) \mathbin{;} \mathcal{M}(\Gamma) \mathbin{;} \square[\![\Gamma^{\mathsf{s}} \vdash e : A]\!] \mathbin{;} \epsilon_{TA}
\end{aligned}$$

$=\langle$ definition $\rangle$

$\rho(\Gamma) \,;\, \mathcal{M}(\Gamma) \,;\, \Box[\![\Gamma^{\mathsf{s}} \vdash e : A]\!] \,;\, \Phi_A \,;\, \epsilon_A \,;\, \eta_A$

$=\langle$ definition $\rangle$

$[\![\Gamma \vdash^{\mathsf{s}} e : A]\!]_p \,;\, \epsilon_A \,;\, \eta_A$

□

Lemma 5.3 (Value interpretation). *If* $\Gamma \vdash v : A$, *then*

$$[\![\Gamma \vdash v : A]\!] = [\![\Gamma \vdash v : A]\!]_v \,;\, \eta_A.$$

Proof. We proceed by induction on $\Gamma \vdash v : A$.

$\diamond$ $\dfrac{}{\Gamma \vdash () : \mathsf{unit}}$ unitI

$[\![\Gamma \vdash () : \mathsf{unit}]\!]$

$=\langle$ definition $\rangle$

$!_\Gamma \,;\, \eta_1$

$=\langle$ definition $\rangle$

$[\![\Gamma \vdash () : \mathsf{unit}]\!]_v \,;\, \eta_1$

$\diamond$ $\dfrac{\Gamma \vdash v_1 : A \qquad \Gamma \vdash v_2 : B}{\Gamma \vdash (v_1\,, v_2) : A \times B}$ $\times$I

$[\![\Gamma \vdash (v_1\,, v_2) : A \times B]\!]$

$=\langle$ definition $\rangle$

$\langle [\![\Gamma \vdash v_1 : A]\!]\,,\, [\![\Gamma \vdash v_2 : B]\!] \rangle \,;\, \beta_{A,B}$

$=\langle$ induction hypothesis $\rangle$

$\langle [\![\Gamma \vdash v_1 : A]\!]_v \,;\, \eta_A\,,\, [\![\Gamma \vdash v_2 : B]\!]_v \,;\, \eta_B \rangle \,;\, \beta_{A,B}$

$=\langle$ tensorial strength of $T$ $\rangle$

$\langle [\![\Gamma \vdash v_1 : A]\!]_v\,,\, [\![\Gamma \vdash v_2 : B]\!]_v \,;\, \eta_B \rangle \,;\, \sigma_{A,B}$

$=\langle$ tensorial strength of $T$ $\rangle$

$\langle [\![\Gamma \vdash v_1 : A]\!]_v\,,\, [\![\Gamma \vdash v_2 : B]\!]_v \rangle \,;\, \eta_{A\times B}$

$=\langle$ definition $\rangle$

$[\![\Gamma \vdash (v_1\,, v_2) : A \times B]\!]_v \,;\, \eta_{A\times B}$

$\diamond$ $\dfrac{x : A^q \in \Gamma}{\Gamma \vdash x : A}$ Var

$$\llbracket \Gamma \vdash x : A \rrbracket$$
$=\langle$ definition $\rangle$
$$\llbracket x : A^q \in \Gamma \rrbracket \mathbin{;} \eta_A$$
$=\langle$ definition $\rangle$
$$\llbracket \Gamma \vdash x : A \rrbracket_v \mathbin{;} \eta_A$$

$\diamond$ $$\frac{\Gamma, x : A^{\mathsf{i}} \vdash e : B}{\Gamma \vdash \lambda x : A.\ e : A \Rightarrow B} \Rightarrow\mathrm{I}$$

$$\llbracket \Gamma \vdash \lambda x.\ e : A \Rightarrow B \rrbracket$$
$=\langle$ definition $\rangle$
$$\mathsf{curry}\,(\llbracket \Gamma, x : A^{\mathsf{i}} \vdash e : B \rrbracket) \mathbin{;} \eta_{A \to TB}$$
$=\langle$ definition $\rangle$
$$\llbracket \Gamma \vdash \lambda x.\ e : A \Rightarrow B \rrbracket_v \mathbin{;} \eta_{A \to TB}$$

$\diamond$ $$\frac{\Gamma \vdash^{\mathsf{s}} e : A}{\Gamma \vdash \mathsf{box}\,\boxed{e} : \Box A} \Box\mathrm{I}$$

$$\llbracket \Gamma \vdash \mathsf{box}\,\boxed{e} : \Box A \rrbracket$$
$=\langle$ definition $\rangle$
$$\llbracket \Gamma \vdash^{\mathsf{s}} e : A \rrbracket_p \mathbin{;} \eta_{\Box A}$$
$=\langle$ definition $\rangle$
$$\llbracket \Gamma \vdash \mathsf{box}\,\boxed{e} : \Box A \rrbracket_v \mathbin{;} \eta_{\Box A}$$

$\square$

Lemma C.9. *If* $\Gamma \vdash \theta : \Delta$ *and* $x : A^q \in \Delta$, *then*

$$\llbracket \Gamma \vdash \theta[x] : A \rrbracket = \llbracket \Gamma \vdash \theta : \Delta \rrbracket \mathbin{;} \llbracket x : A^q \in \Delta \rrbracket \mathbin{;} \eta_A$$

Proof. We proceed by induction on $x : A^q \in \Delta$.

$\diamond$ $$\frac{}{x : A^q \in (\Gamma, x : A^q)} \in\text{-ID}$$
When $q = \mathsf{s}$,

$$\llbracket \Gamma \vdash \langle \phi, e^{\mathsf{s}} / x \rangle [x] : A \rrbracket$$
$=\langle$ definition $\rangle$
$$\llbracket \Gamma \vdash e : A \rrbracket$$

$=\langle$ safe interpretation lemma 5.2 $\rangle$

$$[\![\Gamma \vdash^{s} e : A]\!]_p \,;\, \epsilon_A \,;\, \eta_A$$

$=\langle$ definition of $\pi_2$ $\rangle$

$$\langle [\![\Gamma \vdash \phi : \Delta]\!] \,,\, [\![\Gamma \vdash^{s} e : A]\!]_p \rangle \,;\, \pi_2 \,;\, \epsilon_A \,;\, \eta_A$$

$=\langle$ definition $\rangle$

$$[\![\Gamma \vdash \langle \phi, e^{s}/x \rangle : \Delta, x : A^{s}]\!] \,;\, [\![x : A^{s} \in (\Delta, x : A^{s})]\!] \,;\, \eta_A$$

When $q = \mathbf{i}$,

$$[\![\Gamma \vdash \langle \phi, v^{i}/x \rangle [x] : A]\!]$$

$=\langle$ definition $\rangle$

$$[\![\Gamma \vdash v : A]\!]$$

$=\langle$ value interpretation lemma 5.3 $\rangle$

$$[\![\Gamma \vdash v : A]\!]_v \,;\, \eta_A$$

$=\langle$ definition of $\pi_2$ $\rangle$

$$\langle [\![\Gamma \vdash \phi : \Delta]\!] \,,\, [\![\Gamma \vdash v : A]\!]_v \rangle \,;\, \pi_2 \,;\, \eta_A$$

$=\langle$ definition $\rangle$

$$[\![\Gamma \vdash \langle \phi, v^{i}/x \rangle : \Delta, x : A^{i}]\!] \,;\, [\![x : A^{i} \in (\Delta, x : A^{i})]\!] \,;\, \eta_A$$

$$\diamond\ \frac{x : A^{q} \in \Gamma \qquad (x \neq y)}{x : A^{q} \in (\Gamma, y : B^{r})}\ \in\text{-EX}$$

When $r = \mathbf{s}$

$$[\![\Gamma \vdash \langle \phi, e^{s}/y \rangle [x] : A]\!]$$

$=\langle$ definition $\rangle$

$$[\![\Gamma \vdash \phi[x] : A]\!]$$

$=\langle$ induction hypothesis $\rangle$

$$[\![\Gamma \vdash \phi : \Delta]\!] \,;\, [\![x : A^{q} \in \Delta]\!] \,;\, \eta_A$$

$=\langle$ definition of $\pi_1$ $\rangle$

$$\langle [\![\Gamma \vdash \phi : \Delta]\!] \,,\, [\![\Gamma \vdash^{s} e : B]\!] \rangle \,;\, \pi_1 \,;\, [\![x : A^{q} \in \Delta]\!] \,;\, \eta_A$$

$=\langle$ definition $\rangle$

$$[\![\Gamma \vdash \langle \phi, e^{s}/y \rangle : \Delta, y : B^{s}]\!] \,;\, \pi_1 \,;\, [\![x : A^{q} \in \Delta]\!] \,;\, \eta_A$$

$=\langle$ definition $\rangle$

$$[\![\Gamma \vdash \langle \phi, e^{s}/y \rangle : \Delta, y : B^{s}]\!] \,;\, [\![x : A^{q} \in (\Delta, y : B^{s})]\!] \,;\, \eta_A$$

When $r = \mathsf{i}$,

$$\llbracket \Gamma \vdash \langle \phi, v^{\mathsf{i}}/y \rangle[x] : A \rrbracket$$

$=\langle$ definition $\rangle$

$$\llbracket \Gamma \vdash \phi[x] : A \rrbracket$$

$=\langle$ induction hypothesis $\rangle$

$$\llbracket \Gamma \vdash \phi : \Delta \rrbracket \,;\, \llbracket x : A^q \in \Delta \rrbracket \,;\, \eta_A$$

$=\langle$ definition of $\pi_1$ $\rangle$

$$\langle \llbracket \Gamma \vdash \phi : \Delta \rrbracket, \llbracket \Gamma \vdash v : B \rrbracket \rangle \,;\, \pi_1 \,;\, \llbracket x : A^q \in \Delta \rrbracket \,;\, \eta_A$$

$=\langle$ definition $\rangle$

$$\llbracket \Gamma \vdash \langle \phi, v^{\mathsf{i}}/y \rangle : \Delta, y : B^{\mathsf{i}} \rrbracket \,;\, \pi_1 \,;\, \llbracket x : A^q \in \Delta \rrbracket \,;\, \eta_A$$

$=\langle$ definition $\rangle$

$$\llbracket \Gamma \vdash \langle \phi, v^{\mathsf{i}}/y \rangle : \Delta, y : B^{\mathsf{i}} \rrbracket \,;\, \llbracket x : A^q \in (\Delta, y : B^{\mathsf{i}}) \rrbracket \,;\, \eta_A$$

□

THEOREM 5.4 (SEMANTIC SUBSTITUTION). *If* $\Gamma \vdash \theta : \Delta$ *and* $\Delta \vdash e : A$, *then*

$$\llbracket \Gamma \vdash \theta(e) : A \rrbracket = \llbracket \Gamma \vdash \theta : \Delta \rrbracket \,;\, \llbracket \Delta \vdash e : A \rrbracket.$$

PROOF. Assume $\Gamma \vdash \theta : \Delta$. We proceed by induction on $\Delta \vdash e : A$.

◇ $$\frac{x : A^q \in \Gamma}{\Gamma \vdash x : A}\ \textsc{Var}$$

$$\llbracket \Gamma \vdash \theta(x) : A \rrbracket$$

$=\langle$ definition $\rangle$

$$\llbracket \Gamma \vdash \theta[x] : A \rrbracket$$

$=\langle$ lemma C.9 $\rangle$

$$\llbracket \Gamma \vdash \theta : \Delta \rrbracket \,;\, \llbracket x : A^q \in \Delta \rrbracket \,;\, \eta_A$$

$=\langle$ definition $\rangle$

$$\llbracket \Gamma \vdash \theta : \Delta \rrbracket \,;\, \llbracket \Delta \vdash x : A \rrbracket$$

◇ $$\frac{}{\Gamma \vdash () : \mathsf{unit}}\ \mathsf{unitI}$$

$$\llbracket \Gamma \vdash \theta(()) : \mathsf{unit} \rrbracket$$

$=\langle$ definition $\rangle$

$$
\begin{aligned}
& [\![\Gamma \vdash () : \mathsf{unit}]\!] \\
=\langle & \text{ definition } \rangle \\
& !_\Gamma \,;\, \eta_1 \\
=\langle & \text{ universal property of 1 } \rangle \\
& [\![\Gamma \vdash \theta : \Delta]\!] \,;\, !_\Delta \,;\, \eta_1 \\
=\langle & \text{ definition } \rangle \\
& [\![\Gamma \vdash \theta : \Delta]\!] \,;\, [\![\Delta \vdash () : \mathsf{unit}]\!]
\end{aligned}
$$

$$\diamond \frac{\Gamma \vdash e_1 : A \qquad \Gamma \vdash e_2 : B}{\Gamma \vdash (e_1\,,\,e_2) : A \times B} \times\mathrm{I}$$

$$
\begin{aligned}
& [\![\Gamma \vdash \theta((e_1\,,\,e_2)) : A \times B]\!] \\
=\langle & \text{ definition } \rangle \\
& [\![\Gamma \vdash (\theta(e_1)\,,\,\theta(e_2)) : A \times B]\!] \\
=\langle & \text{ definition } \rangle \\
& \langle [\![\Gamma \vdash \theta(e_1) : A]\!]\,,\,[\![\Gamma \vdash \theta(e_2) : B]\!] \rangle \,;\, \beta_{A,B} \\
=\langle & \text{ induction hypothesis } \rangle \\
& \langle [\![\Gamma \vdash \theta : \Delta]\!] \,;\, [\![\Delta \vdash e_1 : A]\!]\,,\,[\![\Gamma \vdash \theta : \Delta]\!] \,;\, [\![\Delta \vdash e_2 : B]\!] \rangle \,;\, \beta_{A,B} \\
=\langle & \text{ universal property of products } \rangle \\
& [\![\Gamma \vdash \theta : \Delta]\!] \,;\, \langle [\![\Delta \vdash e_1 : A]\!]\,,\,[\![\Delta \vdash e_2 : B]\!] \rangle \,;\, \beta_{A,B} \\
=\langle & \text{ definition } \rangle \\
& [\![\Gamma \vdash \theta : \Delta]\!] \,;\, [\![\Delta \vdash (e_1\,,\,e_2) : A \times B]\!]
\end{aligned}
$$

$$\diamond \frac{\Gamma \vdash e : A \times B}{\Gamma \vdash \mathsf{fst}\, e : A} \times\mathrm{E}_1$$

$$
\begin{aligned}
& [\![\Gamma \vdash \theta(\mathsf{fst}\, e) : A]\!] \\
=\langle & \text{ definition } \rangle \\
& [\![\Gamma \vdash \mathsf{fst}\, \theta(e) : A]\!] \\
=\langle & \text{ definition } \rangle \\
& [\![\Gamma \vdash \theta(e) : A \times B]\!] \,;\, T\pi_1 \\
=\langle & \text{ induction hypothesis } \rangle \\
& [\![\Gamma \vdash \theta : \Delta]\!] \,;\, [\![\Delta \vdash e : A \times B]\!] \,;\, T\pi_1 \\
=\langle & \text{ definition } \rangle \\
& [\![\Gamma \vdash \theta : \Delta]\!] \,;\, [\![\Delta \vdash \mathsf{fst}\, e : A]\!]
\end{aligned}
$$

$$\diamond\ \frac{\Gamma \vdash e : A \times B}{\Gamma \vdash \mathsf{snd}\, e : B}\ \times E_2$$

$$
\begin{array}{ll}
& [\![\Gamma \vdash \theta(\mathsf{snd}\, e) : B]\!] \\
=\langle & \text{definition} \quad \rangle \\
& [\![\Gamma \vdash \mathsf{snd}\, \theta(e) : B]\!] \\
=\langle & \text{definition} \quad \rangle \\
& [\![\Gamma \vdash \theta(e) : A \times B]\!] \mathbin{;} T\pi_2 \\
=\langle & \text{induction hypothesis} \quad \rangle \\
& [\![\Gamma \vdash \theta : \Delta]\!] \mathbin{;} [\![\Delta \vdash e : A \times B]\!] \mathbin{;} T\pi_2 \\
=\langle & \text{definition} \quad \rangle \\
& [\![\Gamma \vdash \theta : \Delta]\!] \mathbin{;} [\![\Delta \vdash \mathsf{snd}\, e : B]\!]
\end{array}
$$

$$\diamond\ \frac{\Gamma \vdash^{s} e : A}{\Gamma \vdash \mathsf{box}\, \boxed{e} : \Box A}\ \Box I$$

$$
\begin{array}{ll}
& [\![\Gamma \vdash \theta(\mathsf{box}\, \boxed{e}) : \Box A]\!] \\
=\langle & \text{definition} \quad \rangle \\
& [\![\Gamma \vdash \mathsf{box}\, \boxed{\theta^{s}(e)} : \Box A]\!] \\
=\langle & \text{definition} \quad \rangle \\
& [\![\Gamma \vdash^{s} \theta^{s}(e) : A]\!]_p \mathbin{;} \eta_{\Box A} \\
=\langle & \text{definition} \quad \rangle \\
& \rho(\Gamma) \mathbin{;} \mathcal{M}(\Gamma) \mathbin{;} \Box[\![\Gamma^{s} \vdash \theta^{s}(e) : A]\!] \mathbin{;} \Phi_A \mathbin{;} \eta_{\Box A} \\
=\langle & \text{induction hypothesis} \quad \rangle \\
& \rho(\Gamma) \mathbin{;} \mathcal{M}(\Gamma) \mathbin{;} \Box([\![\Gamma^{s} \vdash \theta^{s} : \Delta^{s}]\!] \mathbin{;} [\![\Delta^{s} \vdash e : A]\!]) \mathbin{;} \Phi_A \mathbin{;} \eta_{\Box A} \\
=\langle & \Box \text{ preserves composition} \quad \rangle \\
& \rho(\Gamma) \mathbin{;} \mathcal{M}(\Gamma) \mathbin{;} \Box[\![\Gamma^{s} \vdash \theta^{s} : \Delta^{s}]\!] \mathbin{;} \Box[\![\Delta^{s} \vdash e : A]\!] \mathbin{;} \Phi_A \mathbin{;} \eta_{\Box A} \\
=\langle & \text{lemma C.7} \quad \rangle \\
& [\![\Gamma \vdash \theta : \Delta]\!] \mathbin{;} \rho(\Delta) \mathbin{;} \mathcal{M}(\Delta) \mathbin{;} \Box[\![\Delta^{s} \vdash e : A]\!] \mathbin{;} \Phi_A \mathbin{;} \eta_{\Box A} \\
=\langle & \text{definition} \quad \rangle \\
& [\![\Gamma \vdash \theta : \Delta]\!] \mathbin{;} [\![\Delta \vdash^{s} e : A]\!]_p \mathbin{;} \eta_{\Box A} \\
=\langle & \text{definition} \quad \rangle \\
& [\![\Gamma \vdash \theta : \Delta]\!] \mathbin{;} [\![\Delta \vdash \mathsf{box}\, \boxed{e} : \Box A]\!]
\end{array}
$$

$$\diamond\ \frac{\Gamma \vdash e_1 : \Box A \qquad \Gamma, x : A^s \vdash e_2 : B}{\Gamma \vdash \mathsf{let\ box}\ \boxed{x} = e_1\ \mathsf{in}\ e_2 : B}\ \Box\mathrm{E}$$

$$[\![ \Gamma \vdash \theta(\mathsf{let\ box}\ \boxed{x} = e_1\ \mathsf{in}\ e_2) : B ]\!]$$

$=\langle$ definition $\rangle$

$$[\![ \Gamma \vdash \mathsf{let\ box}\ \boxed{y} = \theta(e_1)\ \mathsf{in}\ \langle \theta, y^s/x \rangle(e_2) : B ]\!]$$

$=\langle$ definition $\rangle$

$$\langle id_\Gamma\ ,\ [\![ \Gamma \vdash \theta(e_1) : A ]\!] \rangle \mathbin{;} \tau_{\Gamma,\Box A} \mathbin{;} T[\![ \Gamma, y : A^s \vdash \langle \theta, y^s/x \rangle(e_2) : B ]\!] \mathbin{;} \mu_B$$

$=\langle$ induction hypothesis $\rangle$

$$\langle id_\Gamma\ ,\ [\![ \Gamma \vdash \theta : \Delta ]\!] \mathbin{;} [\![ \Delta \vdash e_1 : A ]\!] \rangle \mathbin{;} \tau_{\Gamma,\Box A}$$
$$\mathbin{;} T([\![ \Gamma, y : A^s \vdash \langle \theta, y^s/x \rangle : \Delta, x : A^s ]\!] \mathbin{;} [\![ \Delta, x : A^s \vdash e_2 : B ]\!]) \mathbin{;} \mu_B$$

$=\langle$ $T$ preserves composition $\rangle$

$$\langle id_\Gamma\ ,\ [\![ \Gamma \vdash \theta : \Delta ]\!] \mathbin{;} [\![ \Delta \vdash e_1 : A ]\!] \rangle \mathbin{;} \tau_{\Gamma,\Box A}$$
$$\mathbin{;} T[\![ \Gamma, y : A^s \vdash \langle \theta, y^s/x \rangle : \Delta, x : A^s ]\!] \mathbin{;} T[\![ \Delta, x : A^s \vdash e_2 : B ]\!] \mathbin{;} \mu_B$$

$=\langle$ definition $\rangle$

$$\langle id_\Gamma\ ,\ [\![ \Gamma \vdash \theta : \Delta ]\!] \mathbin{;} [\![ \Delta \vdash e_1 : A ]\!] \rangle \mathbin{;} \tau_{\Gamma,\Box A}$$
$$\mathbin{;} T\langle [\![ \Gamma, y : A^s \vdash \theta : \Delta ]\!]\ ,\ [\![ \Gamma, y : A^s \vdash^s y : A ]\!]_p \rangle$$
$$\mathbin{;} T[\![ \Delta, x : A^s \vdash e_2 : B ]\!] \mathbin{;} \mu_B$$

$=\langle$ lemma C.5 $\rangle$

$$\langle id_\Gamma\ ,\ [\![ \Gamma \vdash \theta : \Delta ]\!] \mathbin{;} [\![ \Delta \vdash e_1 : A ]\!] \rangle \mathbin{;} \tau_{\Gamma,\Box A}$$
$$\mathbin{;} T\langle \mathsf{Wk}(\Gamma, y : A^s \supseteq \Gamma) \mathbin{;} [\![ \Gamma \vdash \theta : \Delta ]\!]\ ,\ [\![ \Gamma, y : A^s \vdash^s y : A ]\!]_p \rangle$$
$$\mathbin{;} T[\![ \Delta, x : A^s \vdash e_2 : B ]\!] \mathbin{;} \mu_B$$

$=\langle$ definition $\rangle$

$$\langle id_\Gamma\ ,\ [\![ \Gamma \vdash \theta : \Delta ]\!] \mathbin{;} [\![ \Delta \vdash e_1 : A ]\!] \rangle \mathbin{;} \tau_{\Gamma,\Box A}$$
$$\mathbin{;} T\langle \pi_1 \mathbin{;} [\![ \Gamma \vdash \theta : \Delta ]\!]\ ,\ \pi_2 \rangle \mathbin{;} T[\![ \Delta, x : A^s \vdash e_2 : B ]\!] \mathbin{;} \mu_B$$

$=\langle$ universal property of products $\rangle$

$$\langle id_\Gamma\ ,\ [\![ \Gamma \vdash \theta : \Delta ]\!] \mathbin{;} [\![ \Delta \vdash e_1 : A ]\!] \rangle \mathbin{;} \tau_{\Gamma,\Box A}$$
$$\mathbin{;} T[\,[\![ \Gamma \vdash \theta : \Delta ]\!] \times id_{\Box A}\,] \mathbin{;} T[\![ \Delta, x : A^s \vdash e_2 : B ]\!] \mathbin{;} \mu_B$$

$=\langle$ tensorial strength of $T$ $\rangle$

$$\langle [\![ \Gamma \vdash \theta : \Delta ]\!] \mathbin{;} id_\Delta\ ,\ [\![ \Gamma \vdash \theta : \Delta ]\!] \mathbin{;} [\![ \Delta \vdash e_1 : A ]\!] \rangle \mathbin{;} \tau_{\Delta,\Box A}$$
$$\mathbin{;} T[\![ \Delta, x : A^s \vdash e_2 : B ]\!] \mathbin{;} \mu_B$$

$=\langle$ universal property of products $\rangle$

$$[\![ \Gamma \vdash \theta : \Delta ]\!] \mathbin{;} \langle id_\Delta\ ,\ [\![ \Delta \vdash e_1 : A ]\!] \rangle \mathbin{;} \tau_{\Delta,\Box A} \mathbin{;} T[\![ \Delta, x : A^s \vdash e_2 : B ]\!] \mathbin{;} \mu_B$$

$=\langle$ definition $\rangle$

$[\![\Gamma \vdash \theta : \Delta]\!] \mathbin{;} [\![\Delta \vdash \text{let box}\ \boxed{x} = e_1\ \text{in}\ e_2 : B]\!]$

$$\diamond \frac{\Gamma, x : A^{\mathsf{i}} \vdash e : B}{\Gamma \vdash \lambda x : A.\ e : A \Rightarrow B} \Rightarrow\text{I}$$

$[\![\Gamma \vdash \theta(\lambda x.\ e) : A \Rightarrow B]\!]$

$=\langle$ definition $\rangle$

$[\![\Gamma \vdash \lambda y.\ \langle \theta, y^{\mathsf{i}}/x \rangle(e) : A \Rightarrow B]\!]$

$=\langle$ definition $\rangle$

$\text{curry}\,([\![\Gamma, y : A^{\mathsf{i}} \vdash \langle \theta, y^{\mathsf{i}}/x \rangle(e) : B]\!]) \mathbin{;} \eta_{A \to TB}$

$=\langle$ induction hypothesis $\rangle$

$\text{curry}\,([\![\Gamma, y : A^{\mathsf{i}} \vdash \langle \theta, y^{\mathsf{i}}/x \rangle : \Delta, x : A^{\mathsf{i}}]\!] \mathbin{;} [\![\Delta, x : A^{\mathsf{i}} \vdash e : B]\!])$
$\mathbin{;} \eta_{A \to TB}$

$=\langle$ definition $\rangle$

$\text{curry}\,(\langle [\![\Gamma, y : A^{\mathsf{i}} \vdash \theta : \Delta]\!], [\![\Gamma, y : A^{\mathsf{i}} \vdash y : A]\!]_v \rangle \mathbin{;} [\![\Delta, x : A^{\mathsf{i}} \vdash e : B]\!])$
$\mathbin{;} \eta_{A \to TB}$

$=\langle$ lemma C.5 $\rangle$

$\text{curry}\,(\langle \text{Wk}(\Gamma, y : A^{\mathsf{i}} \supseteq \Gamma) \mathbin{;} [\![\Gamma \vdash \theta : \Delta]\!], \pi_2 \rangle \mathbin{;} [\![\Delta, x : A^{\mathsf{i}} \vdash e : B]\!])$
$\mathbin{;} \eta_{A \to TB}$

$=\langle$ definition $\rangle$

$\text{curry}\,(\langle \pi_1 \mathbin{;} [\![\Gamma \vdash \theta : \Delta]\!], \pi_2 \rangle \mathbin{;} [\![\Delta, x : A^{\mathsf{i}} \vdash e : B]\!]) \mathbin{;} \eta_{A \to TB}$

$=\langle$ universal property of products $\rangle$

$\text{curry}\,([[\![\Gamma \vdash \theta : \Delta]\!] \times id_A] \mathbin{;} [\![\Delta, x : A^{\mathsf{i}} \vdash e : B]\!]); \eta_{A \to TB}$

$=\langle$ universal property of exponential $\rangle$

$[\![\Gamma \vdash \theta : \Delta]\!] \mathbin{;} \text{curry}\,([\![\Delta, x : A^{\mathsf{i}} \vdash e : B]\!]) \mathbin{;} \eta_{A \to TB}$

$=\langle$ definition $\rangle$

$[\![\Gamma \vdash \theta : \Delta]\!] \mathbin{;} [\![\Delta \vdash \lambda x.\ e : A \Rightarrow B]\!]$

$$\diamond \frac{\Gamma \vdash e_1 : A \Rightarrow B \qquad \Gamma \vdash e_2 : A}{\Gamma \vdash e_1\, e_2 : B} \Rightarrow\text{E}$$

$[\![\Gamma \vdash \theta(e_1\, e_2) : B]\!]$

$=\langle$ definition $\rangle$

$$[\![\Gamma \vdash \theta(e_1)\,\theta(e_2) : B]\!]$$

$=\langle$ definition $\rangle$

$$\langle [\![\Gamma \vdash \theta(e_1) : A \Rightarrow B]\!] , [\![\Gamma \vdash \theta(e_2) : A]\!] \rangle \,;\, \beta_{A \to TB, A} \,;\, T\,\mathsf{ev}_{A,TB} \,;\, \mu_B$$

$=\langle$ induction hypothesis $\rangle$

$$\langle [\![\Gamma \vdash \theta : \Delta]\!] \,;\, [\![\Delta \vdash e_1 : A \Rightarrow B]\!] , [\![\Gamma \vdash \theta : \Delta]\!] \,;\, [\![\Delta \vdash e_2 : A]\!] \rangle \,;\, \beta_{A \to TB, A} \,;\, T\,\mathsf{ev}_{A,TB} \,;\, \mu_B$$

$=\langle$ universal property of products $\rangle$

$$[\![\Gamma \vdash \theta : \Delta]\!] \,;\, \langle [\![\Delta \vdash e_1 : A \Rightarrow B]\!] , [\![\Delta \vdash e_2 : A]\!] \rangle \,;\, \beta_{A \to TB, A} \,;\, T\,\mathsf{ev}_{A,TB} \,;\, \mu_B$$

$=\langle$ definition $\rangle$

$$[\![\Gamma \vdash \theta : \Delta]\!] \,;\, [\![\Delta \vdash e_1\,e_2 : B]\!]$$

$$\diamond\ \frac{}{\Gamma \vdash s : \mathsf{str}}\ \mathsf{strI}$$

$$[\![\Gamma \vdash \theta(s) : \mathsf{str}]\!]$$

$=\langle$ definition $\rangle$

$$[\![\Gamma \vdash s : \mathsf{str}]\!]$$

$=\langle$ definition $\rangle$

$$[\![\Gamma \vdash \langle\rangle : \ ]\!] \,;\, [\![\ \vdash s : \mathsf{str}]\!]$$

$=\langle$ universal property of 1 $\rangle$

$$[\![\Gamma \vdash \theta : \Delta]\!] \,;\, [\![\Delta \vdash \langle\rangle : \ ]\!] \,;\, [\![\ \vdash s : \mathsf{str}]\!]$$

$=\langle$ definition $\rangle$

$$[\![\Gamma \vdash \theta : \Delta]\!] \,;\, [\![\Delta \vdash s : \mathsf{str}]\!]$$

$$\diamond\ \frac{\Gamma \vdash e_1 : \mathsf{cap} \qquad \Gamma \vdash e_2 : \mathsf{str}}{\Gamma \vdash e_1 . \mathsf{print}(e_2) : \mathsf{unit}}\ \textsc{Print}$$

$$[\![\Gamma \vdash \theta(e_1 . \mathsf{print}(e_2)) : \mathsf{unit}]\!]$$

$=\langle$ definition $\rangle$

$$[\![\Gamma \vdash \theta(e_1) . \mathsf{print}(\theta(e_2)) : \mathsf{unit}]\!]$$

$=\langle$ definition $\rangle$

$$\langle [\![\Gamma \vdash \theta(e_1) : \mathsf{cap}]\!] , [\![\Gamma \vdash \theta(e_2) : \mathsf{str}]\!] \rangle \,;\, \beta_{\mathcal{C}, \Sigma^*} \,;\, Tp \,;\, \mu_1$$

$=\langle$ induction hypothesis $\rangle$

$$\langle [\![\Gamma \vdash \theta : \Delta]\!] \,;\, [\![\Delta \vdash e_1 : \mathsf{cap}]\!] , [\![\Gamma \vdash \theta : \Delta]\!] \,;\, [\![\Delta \vdash e_2 : \mathsf{str}]\!] \rangle \,;\, \beta_{\mathcal{C}, \Sigma^*} \,;\, Tp \,;\, \mu_1$$

$=\langle$ universal property of products $\rangle$

$$[\![\Gamma \vdash \theta : \Delta]\!] ; \langle [\![\Delta \vdash e_1 : \mathsf{cap}]\!] , [\![\Delta \vdash e_2 : \mathsf{str}]\!] \rangle ; \beta_{\mathcal{C},\Sigma^*} ; Tp ; \mu_1$$

$=\langle$ definition $\rangle$

$$[\![\Gamma \vdash \theta : \Delta]\!] ; [\![\Delta \vdash e_1 . \mathsf{print}(e_2) : \mathsf{unit}]\!]$$

□

## D SUPPLEMENTARY MATERIAL FOR SECTION 6 (EQUATIONAL THEORY)

THEOREM 6.1 (SOUNDNESS OF $\approx$). *If* $\Gamma \vdash e_1 \approx e_2 : A$, *then* $[\![\Gamma \vdash e_1 : A]\!] = [\![\Gamma \vdash e_2 : A]\!]$.

PROOF. We proceed by induction on $\Gamma \vdash e_1 \approx e_2 : A$.

$\diamond$ $$\frac{\Gamma \vdash e : A}{\Gamma \vdash e \approx e : A}\ \textsc{refl}$$

$$[\![\Gamma \vdash e : A]\!]$$

$=\langle$ reflexivity $\rangle$

$$[\![\Gamma \vdash e : A]\!]$$

$\diamond$ $$\frac{\Gamma \vdash e_1 \approx e_2 : A}{\Gamma \vdash e_2 \approx e_1 : A}\ \textsc{sym}$$

$$[\![\Gamma \vdash e_2 : A]\!]$$

$=\langle$ induction hypothesis $\rangle$

$$[\![\Gamma \vdash e_1 : A]\!]$$

$\diamond$ $$\frac{\Gamma \vdash e_1 \approx e_2 : A \qquad \Gamma \vdash e_2 \approx e_3 : A}{\Gamma \vdash e_1 \approx e_3 : A}\ \textsc{trans}$$

$$[\![\Gamma \vdash e_1 : A]\!]$$

$=\langle$ induction hypothesis $\rangle$

$$[\![\Gamma \vdash e_2 : A]\!]$$

$=\langle$ induction hypothesis $\rangle$

$$[\![\Gamma \vdash e_3 : A]\!]$$

$\diamond$ $$\frac{\Gamma \vdash e_1 \approx e_2 : A \times B}{\Gamma \vdash \mathsf{fst}\, e_1 \approx \mathsf{fst}\, e_2 : A}\ \mathsf{fst}\textsc{-cong}$$

$$\llbracket \Gamma \vdash \mathsf{fst}\, e_1 : A \rrbracket$$
$=\langle$ definition $\rangle$
$$\llbracket \Gamma \vdash e_1 : A \times B \rrbracket \mathbin{;} T\pi_1$$
$=\langle$ induction hypothesis $\rangle$
$$\llbracket \Gamma \vdash e_2 : A \times B \rrbracket \mathbin{;} T\pi_1$$
$=\langle$ definition $\rangle$
$$\llbracket \Gamma \vdash \mathsf{fst}\, e_2 : A \rrbracket$$

$$\diamond \quad \frac{\Gamma \vdash e_1 \approx e_2 : A \times B}{\Gamma \vdash \mathsf{snd}\, e_1 \approx \mathsf{snd}\, e_2 : B}\ \textsf{snd-}\textsc{cong}$$

$$\llbracket \Gamma \vdash \mathsf{snd}\, e_1 : B \rrbracket$$
$=\langle$ definition $\rangle$
$$\llbracket \Gamma \vdash e_1 : A \times B \rrbracket \mathbin{;} T\pi_2$$
$=\langle$ induction hypothesis $\rangle$
$$\llbracket \Gamma \vdash e_2 : A \times B \rrbracket \mathbin{;} T\pi_2$$
$=\langle$ definition $\rangle$
$$\llbracket \Gamma \vdash \mathsf{snd}\, e_2 : B \rrbracket$$

$$\diamond \quad \frac{\Gamma \vdash e_1 \approx e_2 : A \qquad \Gamma \vdash e_3 \approx e_4 : B}{\Gamma \vdash (e_1\,,e_3) \approx (e_2\,,e_4) : A \times B}\ \textsc{pair-cong}$$

$$\llbracket \Gamma \vdash (e_1\,,e_3) : A \times B \rrbracket$$
$=\langle$ definition $\rangle$
$$\langle \llbracket \Gamma \vdash e_1 : A \rrbracket\,, \llbracket \Gamma \vdash e_3 : B \rrbracket \rangle \mathbin{;} \beta_{A,B}$$
$=\langle$ induction hypothesis $\rangle$
$$\langle \llbracket \Gamma \vdash e_2 : A \rrbracket\,, \llbracket \Gamma \vdash e_4 : B \rrbracket \rangle \mathbin{;} \beta_{A,B}$$
$=\langle$ definition $\rangle$
$$\llbracket \Gamma \vdash (e_2\,,e_4) : A \times B \rrbracket$$

$$\diamond \quad \frac{\Gamma, x : A^{\mathsf{i}} \vdash e_1 \approx e_2 : B}{\Gamma \vdash \lambda x : A.\ e_1 \approx \lambda x : A.\ e_2 : A \Rightarrow B}\ \lambda\textsc{-cong}$$

$$\llbracket \Gamma \vdash \lambda x.\ e_1 : A \Rightarrow B \rrbracket$$
$=\langle$ definition $\rangle$

$$\mathsf{curry}\,(\llbracket \Gamma, x : A^{\mathsf{i}} \vdash e_1 : B \rrbracket) \mathbin{;} \eta_{A \to TB}$$

$=\langle$ induction hypothesis $\rangle$

$$\mathsf{curry}\,(\llbracket \Gamma, x : A^{\mathsf{i}} \vdash e_2 : B \rrbracket) \mathbin{;} \eta_{A \to TB}$$

$=\langle$ definition $\rangle$

$$\llbracket \Gamma \vdash \lambda x.\, e_2 : A \Rightarrow B \rrbracket$$

$$\diamond \quad \frac{\Gamma \vdash e_1 \approx e_2 : A \Rightarrow B \qquad \Gamma \vdash e_3 \approx e_4 : A}{\Gamma \vdash e_1\, e_3 \approx e_2\, e_4 : B}\ \textsc{app-cong}$$

$$\llbracket \Gamma \vdash e_1\, e_3 : B \rrbracket$$

$=\langle$ definition $\rangle$

$$\langle \llbracket \Gamma \vdash e_1 : A \Rightarrow B \rrbracket, \llbracket \Gamma \vdash e_3 : A \rrbracket \rangle \mathbin{;} \beta_{A \to TB, A} \mathbin{;} T\,\mathsf{ev}_{A,TB} \mathbin{;} \mu_B$$

$=\langle$ induction hypothesis $\rangle$

$$\langle \llbracket \Gamma \vdash e_2 : A \Rightarrow B \rrbracket, \llbracket \Gamma \vdash e_4 : A \rrbracket \rangle \mathbin{;} \beta_{A \to TB, A} \mathbin{;} T\,\mathsf{ev}_{A,TB} \mathbin{;} \mu_B$$

$=\langle$ definition $\rangle$

$$\llbracket \Gamma \vdash e_2\, e_4 : B \rrbracket$$

$$\diamond \quad \frac{\Gamma^{\mathsf{s}} \vdash e_1 \approx e_2 : A}{\Gamma \vdash \mathsf{box}\, \boxed{e_1} \approx \mathsf{box}\, \boxed{e_2} : \Box A}\ \textsc{box-cong}$$

$$\llbracket \Gamma \vdash \mathsf{box}\, \boxed{e_1} : \Box A \rrbracket$$

$=\langle$ definition $\rangle$

$$\llbracket \Gamma \vdash^{\mathsf{s}} e_1 : A \rrbracket_p \mathbin{;} \eta_{\Box A}$$

$=\langle$ definition $\rangle$

$$\rho(\Gamma) \mathbin{;} \mathcal{M}(\Gamma) \mathbin{;} \Box \llbracket \Gamma^{\mathsf{s}} \vdash e_1 : A \rrbracket \mathbin{;} \Phi_A \mathbin{;} \eta_{\Box A}$$

$=\langle$ induction hypothesis $\rangle$

$$\rho(\Gamma) \mathbin{;} \mathcal{M}(\Gamma) \mathbin{;} \Box \llbracket \Gamma^{\mathsf{s}} \vdash e_2 : A \rrbracket \mathbin{;} \Phi_A \mathbin{;} \eta_{\Box A}$$

$=\langle$ definition $\rangle$

$$\llbracket \Gamma \vdash^{\mathsf{s}} e_2 : A \rrbracket_p \mathbin{;} \eta_{\Box A}$$

$=\langle$ definition $\rangle$

$$\llbracket \Gamma \vdash \mathsf{box}\, \boxed{e_2} : \Box A \rrbracket$$

$$\diamond \quad \frac{\Gamma \vdash e_1 \approx e_2 : \Box A \qquad \Gamma, x : A^{\mathsf{s}} \vdash e_3 \approx e_4 : B}{\Gamma \vdash (\mathsf{let\ box}\, \boxed{x} = e_1\ \mathsf{in}\ e_3) \approx (\mathsf{let\ box}\, \boxed{x} = e_2\ \mathsf{in}\ e_4) : B}\ \textsc{let box-cong}$$

$$
\begin{aligned}
& [\![\Gamma \vdash \mathsf{let\ box}\ \boxed{x} = e_1\ \mathsf{in}\ e_3 : B]\!] \\
=\langle & \text{ definition } \rangle \\
& \langle id_\Gamma , [\![\Gamma \vdash e_1 : \Box A]\!]\rangle ; \tau_{\Gamma,\Box A} ; T[\![\Gamma, x : A^{\mathsf{s}} \vdash e_3 : B]\!] ; \mu_B \\
=\langle & \text{ induction hypothesis } \rangle \\
& \langle id_\Gamma , [\![\Gamma \vdash e_2 : \Box A]\!]\rangle ; \tau_{\Gamma,\Box A} ; T[\![\Gamma, x : A^{\mathsf{s}} \vdash e_4 : B]\!] ; \mu_B \\
=\langle & \text{ definition } \rangle \\
& [\![\Gamma \vdash \mathsf{let\ box}\ \boxed{x} = e_2\ \mathsf{in}\ e_4 : B]\!]
\end{aligned}
$$

$$
\diamond\ \frac{\Gamma \vdash e_1 \approx e_2 : \mathsf{cap} \qquad \Gamma \vdash e_3 \approx e_4 : \mathsf{str}}{\Gamma \vdash e_1 . \mathsf{print}(e_3) \approx e_2 . \mathsf{print}(e_4) : \mathsf{unit}}\ \text{print-CONG}
$$

$$
\begin{aligned}
& [\![\Gamma \vdash e_1 . \mathsf{print}(e_3) : \mathsf{unit}]\!] \\
=\langle & \text{ definition } \rangle \\
& \langle [\![\Gamma \vdash e_1 : \mathsf{cap}]\!] , [\![\Gamma \vdash e_3 : \mathsf{str}]\!]\rangle ; \beta_{\mathcal{C},\Sigma^*} ; Tp ; \mu_1 \\
=\langle & \text{ induction hypothesis } \rangle \\
& \langle [\![\Gamma \vdash e_2 : \mathsf{cap}]\!] , [\![\Gamma \vdash e_4 : \mathsf{str}]\!]\rangle ; \beta_{\mathcal{C},\Sigma^*} ; Tp ; \mu_1 \\
=\langle & \text{ definition } \rangle \\
& [\![\Gamma \vdash e_2 . \mathsf{print}(e_4) : \mathsf{unit}]\!]
\end{aligned}
$$

$$
\diamond\ \frac{\Gamma \vdash v_1 : A \qquad \Gamma \vdash v_2 : B}{\Gamma \vdash \mathsf{fst}\ (v_1 , v_2) \approx v_1 : A}\ \times_1\beta
$$

$$
\begin{aligned}
& [\![\Gamma \vdash \mathsf{fst}\ (v_1 , v_2) : A]\!] \\
=\langle & \text{ definition } \rangle \\
& [\![\Gamma \vdash (v_1 , v_2) : A \times B]\!] ; T\pi_1 \\
=\langle & \text{ value interpretation lemma 5.3 } \rangle \\
& [\![\Gamma \vdash (v_1 , v_2) : A \times B]\!]_v ; \eta_{A\times B} ; T\pi_1 \\
=\langle & \text{ monad laws } \rangle \\
& [\![\Gamma \vdash (v_1 , v_2) : A \times B]\!]_v ; \pi_1 ; \eta_A \\
=\langle & \text{ definition } \rangle \\
& \langle [\![\Gamma \vdash v_1 : A]\!]_v , [\![\Gamma \vdash v_2 : B]\!]_v\rangle ; \pi_1 ; \eta_A \\
=\langle & \text{ definition of } \pi_1 \rangle \\
& [\![\Gamma \vdash v_1 : A]\!]_v ; \eta_A \\
=\langle & \text{ value interpretation lemma 5.3 } \rangle
\end{aligned}
$$

$$[\![\Gamma \vdash v_1 : A]\!]$$

$$\diamond \frac{\Gamma \vdash v_1 : A \qquad \Gamma \vdash v_2 : B}{\Gamma \vdash \mathsf{snd}\,(v_1\,,v_2) \approx v_2 : B}\ \times_2\beta$$

$$[\![\Gamma \vdash \mathsf{snd}\,(v_1\,,v_2) : B]\!]$$
=⟨ definition ⟩
$$[\![\Gamma \vdash (v_1\,,v_2) : A \times B]\!]\,;\, T\pi_2$$
=⟨ value interpretation lemma 5.3 ⟩
$$[\![\Gamma \vdash (v_1\,,v_2) : A \times B]\!]_v\,;\,\eta_{A\times B}\,;\, T\pi_2$$
=⟨ monad laws ⟩
$$[\![\Gamma \vdash (v_1\,,v_2) : A \times B]\!]_v\,;\,\pi_2\,;\,\eta_B$$
=⟨ definition ⟩
$$\langle [\![\Gamma \vdash v_1 : A]\!]_v\,,[\![\Gamma \vdash v_2 : B]\!]_v\rangle\,;\,\pi_2\,;\,\eta_B$$
=⟨ definition of $\pi_2$ ⟩
$$[\![\Gamma \vdash v_2 : B]\!]_v\,;\,\eta_B$$
=⟨ value interpretation lemma 5.3 ⟩
$$[\![\Gamma \vdash v_2 : B]\!]$$

$$\diamond \frac{\Gamma \vdash v : A \times B}{\Gamma \vdash v \approx (\mathsf{fst}\,v\,,\mathsf{snd}\,v) : A \times B}\ \times\eta$$

$$[\![\Gamma \vdash (\mathsf{fst}\,v\,,\mathsf{snd}\,v) : A \times B]\!]$$
=⟨ definition ⟩
$$\langle [\![\Gamma \vdash \mathsf{fst}\,v : A]\!]\,,[\![\Gamma \vdash \mathsf{snd}\,v : B]\!]\rangle\,;\,\beta_{A,B}$$
=⟨ definition ⟩
$$\langle [\![\Gamma \vdash v : A \times B]\!]\,;\,T\pi_1\,,[\![\Gamma \vdash v : A \times B]\!]\,;\,T\pi_2\rangle\,;\,\beta_{A,B}$$
=⟨ value interpretation lemma 5.3 ⟩
$$\langle [\![\Gamma \vdash v : A \times B]\!]_v\,;\,\eta_{A\times B}\,;\,T\pi_1\,,[\![\Gamma \vdash v : A \times B]\!]_v\,;\,\eta_{A\times B}\,;\,T\pi_2\rangle\,;\,\beta_{A,B}$$
=⟨ monad laws ⟩
$$\langle [\![\Gamma \vdash v : A \times B]\!]_v\,;\,\pi_1\,;\,\eta_A\,,[\![\Gamma \vdash v : A \times B]\!]_v\,;\,\pi_2\,;\,\eta_B\rangle\,;\,\beta_{A,B}$$
=⟨ universal property of products ⟩
$$[\![\Gamma \vdash v : A \times B]\!]_v\,;\,\langle \pi_1\,;\,\eta_A\,,\pi_2\,;\,\eta_B\rangle\,;\,\beta_{A,B}$$
=⟨ universal property of products ⟩
$$[\![\Gamma \vdash v : A \times B]\!]_v\,;\,[\eta_A \times \eta_B]\,;\,\beta_{A,B}$$

=⟨ diagram ⟩

$[\![\Gamma \vdash v : A \times B]\!]_v \mathbin{;} \eta_{A\times B}$

=⟨ value interpretation lemma 5.3 ⟩

$[\![\Gamma \vdash v : A \times B]\!]$

$$\diamond \quad \frac{\Gamma, x : A^{\mathsf{i}} \vdash e : B \qquad \Gamma \vdash v : A}{\Gamma \vdash (\lambda x : A.\ e)\, v \approx [v/x]e : B} \Rightarrow \beta$$

$[\![\Gamma \vdash (\lambda x.\ e)\, v : B]\!]$

=⟨ definition ⟩

$\langle [\![\Gamma \vdash \lambda x.\ e : A \Rightarrow B]\!] \,,\, [\![\Gamma \vdash v : A]\!] \rangle \mathbin{;} \beta_{A\to TB,A} \mathbin{;} T\,\mathsf{ev}_{A,TB} \mathbin{;} \mu_B$

=⟨ definition ⟩

$\langle \mathsf{curry}\,([\![\Gamma, x : A^{\mathsf{i}} \vdash e : B]\!]) \mathbin{;} \eta_{A\to TB} \,,\, [\![\Gamma \vdash v : A]\!] \rangle$
$\mathbin{;} \beta_{A\to TB,A} \mathbin{;} T\,\mathsf{ev}_{A,TB} \mathbin{;} \mu_B$

=⟨ value interpretation lemma 5.3 ⟩

$\langle \mathsf{curry}\,([\![\Gamma, x : A^{\mathsf{i}} \vdash e : B]\!]) \mathbin{;} \eta_{A\to TB} \,,\, [\![\Gamma \vdash v : A]\!]_v \mathbin{;} \eta_A \rangle$
$\mathbin{;} \beta_{A\to TB,A} \mathbin{;} T\,\mathsf{ev}_{A,TB} \mathbin{;} \mu_B$

=⟨ universal property of products ⟩

$\langle \mathsf{curry}\,([\![\Gamma, x : A^{\mathsf{i}} \vdash e : B]\!]) \,,\, [\![\Gamma \vdash v : A]\!]_v \rangle$
$\mathbin{;} [\eta_{A\to TB} \times \eta_A] \mathbin{;} \beta_{A\to TB,A} \mathbin{;} T\,\mathsf{ev}_{A,TB} \mathbin{;} \mu_B$

=⟨ diagram ⟩

$\langle \mathsf{curry}\,([\![\Gamma, x : A^{\mathsf{i}} \vdash e : B]\!]) \,,\, [\![\Gamma \vdash v : A]\!]_v \rangle$
$\mathbin{;} \eta_{(A\to TB)\times A} \mathbin{;} T\,\mathsf{ev}_{A,TB} \mathbin{;} \mu_B$

=⟨ monad laws ⟩

$\langle \mathsf{curry}\,([\![\Gamma, x : A^{\mathsf{i}} \vdash e : B]\!]) \,,\, [\![\Gamma \vdash v : A]\!]_v \rangle \mathbin{;} \mathsf{ev}_{A,TB}$

=⟨ universal property of exponential ⟩

$\langle id_\Gamma \,,\, [\![\Gamma \vdash v : A]\!]_v \rangle \mathbin{;} [\![\Gamma, x : A^{\mathsf{i}} \vdash e : B]\!]$

=⟨ definition ⟩

$\langle [\![\Gamma \vdash \langle \Gamma \rangle : \Gamma]\!] \,,\, [\![\Gamma \vdash v : A]\!]_v \rangle \mathbin{;} [\![\Gamma, x : A^{\mathsf{i}} \vdash e : B]\!]$

=⟨ definition ⟩

$[\![\Gamma \vdash \langle \langle \Gamma \rangle, v^{\mathsf{i}}/x \rangle : \Gamma, x : A^{\mathsf{i}}]\!] \mathbin{;} [\![\Gamma, x : A^{\mathsf{i}} \vdash e : B]\!]$

=⟨ semantic substitution theorem 5.4 ⟩

$[\![\Gamma \vdash \langle \langle \Gamma \rangle, v^{\mathsf{i}}/x \rangle(e) : B]\!]$

$=\langle$ definition $\rangle$

$[\![\Gamma \vdash [v/x]e : B]\!]$

$$\diamond \frac{\Gamma \vdash v : A \Rightarrow B}{\Gamma \vdash v \approx \lambda x : A.\, v\, x : A \Rightarrow B} \Rightarrow \eta\text{-IMPURE}$$

$[\![\Gamma \vdash \lambda x.\, v\, x : A \Rightarrow B]\!]$

$=\langle$ definition $\rangle$

$\mathsf{curry}\,([\![\Gamma, x : A^{\mathsf{i}} \vdash v\, x : B]\!])\, ;\, \eta_{A \to TB}$

$=\langle$ definition $\rangle$

$$\begin{array}{lll} \mathit{let} & h = & \beta_{A\to TB,A}\, ;\, T\,\mathsf{ev}_{A,TB}\, ;\, \mu_B \\ \mathit{in} & \multicolumn{2}{l}{\mathsf{curry}\,(\langle [\![\Gamma, x : A^{\mathsf{i}} \vdash v : A \Rightarrow B]\!]\, ,\, [\![\Gamma, x : A^{\mathsf{i}} \vdash x : A]\!] \rangle\, ;\, h)\, ;\, \eta_{A\to TB}} \end{array}$$

$=\langle$ semantic weakening lemma 5.1 $\rangle$

$$\begin{array}{lll} & f = & \mathsf{Wk}(\Gamma, x : A^{\mathsf{i}} \supseteq \Gamma) \\ \mathit{let} & g = & [\![x : A^{\mathsf{i}} \in \Gamma, x : A^{\mathsf{i}}]\!] \\ & h = & \beta_{A\to TB,A}\, ;\, T\,\mathsf{ev}_{A,TB}\, ;\, \mu_B \\ \mathit{in} & \multicolumn{2}{l}{\mathsf{curry}\,(\langle f\, ;\, [\![\Gamma \vdash v : A \Rightarrow B]\!]\, ,\, g\, ;\, \eta_A \rangle\, ;\, h)\, ;\, \eta_{A\to TB}} \end{array}$$

$=\langle$ definition $\rangle$

$$\begin{array}{lll} \mathit{let} & h = & \beta_{A\to TB,A}\, ;\, T\,\mathsf{ev}_{A,TB}\, ;\, \mu_B \\ \mathit{in} & \multicolumn{2}{l}{\mathsf{curry}\,(\langle \pi_1\, ;\, [\![\Gamma \vdash v : A \Rightarrow B]\!]\, ,\, \pi_2\, ;\, \eta_A \rangle\, ;\, h)\, ;\, \eta_{A\to TB}} \end{array}$$

$=\langle$ value interpretation lemma 5.3 $\rangle$

$$\begin{array}{lll} \mathit{let} & h = & \beta_{A\to TB,A}\, ;\, T\,\mathsf{ev}_{A,TB}\, ;\, \mu_B \\ \mathit{in} & \multicolumn{2}{l}{\mathsf{curry}\,(\langle \pi_1\, ;\, [\![\Gamma \vdash v : A \Rightarrow B]\!]_v\, ;\, \eta_{A\to TB}\, ,\, \pi_2\, ;\, \eta_A \rangle\, ;\, h)\, ;\, \eta_{A\to TB}} \end{array}$$

$=\langle$ strength diagram and monad laws $\rangle$

$\mathsf{curry}\,(\langle \pi_1\, ;\, [\![\Gamma \vdash v : A \Rightarrow B]\!]_v\, ,\, \pi_2 \rangle\, ;\, \mathsf{ev}_{A,TB})\, ;\, \eta_{A\to TB}$

$=\langle$ universal property of products $\rangle$

$\mathsf{curry}\,([[\![\Gamma \vdash v : A \Rightarrow B]\!]_v \times id_A]\, ;\, \mathsf{ev}_{A,TB})\, ;\, \eta_{A\to TB}$

$=\langle$ universal property of exponential $\rangle$

$[\![\Gamma \vdash v : A \Rightarrow B]\!]_v\, ;\, \eta_{A\to TB}$

$=\langle$ value interpretation lemma 5.3 $\rangle$

$[\![\Gamma \vdash v : A \Rightarrow B]\!]$

$$\diamond \frac{\Gamma \vdash^{\mathsf{s}} e : A \Rightarrow B}{\Gamma \vdash e \approx \lambda x : A.\ e\,x : A \Rightarrow B} \Rightarrow \eta\text{-SAFE}$$

$$[\![\Gamma \vdash \lambda x.\ e\,x : A \Rightarrow B]\!]$$

$=\langle$ definition $\rangle$

$$\mathsf{curry}\,([\![\Gamma, x : A^{\mathsf{i}} \vdash e\,x : B]\!])\ ;\ \eta_{A \to TB}$$

$=\langle$ definition $\rangle$

$$\begin{array}{ll} \mathit{let} & h = \beta_{A \to TB, A}\ ;\ T\,\mathsf{ev}_{A,TB}\ ;\ \mu_B \\ \mathit{in} & \mathsf{curry}\,(\langle [\![\Gamma, x : A^{\mathsf{i}} \vdash e : A \Rightarrow B]\!]\ ,\ [\![\Gamma, x : A^{\mathsf{i}} \vdash x : A]\!] \rangle\ ;\ h)\ ;\ \eta_{A \to TB} \end{array}$$

$=\langle$ semantic weakening lemma 5.1 $\rangle$

$$\begin{array}{ll} & f = \mathsf{Wk}(\Gamma, x : A^{\mathsf{i}} \supseteq \Gamma) \\ \mathit{let} & g = [\![x : A^{\mathsf{i}} \in \Gamma, x : A^{\mathsf{i}}]\!] \\ & h = \beta_{A \to TB, A}\ ;\ T\,\mathsf{ev}_{A,TB}\ ;\ \mu_B \\ \mathit{in} & \mathsf{curry}\,(\langle f\ ;\ [\![\Gamma \vdash e : A \Rightarrow B]\!]\ ,\ g\ ;\ \eta_A \rangle\ ;\ h)\ ;\ \eta_{A \to TB} \end{array}$$

$=\langle$ definition $\rangle$

$$\begin{array}{ll} \mathit{let} & h = \beta_{A \to TB, A}\ ;\ T\,\mathsf{ev}_{A,TB}\ ;\ \mu_B \\ \mathit{in} & \mathsf{curry}\,(\langle \pi_1\ ;\ [\![\Gamma \vdash e : A \Rightarrow B]\!]\ ,\ \pi_2\ ;\ \eta_A \rangle\ ;\ h)\ ;\ \eta_{A \to TB} \end{array}$$

$=\langle$ safe interpretation lemma 5.2 $\rangle$

$$\begin{array}{ll} \mathit{let} & h = \beta_{A \to TB, A}\ ;\ T\,\mathsf{ev}_{A,TB}\ ;\ \mu_B \\ \mathit{in} & \mathsf{curry}\,(\langle \pi_1\ ;\ [\![\Gamma \vdash^{\mathsf{s}} e : A \Rightarrow B]\!]_p\ ;\ \epsilon_{A \to TB}\ ;\ \eta_{A \to TB}\ ,\ \pi_2\ ;\ \eta_A \rangle\ ;\ h)\ ;\ \eta_{A \to TB} \end{array}$$

$=\langle$ diagram and monad laws $\rangle$

$$\mathsf{curry}\,(\langle \pi_1\ ;\ [\![\Gamma \vdash^{\mathsf{s}} e : A \Rightarrow B]\!]_p\ ;\ \epsilon_{A \to TB}\ ,\ \pi_2 \rangle\ ;\ \mathsf{ev}_{A,TB})\ ;\ \eta_{A \to TB}$$

$=\langle$ universal property of products $\rangle$

$$\mathsf{curry}\,([[\![\Gamma \vdash^{\mathsf{s}} e : A \Rightarrow B]\!]_p\ ;\ \epsilon_{A \to TB} \times id_A]\ ;\ \mathsf{ev}_{A,TB})\ ;\ \eta_{A \to TB}$$

$=\langle$ universal property of exponential $\rangle$

$$[\![\Gamma \vdash^{\mathsf{s}} e : A \Rightarrow B]\!]_p\ ;\ \epsilon_{A \to TB}\ ;\ \eta_{A \to TB}$$

$=\langle$ safe interpretation lemma 5.2 $\rangle$

$$[\![\Gamma \vdash e : A \Rightarrow B]\!]$$

$$\diamond \frac{\Gamma^{\mathsf{s}} \vdash e_1 : A \qquad \Gamma, x : A^{\mathsf{s}} \vdash e_2 : B}{\Gamma \vdash \mathsf{let\ box}\ \boxed{x} = \mathsf{box}\ \boxed{e_1}\ \mathsf{in}\ e_2 \approx [e_1/x]e_2 : B} \square\beta$$

$$[\![\Gamma \vdash \mathsf{let\ box}\ \boxed{x} = \mathsf{box}\ \boxed{e_1}\ \mathsf{in}\ e_2 : B]\!]$$

$=\langle$ definition $\rangle$

$\langle id_\Gamma \,,\, [\![ \Gamma \vdash \mathsf{box}\,\boxed{e_1} : \Box A ]\!] \rangle \,;\, \tau_{\Gamma,\Box A} \,;\, T[\![ \Gamma, x : A^{\mathsf{s}} \vdash e_2 : B ]\!] \,;\, \mu_B$

$=\langle$ definition $\rangle$

$\langle id_\Gamma \,,\, [\![ \Gamma \vdash^{\mathsf{s}} e_1 : A ]\!]_p \,;\, \eta_{\Box A} \rangle \,;\, \tau_{\Gamma,\Box A} \,;\, T[\![ \Gamma, x : A^{\mathsf{s}} \vdash e_2 : B ]\!] \,;\, \mu_B$

$=\langle$ strength commutes with unit $\rangle$

$\langle id_\Gamma \,,\, [\![ \Gamma \vdash^{\mathsf{s}} e_1 : A ]\!]_p \rangle \,;\, \eta_{\Gamma\times\Box A} \,;\, T[\![ \Gamma, x : A^{\mathsf{s}} \vdash e_2 : B ]\!] \,;\, \mu_B$

$=\langle$ monad laws $\rangle$

$\langle id_\Gamma \,,\, [\![ \Gamma \vdash^{\mathsf{s}} e_1 : A ]\!]_p \rangle \,;\, [\![ \Gamma, x : A^{\mathsf{s}} \vdash e_2 : B ]\!] \,;\, \eta_{TB} \,;\, \mu_B$

$=\langle$ monad laws $\rangle$

$\langle id_\Gamma \,,\, [\![ \Gamma \vdash^{\mathsf{s}} e_1 : A ]\!]_p \rangle \,;\, [\![ \Gamma, x : A^{\mathsf{s}} \vdash e_2 : B ]\!]$

$=\langle$ definition $\rangle$

$\langle [\![ \Gamma \vdash \langle \Gamma \rangle : \Gamma ]\!] \,,\, [\![ \Gamma \vdash^{\mathsf{s}} e_1 : A ]\!]_p \rangle \,;\, [\![ \Gamma, x : A^{\mathsf{s}} \vdash e_2 : B ]\!]$

$=\langle$ definition $\rangle$

$[\![ \Gamma \vdash \langle \langle \Gamma \rangle, {e_1}^{\mathsf{s}}/x \rangle : \Gamma, x : A^{\mathsf{s}} ]\!] \,;\, [\![ \Gamma, x : A^{\mathsf{s}} \vdash e_2 : B ]\!]$

$=\langle$ semantic substitution theorem 5.4 $\rangle$

$[\![ \Gamma \vdash \langle \langle \Gamma \rangle, {e_1}^{\mathsf{s}}/x \rangle (e_2) : B ]\!]$

$=\langle$ definition $\rangle$

$\Gamma \vdash [e_1/x]\, e_2 : B$

$$\diamond\ \frac{\Gamma \vdash^{\mathsf{s}} e : \Box A \qquad \Gamma \vdash \mathcal{C}\,\langle\!\langle e \rangle\!\rangle : B \qquad \Gamma \vdash \mathsf{let\ box}\,\boxed{x} = e\ \mathsf{in}\ \mathcal{C}\,\langle\!\langle \mathsf{box}\,\boxed{x} \rangle\!\rangle : B}{\Gamma \vdash \mathcal{C}\,\langle\!\langle e \rangle\!\rangle \approx \mathsf{let\ box}\,\boxed{x} = e\ \mathsf{in}\ \mathcal{C}\,\langle\!\langle \mathsf{box}\,\boxed{x} \rangle\!\rangle : B}\ \Box\eta\text{-SAFE}$$

We first make the following observation.

*Observation.*

$[\![ \Gamma \vdash \mathsf{let\ box}\,\boxed{x} = e\ \mathsf{in}\ \mathcal{C}\,\langle\!\langle \mathsf{box}\,\boxed{x} \rangle\!\rangle : B ]\!]$

$=\langle$ definition $\rangle$

$$\begin{array}{lll} \mathit{let} & f = & [\![ \Gamma \vdash e : \Box A ]\!] \\ & g = & [\![ \Gamma, x : A^{\mathsf{s}} \vdash \mathcal{C}\,\langle\!\langle \mathsf{box}\,\boxed{x} \rangle\!\rangle : B ]\!] \\ \mathit{in} & \multicolumn{2}{l}{\langle id_\Gamma \,,\, f \rangle \,;\, \tau_{\Gamma,\Box A} \,;\, Tg \,;\, \mu_B} \end{array}$$

$=\langle$ safe interpretation lemma 5.2 $\rangle$

$$\begin{array}{lll} \mathit{let} & f = & [\![ \Gamma \vdash^{\mathsf{s}} e : \Box A ]\!]_p \,;\, \epsilon_{\Box A} \,;\, \eta_{\Box A} \\ & g = & [\![ \Gamma, x : A^{\mathsf{s}} \vdash \mathcal{C}\,\langle\!\langle \mathsf{box}\,\boxed{x} \rangle\!\rangle : B ]\!] \\ \mathit{in} & \multicolumn{2}{l}{\langle id_\Gamma \,,\, f \rangle \,;\, \tau_{\Gamma,\Box A} \,;\, Tg \,;\, \mu_B} \end{array}$$

$$=\langle \text{ simplification } \rangle$$

$$\begin{array}{ll} \mathit{let} & f = [\![\Gamma \vdash^s e : \square A]\!]_p \mathbin{;} \epsilon_{\square A} \\ & g = [\![\Gamma, x : A^s \vdash \mathcal{C}\langle\!\langle \mathsf{box}\, \boxed{x} \rangle\!\rangle : B]\!] \\ \mathit{in} & \langle id_\Gamma , f \mathbin{;} \eta_{\square A} \rangle \mathbin{;} \tau_{\Gamma,\square A} \mathbin{;} Tg \mathbin{;} \mu_B \end{array}$$

$$=\langle \text{ strength commutes with unit } \rangle$$

$$\begin{array}{ll} \mathit{let} & f = [\![\Gamma \vdash^s e : \square A]\!]_p \mathbin{;} \epsilon_{\square A} \\ & g = [\![\Gamma, x : A^s \vdash \mathcal{C}\langle\!\langle \mathsf{box}\, \boxed{x} \rangle\!\rangle : B]\!] \\ \mathit{in} & \langle id_\Gamma , f \rangle \mathbin{;} \eta_{\Gamma\times\square A} \mathbin{;} Tg \mathbin{;} \mu_B \end{array}$$

$$=\langle \text{ monad laws } \rangle$$

$$\begin{array}{ll} \mathit{let} & f = [\![\Gamma \vdash^s e : \square A]\!]_p \mathbin{;} \epsilon_{\square A} \\ & g = [\![\Gamma, x : A^s \vdash \mathcal{C}\langle\!\langle \mathsf{box}\, \boxed{x} \rangle\!\rangle : B]\!] \\ \mathit{in} & \langle id_\Gamma , f \rangle \mathbin{;} g \mathbin{;} T\eta_B \mathbin{;} \mu_B \end{array}$$

$$=\langle \text{ monad laws } \rangle$$

$$\begin{array}{ll} \mathit{let} & f = [\![\Gamma \vdash^s e : \square A]\!]_p \mathbin{;} \epsilon_{\square A} \\ & g = [\![\Gamma, x : A^s \vdash \mathcal{C}\langle\!\langle \mathsf{box}\, \boxed{x} \rangle\!\rangle : B]\!] \\ \mathit{in} & \langle id_\Gamma , f \rangle \mathbin{;} g \end{array}$$

Fixing $f$, we proceed by cases on $\mathcal{C}$.

$\diamond\, \mathcal{C} = []$

$$[\![\Gamma \vdash \mathsf{let\ box}\, \boxed{x} = e \,\mathsf{in\ box}\, \boxed{x} : \square A]\!]$$

$$=\langle \text{ observation } \rangle$$

$$\langle id_\Gamma , f \rangle \mathbin{;} [\![\Gamma, x : A^s \vdash \mathsf{box}\, \boxed{x} : \square A]\!]$$

$$=\langle \text{ definition } \rangle$$

$$\langle id_\Gamma , f \rangle \mathbin{;} [\![\Gamma, x : A^s \vdash^s x : A]\!]_p \mathbin{;} \eta_{\square A}$$

$$=\langle \text{ definition } \rangle$$

$$\langle id_\Gamma , f \rangle \mathbin{;} \pi_2 \mathbin{;} \eta_{\square A}$$

$$=\langle \text{ applying } \pi_2 \rangle$$

$$f \mathbin{;} \eta_{\square A}$$

$$=\langle \text{ definition } \rangle$$

$$[\![\Gamma \vdash e : \square A]\!]$$

$\diamond \mathcal{C} = e_1\ \mathcal{C}_1$

$$[\![\Gamma \vdash \text{let box}\ \boxed{x} = e \text{ in } e_1\ \mathcal{C}_1 \langle\!\langle \text{box}\ \boxed{x} \rangle\!\rangle : B]\!]$$

$=\langle$ observation $\rangle$

$$\langle id_\Gamma , f \rangle \mathbin{;} [\![\Gamma, x : A^s \vdash e_1\ \mathcal{C}_1 \langle\!\langle \text{box}\ \boxed{x} \rangle\!\rangle : B]\!]$$

$=\langle$ definition $\rangle$

$$\begin{array}{ll} \mathit{let} & \begin{array}{lcl} h_1 & = & [\![\Gamma, x : A^s \vdash e_1 : C \Rightarrow B]\!] \\ h_2 & = & [\![\Gamma, x : A^s \vdash \mathcal{C}_1 \langle\!\langle \text{box}\ \boxed{x} \rangle\!\rangle : C]\!] \end{array} \\ \mathit{in} & \langle id_\Gamma , f \rangle \mathbin{;} \langle h_1 , h_2 \rangle \mathbin{;} \beta_{C \to TB, C} \mathbin{;} T\,\text{ev}_{C,TB} \mathbin{;} \mu_B \end{array}$$

$=\langle$ semantic weakening lemma 5.1 $\rangle$

$$\begin{array}{ll} \mathit{let} & \begin{array}{lcl} h_1 & = & [\![\Gamma \vdash e_1 : C \Rightarrow B]\!] \\ h_2 & = & [\![\Gamma, x : A^s \vdash \mathcal{C}_1 \langle\!\langle \text{box}\ \boxed{x} \rangle\!\rangle : C]\!] \end{array} \\ \mathit{in} & \langle id_\Gamma , f \rangle \mathbin{;} \langle \pi_1 \mathbin{;} h_1 , h_2 \rangle \mathbin{;} \beta_{C \to TB, C} \mathbin{;} T\,\text{ev}_{C,TB} \mathbin{;} \mu_B \end{array}$$

$=\langle$ simplification $\rangle$

$$\begin{array}{ll} \mathit{let} & \begin{array}{lcl} h_1 & = & [\![\Gamma \vdash e_1 : C \Rightarrow B]\!] \\ h_2 & = & [\![\Gamma, x : C^s \vdash \mathcal{C}_1 \langle\!\langle \text{box}\ \boxed{x} \rangle\!\rangle : C]\!] \end{array} \\ \mathit{in} & \langle \langle id_\Gamma , f \rangle \mathbin{;} \pi_1 \mathbin{;} h_1 , \langle id_\Gamma , f \rangle \mathbin{;} h_2 \rangle \mathbin{;} \beta_{C \to TB, C} \mathbin{;} T\,\text{ev}_{C,TB} \mathbin{;} \mu_B \end{array}$$

$=\langle$ simplification $\rangle$

$$\begin{array}{ll} \mathit{let} & \begin{array}{lcl} h_1 & = & [\![\Gamma \vdash e_1 : C \Rightarrow B]\!] \\ h_2 & = & [\![\Gamma, x : A^s \vdash \mathcal{C}_1 \langle\!\langle \text{box}\ \boxed{x} \rangle\!\rangle : C]\!] \end{array} \\ \mathit{in} & \langle h_1 , \langle id_\Gamma , f \rangle \mathbin{;} h_2 \rangle \mathbin{;} \beta_{C \to TB, C} \mathbin{;} T\,\text{ev}_{C,TB} \mathbin{;} \mu_B \end{array}$$

$=\langle$ observation $\rangle$

$$\begin{array}{ll} \mathit{let} & \begin{array}{lcl} h_1 & = & [\![\Gamma \vdash e_1 : C \Rightarrow B]\!] \\ h_2 & = & [\![\Gamma \vdash \text{let box}\ \boxed{x} = e \text{ in } \mathcal{C}_1 \langle\!\langle \text{box}\ \boxed{x} \rangle\!\rangle : C]\!] \end{array} \\ \mathit{in} & \langle h_1 , h_2 \rangle \mathbin{;} \beta_{C \to TB, C} \mathbin{;} T\,\text{ev}_{C,TB} \mathbin{;} \mu_B \end{array}$$

$=\langle$ induction hypothesis $\rangle$

$$\begin{array}{ll} \mathit{let} & \begin{array}{lcl} h_1 & = & [\![\Gamma \vdash e_1 : C \Rightarrow B]\!] \\ h_2 & = & [\![\Gamma \vdash \mathcal{C}_1 \langle\!\langle e \rangle\!\rangle : C]\!] \end{array} \\ \mathit{in} & \langle h_1 , h_2 \rangle \mathbin{;} \beta_{C \to TB, C} \mathbin{;} T\,\text{ev}_{C,TB} \mathbin{;} \mu_B \end{array}$$

$=\langle$ definition $\rangle$

$$[\![\Gamma \vdash e_1\ \mathcal{C}_1 \langle\!\langle e \rangle\!\rangle : B]\!]$$

$\diamond\, \mathcal{C} = \mathcal{C}_1\ e_1$

$$\llbracket \Gamma \vdash \text{let box}\, \boxed{x} = e \text{ in } \mathcal{C}_1 \langle\!\langle \text{box}\, \boxed{x} \rangle\!\rangle\ e_1 : B \rrbracket$$

=⟨ observation ⟩

$$\langle id_\Gamma , f \rangle \mathbin{;} \llbracket \Gamma, x : A^s \vdash \mathcal{C}_1 \langle\!\langle \text{box}\, \boxed{x} \rangle\!\rangle\ e_1 : B \rrbracket$$

=⟨ definition ⟩

$$\begin{array}{ll} let & h_1 = \llbracket \Gamma, x : A^s \vdash \mathcal{C}_1 \langle\!\langle \text{box}\, \boxed{x} \rangle\!\rangle : C \Rightarrow B \rrbracket \\ & h_2 = \llbracket \Gamma, x : A^s \vdash e_1 : C \rrbracket \\ in & \langle id_\Gamma , f \rangle \mathbin{;} \langle h_1 , h_2 \rangle \mathbin{;} \beta_{C \to TB, C} \mathbin{;} T\, \text{ev}_{C,TB} \mathbin{;} \mu_B \end{array}$$

=⟨ semantic weakening lemma 5.1 ⟩

$$\begin{array}{ll} let & h_1 = \llbracket \Gamma, x : A^s \vdash \mathcal{C}_1 \langle\!\langle \text{box}\, \boxed{x} \rangle\!\rangle : C \Rightarrow B \rrbracket \\ & h_2 = \llbracket \Gamma \vdash e_1 : C \rrbracket \\ in & \langle id_\Gamma , f \rangle \mathbin{;} \langle h_1 , \pi_1 \mathbin{;} h_2 \rangle \mathbin{;} \beta_{C \to TB, C} \mathbin{;} T\, \text{ev}_{C,TB} \mathbin{;} \mu_B \end{array}$$

=⟨ simplification ⟩

$$\begin{array}{ll} let & h_1 = \llbracket \Gamma, x : A^s \vdash \mathcal{C}_1 \langle\!\langle \text{box}\, \boxed{x} \rangle\!\rangle : C \Rightarrow B \rrbracket \\ & h_2 = \llbracket \Gamma \vdash e_1 : C \rrbracket \\ in & \langle \langle id_\Gamma , f \rangle \mathbin{;} h_1 , \langle id_\Gamma , f \rangle \mathbin{;} \pi_1 \mathbin{;} h_2 \rangle \mathbin{;} \beta_{C \to TB, C} \mathbin{;} T\, \text{ev}_{C,TB} \mathbin{;} \mu_B \end{array}$$

=⟨ simplification ⟩

$$\begin{array}{ll} let & h_1 = \llbracket \Gamma, x : A^s \vdash \mathcal{C}_1 \langle\!\langle \text{box}\, \boxed{x} \rangle\!\rangle : C \Rightarrow B \rrbracket \\ & h_2 = \llbracket \Gamma \vdash e_1 : C \rrbracket \\ in & \langle \langle id_\Gamma , f \rangle \mathbin{;} h_1 , h_2 \rangle \mathbin{;} \beta_{C \to TB, C} \mathbin{;} T\, \text{ev}_{C,TB} \mathbin{;} \mu_B \end{array}$$

=⟨ observation ⟩

$$\begin{array}{ll} let & h_1 = \llbracket \Gamma \vdash \text{let box}\, \boxed{x} = e \text{ in } \mathcal{C}_1 \langle\!\langle \text{box}\, \boxed{x} \rangle\!\rangle : C \Rightarrow B \rrbracket \\ & h_2 = \llbracket \Gamma \vdash e_1 : C \rrbracket \\ in & \langle h_1 , h_2 \rangle \mathbin{;} \beta_{C \to TB, C} \mathbin{;} T\, \text{ev}_{C,TB} \mathbin{;} \mu_B \end{array}$$

=⟨ induction hypothesis ⟩

$$\begin{array}{ll} let & h_1 = \llbracket \Gamma \vdash \mathcal{C}_1 \langle\!\langle e \rangle\!\rangle : C \Rightarrow B \rrbracket \\ & h_2 = \llbracket \Gamma \vdash e_1 : C \rrbracket \\ in & \langle h_1 , h_2 \rangle \mathbin{;} \beta_{C \to TB, C} \mathbin{;} T\, \text{ev}_{C,TB} \mathbin{;} \mu_B \end{array}$$

=⟨ definition ⟩

$$\llbracket \Gamma \vdash \mathcal{C}_1 \langle\!\langle e \rangle\!\rangle\ e_1 : B \rrbracket$$

$\diamond \mathcal{C} = \lambda z : C.\ \mathcal{C}_1$

$$\llbracket \Gamma \vdash \mathsf{let\ box}\ \boxed{x} = e\ \mathsf{in}\ \lambda z : C.\ \mathcal{C}_1 \langle\!\langle \mathsf{box}\ \boxed{x} \rangle\!\rangle : C \Rightarrow B \rrbracket$$

=⟨ observation ⟩

$$\langle id_\Gamma\ ,\ f \rangle \,;\, \llbracket \Gamma, x : A^{\mathsf{s}} \vdash \lambda z : C.\ \mathcal{C}_1 \langle\!\langle \mathsf{box}\ \boxed{x} \rangle\!\rangle : C \Rightarrow B \rrbracket$$

=⟨ definition ⟩

$$\begin{array}{lll} \mathit{let} & h = & \llbracket \Gamma, x : A^{\mathsf{s}}, z : C^{\mathsf{i}} \vdash \mathcal{C}_1 \langle\!\langle \mathsf{box}\ \boxed{x} \rangle\!\rangle : B \rrbracket \\ \mathit{in} & \multicolumn{2}{l}{\langle id_\Gamma\ ,\ f \rangle \,;\, \mathsf{curry}\ (h) \,;\, \eta_{C \to TB}} \end{array}$$

=⟨ semantic substitution theorem 5.4 and semantic weakening lemma 5.1 ⟩

$$\begin{array}{lll} \mathit{let} & s = & \llbracket \Gamma, x : A^{\mathsf{s}}, z : C^{\mathsf{i}} \vdash \theta : \Gamma, z : C^{\mathsf{i}}, x : A^{\mathsf{s}} \rrbracket \\ & h = & s \,;\, \llbracket \Gamma, z : C^{\mathsf{i}}, x : A^{\mathsf{s}} \vdash \mathcal{C}_1 \langle\!\langle \mathsf{box}\ \boxed{x} \rangle\!\rangle : B \rrbracket \\ \mathit{in} & \multicolumn{2}{l}{\langle id_\Gamma\ ,\ f \rangle \,;\, \mathsf{curry}\ (h) \,;\, \eta_{C \to TB}} \end{array}$$

=⟨ simplification ⟩

$$\begin{array}{lll} \mathit{let} & h = & \llbracket \Gamma, z : C^{\mathsf{i}}, x : A^{\mathsf{s}} \vdash \mathcal{C}_1 \langle\!\langle \mathsf{box}\ \boxed{x} \rangle\!\rangle : B \rrbracket \\ \mathit{in} & \multicolumn{2}{l}{\langle id_\Gamma\ ,\ f \rangle \,;\, \mathsf{curry}\ (\langle \pi_1 \,;\, \pi_1\ ,\ \pi_2\ ,\ \pi_1 \,;\, \pi_2 \rangle \,;\, h) \,;\, \eta_{C \to TB}} \end{array}$$

=⟨ universal property of exponential ⟩

$$\begin{array}{lll} \mathit{let} & h = & \llbracket \Gamma, z : C^{\mathsf{i}}, x : A^{\mathsf{s}} \vdash \mathcal{C}_1 \langle\!\langle \mathsf{box}\ \boxed{x} \rangle\!\rangle : B \rrbracket \\ \mathit{in} & \multicolumn{2}{l}{\mathsf{curry}\ (\langle id_{\Gamma \times C}\ ,\ \pi_1 \,;\, f \rangle \,;\, h) \,;\, \eta_{C \to TB}} \end{array}$$

=⟨ observation ⟩

$$\begin{array}{lll} \mathit{let} & h = & \llbracket \Gamma, z : C^{\mathsf{i}} \vdash \mathsf{let\ box}\ \boxed{x} = e\ \mathsf{in}\ \mathcal{C}_1 \langle\!\langle \mathsf{box}\ \boxed{x} \rangle\!\rangle : B \rrbracket \\ \mathit{in} & \multicolumn{2}{l}{\mathsf{curry}\ (h) \,;\, \eta_{C \to TB}} \end{array}$$

=⟨ induction hypothesis ⟩

$$\begin{array}{lll} \mathit{let} & h = & \llbracket \Gamma, z : C^{\mathsf{i}} \vdash \mathcal{C}_1 \langle\!\langle e \rangle\!\rangle : B \rrbracket \\ \mathit{in} & \multicolumn{2}{l}{\mathsf{curry}\ (h) \,;\, \eta_{C \to TB}} \end{array}$$

=⟨ definition ⟩

$$\llbracket \Gamma \vdash \lambda z.\ \mathcal{C}_1 \langle\!\langle e \rangle\!\rangle : C \Rightarrow B \rrbracket$$

$\diamond \mathcal{C} = \mathsf{fst}\ \mathcal{C}_1$

$$\llbracket \Gamma \vdash \mathsf{let\ box}\ \boxed{x} = e\ \mathsf{in\ fst}\ \mathcal{C}_1 \langle\!\langle \mathsf{box}\ \boxed{x} \rangle\!\rangle : B \rrbracket$$

$=\langle$ observation $\rangle$

$\langle id_\Gamma , f\rangle ; [\![\Gamma, x : A^s \vdash \mathsf{fst}\, \mathcal{C}_1 \langle\!\langle \mathsf{box}\, \boxed{x} \rangle\!\rangle : B]\!]$

$=\langle$ definition $\rangle$

$\langle id_\Gamma , f\rangle ; [\![\Gamma, x : A^s \vdash \mathcal{C}_1 \langle\!\langle \mathsf{box}\, \boxed{x} \rangle\!\rangle : B \times C]\!] ; T\pi_1$

$=\langle$ observation $\rangle$

$[\![\Gamma \vdash \mathsf{let\ box}\, \boxed{x} = e \,\mathsf{in}\, \mathcal{C}_1 \langle\!\langle \mathsf{box}\, \boxed{x} \rangle\!\rangle : B \times C]\!] ; T\pi_1$

$=\langle$ induction hypothesis $\rangle$

$[\![\Gamma \vdash \mathcal{C}_1 \langle\!\langle e \rangle\!\rangle : B \times C]\!] ; T\pi_1$

$=\langle$ definition $\rangle$

$[\![\Gamma \vdash \mathsf{fst}\, \mathcal{C}_1 \langle\!\langle e \rangle\!\rangle : B]\!]$

$\diamond\, \mathcal{C} = \mathsf{snd}\, \mathcal{C}_1$

$[\![\Gamma \vdash \mathsf{let\ box}\, \boxed{x} = e \,\mathsf{in\ snd}\, \mathcal{C}_1 \langle\!\langle \mathsf{box}\, \boxed{x} \rangle\!\rangle : B]\!]$

$=\langle$ observation $\rangle$

$\langle id_\Gamma , f\rangle ; [\![\Gamma, x : A^s \vdash \mathsf{snd}\, \mathcal{C}_1 \langle\!\langle \mathsf{box}\, \boxed{x} \rangle\!\rangle : B]\!]$

$=\langle$ definition $\rangle$

$\langle id_\Gamma , f\rangle ; [\![\Gamma, x : A^s \vdash \mathcal{C}_1 \langle\!\langle \mathsf{box}\, \boxed{x} \rangle\!\rangle : C \times B]\!] ; T\pi_2$

$=\langle$ observation $\rangle$

$[\![\Gamma \vdash \mathsf{let\ box}\, \boxed{x} = e \,\mathsf{in}\, \mathcal{C}_1 \langle\!\langle \mathsf{box}\, \boxed{x} \rangle\!\rangle : C \times B]\!] ; T\pi_2$

$=\langle$ induction hypothesis $\rangle$

$[\![\Gamma \vdash \mathcal{C}_1 \langle\!\langle e \rangle\!\rangle : C \times B]\!] ; T\pi_2$

$=\langle$ definition $\rangle$

$[\![\Gamma \vdash \mathsf{snd}\, \mathcal{C}_1 \langle\!\langle e \rangle\!\rangle : B]\!]$

$\diamond\, \mathcal{C} = (e_1 , \mathcal{C}_1)$

$[\![\Gamma \vdash \mathsf{let\ box}\, \boxed{x} = e \,\mathsf{in}\, (e_1 , \mathcal{C}_1 \langle\!\langle \mathsf{box}\, \boxed{x} \rangle\!\rangle) : B \times C]\!]$

$=\langle$ observation $\rangle$

$\langle id_\Gamma , f\rangle ; [\![\Gamma, x : A^s \vdash (e_1 , \mathcal{C}_1 \langle\!\langle \mathsf{box}\, \boxed{x} \rangle\!\rangle) : B \times C]\!]$

$=\langle$ definition $\rangle$

$$\langle id_\Gamma , f \rangle ; \langle [\![ \Gamma, x : A^s \vdash e_1 : B ]\!] , [\![ \Gamma, x : A^s \vdash \mathcal{C}_1 \langle\!\langle \mathsf{box}\, \boxed{x} \rangle\!\rangle : C ]\!] \rangle ; \beta_{B,C}$$

$=\langle$ semantic weakening lemma 5.1 $\rangle$

$$\langle id_\Gamma , f \rangle ; \langle \pi_1 ; [\![ \Gamma \vdash e_1 : B ]\!] , [\![ \Gamma, x : A^s \vdash \mathcal{C}_1 \langle\!\langle \mathsf{box}\, \boxed{x} \rangle\!\rangle : C ]\!] \rangle ; \beta_{B,C}$$

$=\langle$ universal property of products $\rangle$

$$\langle \langle id_\Gamma , f \rangle ; \pi_1 ; [\![ \Gamma \vdash e_1 : B ]\!] , \langle id_\Gamma , f \rangle ; [\![ \Gamma, x : A^s \vdash \mathcal{C}_1 \langle\!\langle \mathsf{box}\, \boxed{x} \rangle\!\rangle : C ]\!] \rangle ; \beta_{B,C}$$

$=\langle$ definition of $\pi_1$ $\rangle$

$$\langle [\![ \Gamma \vdash e_1 : B ]\!] , \langle id_\Gamma , f \rangle ; [\![ \Gamma, x : A^s \vdash \mathcal{C}_1 \langle\!\langle \mathsf{box}\, \boxed{x} \rangle\!\rangle : C ]\!] \rangle ; \beta_{B,C}$$

$=\langle$ observation $\rangle$

$$\langle [\![ \Gamma \vdash e_1 : B ]\!] , [\![ \Gamma \vdash \mathsf{let\ box}\, \boxed{x} = e \mathsf{\ in\ } \mathcal{C}_1 \langle\!\langle \mathsf{box}\, \boxed{x} \rangle\!\rangle : C ]\!] \rangle ; \beta_{B,C}$$

$=\langle$ induction hypothesis $\rangle$

$$\langle [\![ \Gamma \vdash e_1 : B ]\!] , [\![ \Gamma \vdash \mathcal{C}_1 \langle\!\langle e \rangle\!\rangle : C ]\!] \rangle ; \beta_{B,C}$$

$=\langle$ definition $\rangle$

$$[\![ \Gamma \vdash (e_1 , \mathcal{C}_1 \langle\!\langle e \rangle\!\rangle) : B \times C ]\!]$$

$\diamond\, \mathcal{C} = (\mathcal{C}_1 , e_1)$

$$[\![ \Gamma \vdash \mathsf{let\ box}\, \boxed{x} = e \mathsf{\ in\ } (\mathcal{C}_1 \langle\!\langle \mathsf{box}\, \boxed{x} \rangle\!\rangle , e_1) : C \times B ]\!]$$

$=\langle$ observation $\rangle$

$$\langle id_\Gamma , f \rangle ; [\![ \Gamma, x : A^s \vdash (\mathcal{C}_1 \langle\!\langle \mathsf{box}\, \boxed{x} \rangle\!\rangle , e_1) : C \times B ]\!]$$

$=\langle$ definition $\rangle$

$$\langle id_\Gamma , f \rangle ; \langle [\![ \Gamma, x : A^s \vdash \mathcal{C}_1 \langle\!\langle \mathsf{box}\, \boxed{x} \rangle\!\rangle : C ]\!] , [\![ \Gamma, x : A^s \vdash e_1 : B ]\!] \rangle ; \beta_{C,B}$$

$=\langle$ semantic weakening lemma 5.1 $\rangle$

$$\langle id_\Gamma , f \rangle ; \langle [\![ \Gamma, x : A^s \vdash \mathcal{C}_1 \langle\!\langle \mathsf{box}\, \boxed{x} \rangle\!\rangle : C ]\!] , \pi_1 ; [\![ \Gamma \vdash e_1 : B ]\!] \rangle ; \beta_{C,B}$$

$=\langle$ universal property of products $\rangle$

$$\langle \langle id_\Gamma , f \rangle ; [\![ \Gamma, x : A^s \vdash \mathcal{C}_1 \langle\!\langle \mathsf{box}\, \boxed{x} \rangle\!\rangle : C ]\!] , \langle id_\Gamma , f \rangle ; \pi_1 ; [\![ \Gamma \vdash e_1 : B ]\!] \rangle ; \beta_{C,B}$$

$=\langle$ definition of $\pi_1$ $\rangle$

$$\langle \langle id_\Gamma , f \rangle ; [\![ \Gamma, x : A^s \vdash \mathcal{C}_1 \langle\!\langle \mathsf{box}\, \boxed{x} \rangle\!\rangle : C ]\!] , [\![ \Gamma \vdash e_1 : B ]\!] \rangle ; \beta_{C,B}$$

$=\langle$ observation $\rangle$

$$\langle [\![ \Gamma \vdash \mathsf{let\ box}\, \boxed{x} = e \mathsf{\ in\ } \mathcal{C}_1 \langle\!\langle \mathsf{box}\, \boxed{x} \rangle\!\rangle : C ]\!] , [\![ \Gamma \vdash e_1 : B ]\!] \rangle ; \beta_{C,B}$$

$=\langle$ induction hypothesis $\rangle$

$$\langle [\![\Gamma \vdash \mathcal{C}_1 \langle\!\langle e \rangle\!\rangle : C]\!] , [\![\Gamma \vdash e_1 : B]\!] \rangle ; \beta_{C,B}$$

$=\langle$ definition $\rangle$

$$[\![\Gamma \vdash (\mathcal{C}_1 \langle\!\langle e \rangle\!\rangle , e_1) : C \times B]\!]$$

$\diamond\, \mathcal{C} = \mathsf{box}\, \boxed{\mathcal{C}_1}$

$$[\![\Gamma \vdash \mathsf{let\ box}\, \boxed{x} = e \ \mathsf{in\ box}\, \boxed{\mathcal{C}_1 \langle\!\langle \mathsf{box}\, \boxed{x} \rangle\!\rangle} : \Box B]\!]$$

$=\langle$ observation $\rangle$

$$\langle id_\Gamma , f \rangle ; [\![\Gamma, x : A^s \vdash \mathsf{box}\, \boxed{\mathcal{C}_1 \langle\!\langle \mathsf{box}\, \boxed{x} \rangle\!\rangle} : \Box B]\!]$$

$=\langle$ definition $\rangle$

$$\langle id_\Gamma , f \rangle ; [\![\Gamma, x : A^s \vdash^s \mathcal{C}_1 \langle\!\langle \mathsf{box}\, \boxed{x} \rangle\!\rangle : B]\!]_p ; \eta_{\Box Y}$$

$=\langle$ observation $\rangle$

$$[\![\Gamma \vdash^s \mathsf{let\ box}\, \boxed{x} = e \ \mathsf{in}\ \mathcal{C}_1 \langle\!\langle \mathsf{box}\, \boxed{x} \rangle\!\rangle : B]\!]_p ; \eta_{\Box Y}$$

$=\langle$ induction hypothesis $\rangle$

$$[\![\Gamma \vdash^s \mathcal{C}_1 \langle\!\langle e \rangle\!\rangle : B]\!]_p ; \eta_{\Box Y}$$

$=\langle$ definition $\rangle$

$$[\![\Gamma \vdash \mathsf{box}\, \boxed{\mathcal{C}_1 \langle\!\langle e \rangle\!\rangle} : \Box B]\!]$$

$\diamond\, \mathcal{C} = \mathsf{let\ box}\, \boxed{z} = \mathcal{C}_1 \ \mathsf{in}\ e_1$

$$[\![\Gamma \vdash \mathsf{let\ box}\, \boxed{x} = e \ \mathsf{in}\ (\mathsf{let\ box}\, \boxed{z} = \mathcal{C}_1 \langle\!\langle \mathsf{box}\, \boxed{x} \rangle\!\rangle \ \mathsf{in}\ e_1) : B]\!]$$

$=\langle$ observation $\rangle$

$$\langle id_\Gamma , f \rangle ; [\![\Gamma, x : A^s \vdash \mathsf{let\ box}\, \boxed{z} = \mathcal{C}_1 \langle\!\langle \mathsf{box}\, \boxed{x} \rangle\!\rangle \ \mathsf{in}\ e_1 : B]\!]$$

$=\langle$ definition $\rangle$

$$\begin{array}{lll} \mathit{let} & g = & [\![\Gamma, x : A^s \vdash \mathcal{C}_1 \langle\!\langle \mathsf{box}\, \boxed{x} \rangle\!\rangle : \Box C]\!] \\ & h = & [\![\Gamma, x : A^s, z : C^s \vdash e_1 : B]\!] \\ \mathit{in} & \multicolumn{2}{l}{\langle id_\Gamma , f \rangle ; \langle id_{\Gamma \times \Box A} , g \rangle ; \tau_{\Gamma \times \Box A, \Box C} ; Th ; \mu_B} \end{array}$$

$=\langle$ semantic substitution theorem 5.4 and semantic weakening lemma 5.1 $\rangle$

$$\begin{array}{lll} \mathit{let} & g = & [\![\Gamma, x : A^{\mathsf{s}} \vdash \mathcal{C}_1 \langle\!\langle \mathsf{box}\, \boxed{x} \rangle\!\rangle : \Box C]\!] \\ & h = & \langle \pi_1 \mathbin{;} \pi_1 , \pi_2 \rangle \mathbin{;} [\![\Gamma, z : C^{\mathsf{s}} \vdash e_1 : B]\!] \\ \mathit{in} & \multicolumn{2}{l}{\langle id_\Gamma , f \rangle \mathbin{;} \langle id_{\Gamma \times \Box A} , g \rangle \mathbin{;} \tau_{\Gamma \times \Box A, \Box C} \mathbin{;} Th \mathbin{;} \mu_B} \end{array}$$

$=\langle$ simplification $\rangle$

$$\begin{array}{lll} \mathit{let} & g = & [\![\Gamma, x : A^{\mathsf{s}} \vdash \mathcal{C}_1 \langle\!\langle \mathsf{box}\, \boxed{x} \rangle\!\rangle : \Box C]\!] \\ & h = & [\![\Gamma, z : C^{\mathsf{s}} \vdash e_1 : B]\!] \\ \mathit{in} & \multicolumn{2}{l}{\langle \langle id_\Gamma , f \rangle , \langle id_\Gamma , f \rangle \mathbin{;} g \rangle \mathbin{;} \tau_{\Gamma \times \Box A, \Box C} \mathbin{;} T \langle \pi_1 \mathbin{;} \pi_1 , \pi_2 \rangle \mathbin{;} Th \mathbin{;} \mu_B} \end{array}$$

$=\langle$ simplification $\rangle$

$$\begin{array}{lll} \mathit{let} & g = & [\![\Gamma, x : A^{\mathsf{s}} \vdash \mathcal{C}_1 \langle\!\langle \mathsf{box}\, \boxed{x} \rangle\!\rangle : \Box C]\!] \\ & h = & [\![\Gamma, z : C^{\mathsf{s}} \vdash e_1 : B]\!] \\ \mathit{in} & \multicolumn{2}{l}{\langle id_\Gamma , \langle id_\Gamma , f \rangle \mathbin{;} g \rangle \mathbin{;} \tau_{\Gamma, \Box C} \mathbin{;} Th \mathbin{;} \mu_B} \end{array}$$

$=\langle$ observation $\rangle$

$$\begin{array}{lll} \mathit{let} & g = & [\![\Gamma \vdash \mathsf{let\ box}\, \boxed{x} = e \,\mathsf{in}\, \mathcal{C}_1 \langle\!\langle \mathsf{box}\, \boxed{x} \rangle\!\rangle : \Box C]\!] \\ & h = & [\![\Gamma, z : C^{\mathsf{s}} \vdash e_1 : B]\!] \\ \mathit{in} & \multicolumn{2}{l}{\langle id_\Gamma , g \rangle \mathbin{;} \tau_{\Gamma, \Box C} \mathbin{;} Th \mathbin{;} \mu_B} \end{array}$$

$=\langle$ induction hypothesis $\rangle$

$$\begin{array}{lll} \mathit{let} & g = & [\![\Gamma \vdash \mathcal{C}_1 \langle\!\langle e \rangle\!\rangle : \Box C]\!] \\ & h = & [\![\Gamma, z : C^{\mathsf{s}} \vdash e_1 : B]\!] \\ \mathit{in} & \multicolumn{2}{l}{\langle id_\Gamma , g \rangle \mathbin{;} \tau_{\Gamma, \Box C} \mathbin{;} Th \mathbin{;} \mu_B} \end{array}$$

$=\langle$ definition $\rangle$

$$[\![\Gamma \vdash \mathsf{let\ box}\, \boxed{z} = \mathcal{C}_1 \langle\!\langle e \rangle\!\rangle \,\mathsf{in}\, e_1 : B]\!]$$

$\diamond\, \mathcal{C} = \mathsf{let\ box}\, \boxed{z} = e_1 \,\mathsf{in}\, \mathcal{C}_1$

$$[\![\Gamma \vdash \mathsf{let\ box}\, \boxed{x} = e \,\mathsf{in}\, (\mathsf{let\ box}\, \boxed{z} = e_1 \,\mathsf{in}\, \mathcal{C}_1 \langle\!\langle \mathsf{box}\, \boxed{x} \rangle\!\rangle) : B]\!]$$

$=\langle$ observation $\rangle$

$$\langle id_\Gamma , f \rangle \mathbin{;} [\![\Gamma, x : A^{\mathsf{s}} \vdash \mathsf{let\ box}\, \boxed{z} = e_1 \,\mathsf{in}\, \mathcal{C}_1 \langle\!\langle \mathsf{box}\, \boxed{x} \rangle\!\rangle : B]\!]$$

$=\langle$ definition $\rangle$

$$\begin{array}{lll} \mathit{let} & h_1 = & [\![\Gamma, x : A^{\mathsf{s}} \vdash e_1 : \Box C]\!] \\ & h_2 = & [\![\Gamma, x : A^{\mathsf{s}}, z : C^{\mathsf{s}} \vdash \mathcal{C}_1 \langle\!\langle \mathsf{box}\, \boxed{x} \rangle\!\rangle : B]\!] \\ \mathit{in} & \multicolumn{2}{l}{\langle id_\Gamma , f \rangle \mathbin{;} \langle id_{\Gamma \times \Box A} , h_1 \rangle \mathbin{;} \tau_{\Gamma \times \Box A, \Box C} \mathbin{;} Th_2 \mathbin{;} \mu_B} \end{array}$$

$=\langle$ semantic weakening lemma 5.1 $\rangle$

$$\begin{array}{ll} let & h_1 = [\![\Gamma \vdash e_1 : \Box C]\!] \\ & h_2 = [\![\Gamma, x : A^s, z : C^s \vdash \mathcal{C}_1 \langle\!\langle \mathsf{box}\, \boxed{x} \rangle\!\rangle : B]\!] \\ in & \langle id_\Gamma , f \rangle ; \langle id_{\Gamma\times\Box A} , \pi_1 ; h_1 \rangle ; \tau_{\Gamma\times\Box A,\Box C} ; Th_2 ; \mu_B \end{array}$$

$=\langle$ simplification $\rangle$

$$\begin{array}{ll} let & h_1 = [\![\Gamma \vdash e_1 : \Box C]\!] \\ & h_2 = [\![\Gamma, x : A^s, z : C^s \vdash \mathcal{C}_1 \langle\!\langle \mathsf{box}\, \boxed{x} \rangle\!\rangle : B]\!] \\ in & \langle \langle id_\Gamma , f \rangle , h_1 \rangle ; \tau_{\Gamma\times\Box A,\Box C} ; Th_2 ; \mu_B \end{array}$$

$=\langle$ semantic substitution theorem 5.4 and semantic weakening lemma 5.1 $\rangle$

$$\begin{array}{ll} let & h_1 = [\![\Gamma \vdash e_1 : \Box C]\!] \\ & h_2 = [\![\Gamma, z : C^s, x : A^s \vdash \mathcal{C}_1 \langle\!\langle \mathsf{box}\, \boxed{x} \rangle\!\rangle : B]\!] \\ in & \langle \langle id_\Gamma , f \rangle , h_1 \rangle ; \tau_{\Gamma\times\Box A,\Box C} ; T\langle \pi_1 ; \pi_1 , \pi_2 , \pi_1 ; \pi_2 \rangle ; Th_2 ; \mu_B \end{array}$$

$=\langle$ simplification $\rangle$

$$\begin{array}{ll} let & h_1 = [\![\Gamma \vdash e_1 : \Box C]\!] \\ & h_2 = [\![\Gamma, z : C^s, x : A^s \vdash \mathcal{C}_1 \langle\!\langle \mathsf{box}\, \boxed{x} \rangle\!\rangle : B]\!] \\ in & \langle id_\Gamma , h_1 \rangle ; \tau_{\Gamma,\Box C} ; T\langle id_{\Gamma\times\Box C} , \pi_1 ; f \rangle ; Th_2 ; \mu_B \end{array}$$

$=\langle$ observation $\rangle$

$$\begin{array}{ll} let & h_1 = [\![\Gamma \vdash e_1 : \Box C]\!] \\ & h_2 = [\![\Gamma, z : C^s \vdash \mathsf{let\ box}\, \boxed{x} = e\ \mathsf{in}\ \mathcal{C}_1 \langle\!\langle \mathsf{box}\, \boxed{x} \rangle\!\rangle : B]\!] \\ in & \langle id_\Gamma , h_1 \rangle ; \tau_{\Gamma,\Box C} ; Th_2 ; \mu_B \end{array}$$

$=\langle$ induction hypothesis $\rangle$

$$\begin{array}{ll} let & h_1 = [\![\Gamma \vdash e_1 : \Box C]\!] \\ & h_2 = [\![\Gamma, z : C^s \vdash \mathcal{C}_1 \langle\!\langle e \rangle\!\rangle : B]\!] \\ in & \langle id_\Gamma , h_1 \rangle ; \tau_{\Gamma,\Box C} ; Th_2 ; \mu_B \end{array}$$

$=\langle$ definition $\rangle$

$$[\![\Gamma \vdash \mathsf{let\ box}\, \boxed{z} = e_1\ \mathsf{in}\ \mathcal{C}_1 \langle\!\langle e \rangle\!\rangle : B]\!]$$

$$\diamond\ \frac{\Gamma \vdash e : \Box A \qquad \Gamma \vdash \mathcal{E} \langle\!\langle e \rangle\!\rangle : B \qquad \Gamma \vdash \mathsf{let\ box}\, \boxed{x} = e\ \mathsf{in}\ \mathcal{E} \langle\!\langle \mathsf{box}\, \boxed{x} \rangle\!\rangle : B}{\Gamma \vdash \mathcal{E} \langle\!\langle e \rangle\!\rangle \approx \mathsf{let\ box}\, \boxed{x} = e\ \mathsf{in}\ \mathcal{E} \langle\!\langle \mathsf{box}\, \boxed{x} \rangle\!\rangle : B}\ \Box\eta\text{-IMPURE}$$

We proceed by cases on $\mathcal{E}$.

$\diamond\ \mathcal{E} = [\,]$

$$\llbracket \Gamma \vdash \mathsf{let\ box}\ \boxed{x} = e\ \mathsf{in\ box}\ \boxed{x} : \square A \rrbracket$$

$=\langle$ definition $\rangle$

$$\langle id_\Gamma , \llbracket \Gamma \vdash e : \square A \rrbracket \rangle \mathbin{;} \tau_{\Gamma,\square A} \mathbin{;} T \llbracket \Gamma, x : A^{\mathsf{s}} \vdash \mathsf{box}\ \boxed{x} : \square A \rrbracket \mathbin{;} \mu_{\square A}$$

$=\langle$ definition $\rangle$

$$\langle id_\Gamma , \llbracket \Gamma \vdash e : \square A \rrbracket \rangle \mathbin{;} \tau_{\Gamma,\square A} \mathbin{;} T \llbracket \Gamma, x : A^{\mathsf{s}} \vdash^{\mathsf{s}} x : A \rrbracket_p \mathbin{;} T\eta_{\square A} \mathbin{;} \mu_{\square A}$$

$=\langle$ definition $\rangle$

$$\langle id_\Gamma , \llbracket \Gamma \vdash e : \square A \rrbracket \rangle \mathbin{;} \tau_{\Gamma,\square A} \mathbin{;} T\pi_2 \mathbin{;} T\eta_{\square A} \mathbin{;} \mu_{\square A}$$

$=\langle$ monad laws $\rangle$

$$\langle id_\Gamma , \llbracket \Gamma \vdash e : \square A \rrbracket \rangle \mathbin{;} \tau_{\Gamma,\square A} \mathbin{;} T\pi_2 \mathbin{;} id_{T\square A}$$

$=\langle$ tensorial action of $T$ $\rangle$

$$\langle id_\Gamma , \llbracket \Gamma \vdash e : \square A \rrbracket \rangle \mathbin{;} \pi_2$$

$=\langle$ applying $\pi_2$ $\rangle$

$$\llbracket \Gamma \vdash e : \square A \rrbracket$$

$\diamond\ \mathcal{E} = e_1\ \mathcal{E}_1$

$$\llbracket \Gamma \vdash \mathsf{let\ box}\ \boxed{x} = e\ \mathsf{in}\ e_1\ \mathcal{E}_1 \langle\!\langle \mathsf{box}\ \boxed{x} \rangle\!\rangle : B \rrbracket$$

$=\langle$ definition $\rangle$

$$\begin{array}{ll} \mathit{let} & \begin{array}{lcl} f & = & \llbracket \Gamma \vdash e : \square A \rrbracket \\ g & = & \llbracket \Gamma, x : A^{\mathsf{s}} \vdash e_1\ \mathcal{E}_1 \langle\!\langle \mathsf{box}\ \boxed{x} \rangle\!\rangle : B \rrbracket \end{array} \\ \mathit{in} & \langle id_\Gamma , f \rangle \mathbin{;} \tau_{\Gamma,\square A} \mathbin{;} Tg \mathbin{;} \mu_B \end{array}$$

$=\langle$ definition $\rangle$

$$\begin{array}{ll} \mathit{let} & \begin{array}{lcl} f & = & \llbracket \Gamma \vdash e : \square A \rrbracket \\ g_1 & = & \llbracket \Gamma, x : A^{\mathsf{s}} \vdash e_1 : C \Rightarrow B \rrbracket \\ g_2 & = & \llbracket \Gamma, x : A^{\mathsf{s}} \vdash \mathcal{E}_1 \langle\!\langle \mathsf{box}\ \boxed{x} \rangle\!\rangle : C \rrbracket \\ g & = & \langle g_1 , g_2 \rangle \mathbin{;} \beta_{C \to TB, C} \mathbin{;} T\,\mathsf{ev}_{C,TY} \mathbin{;} \mu_B \end{array} \\ \mathit{in} & \langle id_\Gamma , f \rangle \mathbin{;} \tau_{\Gamma,\square A} \mathbin{;} Tg \mathbin{;} \mu_B \end{array}$$

$=\langle$ functoriality of $T$ $\rangle$

$$\begin{array}{ll} \mathit{let} & \begin{array}{lcl} f & = & \llbracket \Gamma \vdash e : \square A \rrbracket \\ g_1 & = & \llbracket \Gamma, x : A^{\mathsf{s}} \vdash e_1 : C \Rightarrow B \rrbracket \\ g_2 & = & \llbracket \Gamma, x : A^{\mathsf{s}} \vdash \mathcal{E}_1 \langle\!\langle \mathsf{box}\ \boxed{x} \rangle\!\rangle : C \rrbracket \end{array} \\ \mathit{in} & \langle id_\Gamma , f \rangle \mathbin{;} \tau_{\Gamma,\square A} \mathbin{;} T\langle g_1 , g_2 \rangle \mathbin{;} T\beta_{C \to TB, C} \mathbin{;} T^2\,\mathsf{ev}_{C,TY} \mathbin{;} T\mu_B \mathbin{;} \mu_B \end{array}$$

$=\langle$ semantic weakening lemma 5.1 $\rangle$

$$\begin{array}{llcl} & f & = & [\![\Gamma \vdash e : \Box A]\!] \\ let & g_1 & = & \pi_1 \,;\, [\![\Gamma \vdash e_1 : C \Rightarrow B]\!] \\ & g_2 & = & [\![\Gamma, x : A^s \vdash \mathcal{E}_1 \langle\!\langle \mathsf{box}\, \boxed{x} \rangle\!\rangle : C]\!] \\ in & \multicolumn{3}{l}{\langle id_\Gamma \,,\, f\rangle \,;\, \tau_{\Gamma,\Box A} \,;\, T\langle g_1 \,,\, g_2\rangle \,;\, T\beta_{C\to TB,C} \,;\, T^2\,\mathsf{ev}_{C,TY} \,;\, T\mu_B \,;\, \mu_B} \end{array}$$

$=\langle$ simplification $\rangle$

$$\begin{array}{llcl} & f & = & [\![\Gamma \vdash e : \Box A]\!] \\ let & g_1 & = & [\![\Gamma \vdash e_1 : C \Rightarrow B]\!] \\ & g_2 & = & [\![\Gamma, x : A^s \vdash \mathcal{E}_1 \langle\!\langle \mathsf{box}\, \boxed{x} \rangle\!\rangle : C]\!] \\ in & \multicolumn{3}{l}{\langle id_\Gamma \,,\, f\rangle \,;\, \tau_{\Gamma,\Box A} \,;\, T\langle \pi_1 \,;\, g_1 \,,\, g_2\rangle \,;\, T\beta_{C\to TB,C} \,;\, T^2\,\mathsf{ev}_{C,TY} \,;\, T\mu_B \,;\, \mu_B} \end{array}$$

$=\langle$ simplification $\rangle$

$$\begin{array}{llcl} & f & = & [\![\Gamma \vdash e : \Box A]\!] \\ let & g_1 & = & [\![\Gamma \vdash e_1 : C \Rightarrow B]\!] \\ & g_2 & = & [\![\Gamma, x : A^s \vdash \mathcal{E}_1 \langle\!\langle \mathsf{box}\, \boxed{x} \rangle\!\rangle : C]\!] \\ in & \multicolumn{3}{l}{\langle g_1 \,,\, \langle id_\Gamma \,,\, f\rangle \,;\, \tau_{\Gamma,\Box A} \,;\, Tg_2 \,;\, \mu_Z\rangle \,;\, \beta_{C\to TB,C} \,;\, T\,\mathsf{ev}_{C,TY} \,;\, \mu_B} \end{array}$$

$=\langle$ definition $\rangle$

$$\begin{array}{llcl} let & g_1 & = & [\![\Gamma \vdash e_1 : C \Rightarrow B]\!] \\ & g_2 & = & [\![\Gamma \vdash \mathsf{let\ box}\, \boxed{x} = e \,\mathsf{in}\, \mathcal{E}_1 \langle\!\langle \mathsf{box}\, \boxed{x} \rangle\!\rangle : C]\!] \\ in & \multicolumn{3}{l}{\langle g_1 \,,\, g_2\rangle \,;\, \beta_{C\to TB,C} \,;\, T\,\mathsf{ev}_{C,TY} \,;\, \mu_B} \end{array}$$

$=\langle$ induction hypothesis $\rangle$

$$\begin{array}{llcl} let & g_1 & = & [\![\Gamma \vdash e_1 : C \Rightarrow B]\!] \\ & g_2 & = & [\![\Gamma \vdash \mathcal{E}_1 \langle\!\langle e \rangle\!\rangle : C]\!] \\ in & \multicolumn{3}{l}{\langle g_1 \,,\, g_2\rangle \,;\, \beta_{C\to TB,C} \,;\, T\,\mathsf{ev}_{C,TY} \,;\, \mu_B} \end{array}$$

$=\langle$ definition $\rangle$

$$[\![\Gamma \vdash e_1\ \mathcal{E}_1 \langle\!\langle e \rangle\!\rangle : B]\!]$$

$\diamond\ \mathcal{E} = \mathcal{E}_1\ v$

$$[\![\Gamma \vdash \mathsf{let\ box}\, \boxed{x} = e \,\mathsf{in}\, \mathcal{E}_1 \langle\!\langle \mathsf{box}\, \boxed{x} \rangle\!\rangle\ v : B]\!]$$

$=\langle$ definition $\rangle$

$$\begin{array}{ll} \textit{let} & \begin{array}{lcl} f & = & [\![\Gamma \vdash e : \Box A]\!] \\ g & = & [\![\Gamma, x : A^{\mathsf{s}} \vdash \mathcal{E}_1 \langle\!\langle \mathsf{box}\, \boxed{x} \rangle\!\rangle\, v : B]\!] \end{array} \\ \textit{in} & \langle id_\Gamma\,,\, f\rangle \,;\, \tau_{\Gamma,\Box A} \,;\, Tg \,;\, \mu_B \end{array}$$

=⟨ definition ⟩

$$\begin{array}{ll} \textit{let} & \begin{array}{lcl} f & = & [\![\Gamma \vdash e : \Box A]\!] \\ g_1 & = & [\![\Gamma, x : A^{\mathsf{s}} \vdash \mathcal{E}_1 \langle\!\langle \mathsf{box}\, \boxed{x} \rangle\!\rangle : C \Rightarrow B]\!] \\ g_2 & = & [\![\Gamma, x : A^{\mathsf{s}} \vdash v : C]\!] \\ g & = & \langle g_1\,,\, g_2\rangle \,;\, \beta_{(C\to TB),C} \,;\, T\,\mathsf{ev}_{C,TY} \,;\, \mu_B \end{array} \\ \textit{in} & \langle id_\Gamma\,,\, f\rangle \,;\, \tau_{\Gamma,\Box A} \,;\, Tg \,;\, \mu_B \end{array}$$

=⟨ functoriality of $T$ ⟩

$$\begin{array}{ll} \textit{let} & \begin{array}{lcl} f & = & [\![\Gamma \vdash e : \Box A]\!] \\ g_1 & = & [\![\Gamma, x : A^{\mathsf{s}} \vdash \mathcal{E}_1 \langle\!\langle \mathsf{box}\, \boxed{x} \rangle\!\rangle : C \Rightarrow B]\!] \\ g_2 & = & [\![\Gamma, x : A^{\mathsf{s}} \vdash v : C]\!] \end{array} \\ \textit{in} & \langle id_\Gamma\,,\, f\rangle \,;\, \tau_{\Gamma,\Box A} \,;\, T\langle g_1\,,\, g_2\rangle \,;\, T\beta_{C\to TB,C} \,;\, T^2\,\mathsf{ev}_{C,TY} \,;\, T\mu_B \,;\, \mu_B \end{array}$$

=⟨ semantic weakening lemma 5.1 ⟩

$$\begin{array}{ll} \textit{let} & \begin{array}{lcl} f & = & [\![\Gamma \vdash e : \Box A]\!] \\ g_1 & = & [\![\Gamma, x : A^{\mathsf{s}} \vdash \mathcal{E}_1 \langle\!\langle \mathsf{box}\, \boxed{x} \rangle\!\rangle : C \Rightarrow B]\!] \\ g_2 & = & \pi_1 \,;\, [\![\Gamma \vdash v : C]\!] \end{array} \\ \textit{in} & \langle id_\Gamma\,,\, f\rangle \,;\, \tau_{\Gamma,\Box A} \,;\, T\langle g_1\,,\, g_2\rangle \,;\, T\beta_{C\to TB,C} \,;\, T^2\,\mathsf{ev}_{C,TY} \,;\, T\mu_B \,;\, \mu_B \end{array}$$

=⟨ simplification ⟩

$$\begin{array}{ll} \textit{let} & \begin{array}{lcl} f & = & [\![\Gamma \vdash e : \Box A]\!] \\ g_1 & = & [\![\Gamma, x : A^{\mathsf{s}} \vdash \mathcal{E}_1 \langle\!\langle \mathsf{box}\, \boxed{x} \rangle\!\rangle : C \Rightarrow B]\!] \\ g_2 & = & [\![\Gamma \vdash v : C]\!] \end{array} \\ \textit{in} & \langle id_\Gamma\,,\, f\rangle \,;\, \tau_{\Gamma,\Box A} \,;\, T\langle g_1\,,\, \pi_1 \,;\, g_2\rangle \,;\, T\beta_{C\to TB,C} \,;\, T^2\,\mathsf{ev}_{C,TY} \,;\, T\mu_B \,;\, \mu_B \end{array}$$

=⟨ simplification ⟩

$$\begin{array}{ll} \textit{let} & \begin{array}{lcl} f & = & [\![\Gamma \vdash e : \Box A]\!] \\ g_1 & = & [\![\Gamma, x : A^{\mathsf{s}} \vdash \mathcal{E}_1 \langle\!\langle \mathsf{box}\, \boxed{x} \rangle\!\rangle : C \Rightarrow B]\!] \\ g_2 & = & [\![\Gamma \vdash v : C]\!] \end{array} \\ \textit{in} & \langle \langle id_\Gamma\,,\, f\rangle \,;\, \tau_{\Gamma,\Box A} \,;\, Tg_1 \,;\, \mu_{C\to TB}\,,\, g_2\rangle \,;\, \beta_{C\to TB,C} \,;\, T\,\mathsf{ev}_{C,TY} \,;\, \mu_B \end{array}$$

=⟨ definition ⟩

$$\begin{array}{l} \mathit{let} \quad g_1 = [\![\Gamma \vdash \text{let box}\,\boxed{x} = e \text{ in } \mathcal{E}_1 \langle\!\langle \text{box}\,\boxed{x} \rangle\!\rangle : C \Rightarrow B]\!] \\ \qquad\;\; g_2 = [\![\Gamma \vdash v : C]\!] \\ \mathit{in} \quad \langle g_1 , g_2 \rangle \,;\, \beta_{C\to TB,C} \,;\, T\,\text{ev}_{C,TY} \,;\, \mu_B \end{array}$$

$=\langle$ induction hypothesis $\rangle$

$$\begin{array}{l} \mathit{let} \quad g_1 = [\![\Gamma \vdash \mathcal{E}_1 \langle\!\langle e \rangle\!\rangle : C \Rightarrow B]\!] \\ \qquad\;\; g_2 = [\![\Gamma \vdash v : C]\!] \\ \mathit{in} \quad \langle g_1 , g_2 \rangle \,;\, \beta_{C\to TB,C} \,;\, T\,\text{ev}_{C,TY} \,;\, \mu_B \end{array}$$

$=\langle$ definition $\rangle$

$$[\![\Gamma \vdash \mathcal{E}_1 \langle\!\langle e \rangle\!\rangle\, v : B]\!]$$

◇ $\mathcal{E} = \text{fst}\ \mathcal{E}_1$

$$[\![\Gamma \vdash \text{let box}\,\boxed{x} = e \text{ in fst } \mathcal{E}_1 \langle\!\langle \text{box}\,\boxed{x} \rangle\!\rangle : B]\!]$$

$=\langle$ definition $\rangle$

$$\begin{array}{l} \mathit{let} \quad f = [\![\Gamma \vdash e : \Box A]\!] \\ \qquad\;\; g = [\![\Gamma, x : A^{s} \vdash \text{fst } \mathcal{E}_1 \langle\!\langle \text{box}\,\boxed{x} \rangle\!\rangle : B]\!] \\ \mathit{in} \quad \langle id_\Gamma , f \rangle \,;\, \tau_{\Gamma,\Box A} \,;\, Tg \,;\, \mu_B \end{array}$$

$=\langle$ definition $\rangle$

$$\begin{array}{l} \mathit{let} \quad f = [\![\Gamma \vdash e : \Box A]\!] \\ \qquad\;\; g = [\![\Gamma, x : A^{s} \vdash \mathcal{E}_1 \langle\!\langle \text{box}\,\boxed{x} \rangle\!\rangle : B \times C]\!] \\ \mathit{in} \quad \langle id_\Gamma , f \rangle \,;\, \tau_{\Gamma,\Box A} \,;\, Tg \,;\, T^2\pi_1 \,;\, \mu_B \end{array}$$

$=\langle$ monad laws $\rangle$

$$\begin{array}{l} \mathit{let} \quad f = [\![\Gamma \vdash e : \Box A]\!] \\ \qquad\;\; g = [\![\Gamma, x : A^{s} \vdash \mathcal{E}_1 \langle\!\langle \text{box}\,\boxed{x} \rangle\!\rangle : B \times C]\!] \\ \mathit{in} \quad \langle id_\Gamma , f \rangle \,;\, \tau_{\Gamma,\Box A} \,;\, Tg \,;\, \mu_B \,;\, T\pi_1 \end{array}$$

$=\langle$ definition $\rangle$

$$[\![\Gamma \vdash \text{let box}\,\boxed{x} = e \text{ in } \mathcal{E}_1 \langle\!\langle \text{box}\,\boxed{x} \rangle\!\rangle : B \times C]\!] \,;\, T\pi_1$$

$=\langle$ induction hypothesis $\rangle$

$$[\![\Gamma \vdash \mathcal{E}_1 \langle\!\langle e \rangle\!\rangle : B \times C]\!] \,;\, T\pi_1$$

$=\langle$ definition $\rangle$

$$[\![\Gamma \vdash \text{fst } \mathcal{E}_1 \langle\!\langle e \rangle\!\rangle : B]\!]$$

$\diamond\ \mathcal{E} = \mathsf{snd}\ \mathcal{E}_1$

$$\llbracket \Gamma \vdash \mathsf{let\ box}\ \boxed{x} = e\ \mathsf{in\ snd}\ \mathcal{E}_1 \langle\!\langle \mathsf{box}\ \boxed{x} \rangle\!\rangle : B \rrbracket$$

$=\langle$ definition $\rangle$

$$\begin{array}{ll} \mathit{let} & \begin{array}{lll} f & = & \llbracket \Gamma \vdash e : \Box A \rrbracket \\ g & = & \llbracket \Gamma, x : A^s \vdash \mathsf{snd}\ \mathcal{E}_1 \langle\!\langle \mathsf{box}\ \boxed{x} \rangle\!\rangle : B \rrbracket \end{array} \\ \mathit{in} & \langle id_\Gamma , f \rangle \,;\, \tau_{\Gamma,\Box A} \,;\, Tg \,;\, \mu_B \end{array}$$

$=\langle$ definition $\rangle$

$$\begin{array}{ll} \mathit{let} & \begin{array}{lll} f & = & \llbracket \Gamma \vdash e : \Box A \rrbracket \\ g & = & \llbracket \Gamma, x : A^s \vdash \mathcal{E}_1 \langle\!\langle \mathsf{box}\ \boxed{x} \rangle\!\rangle : C \times B \rrbracket \end{array} \\ \mathit{in} & \langle id_\Gamma , f \rangle \,;\, \tau_{\Gamma,\Box A} \,;\, Tg \,;\, T^2\pi_2 \,;\, \mu_B \end{array}$$

$=\langle$ monad laws $\rangle$

$$\begin{array}{ll} \mathit{let} & \begin{array}{lll} f & = & \llbracket \Gamma \vdash e : \Box A \rrbracket \\ g & = & \llbracket \Gamma, x : A^s \vdash \mathcal{E}_1 \langle\!\langle \mathsf{box}\ \boxed{x} \rangle\!\rangle : C \times B \rrbracket \end{array} \\ \mathit{in} & \langle id_\Gamma , f \rangle \,;\, \tau_{\Gamma,\Box A} \,;\, Tg \,;\, \mu_B \,;\, T\pi_2 \end{array}$$

$=\langle$ definition $\rangle$

$$\llbracket \Gamma \vdash \mathsf{let\ box}\ \boxed{x} = e\ \mathsf{in}\ \mathcal{E}_1 \langle\!\langle \mathsf{box}\ \boxed{x} \rangle\!\rangle : C \times B \rrbracket \,;\, T\pi_2$$

$=\langle$ induction hypothesis $\rangle$

$$\llbracket \Gamma \vdash \mathcal{E}_1 \langle\!\langle e \rangle\!\rangle : C \times B \rrbracket \,;\, T\pi_2$$

$=\langle$ definition $\rangle$

$$\llbracket \Gamma \vdash \mathsf{snd}\ \mathcal{E}_1 \langle\!\langle e \rangle\!\rangle : B \rrbracket$$

$\diamond\ \mathcal{E} = (e_1 , \mathcal{E}_1)$

$$\llbracket \Gamma \vdash \mathsf{let\ box}\ \boxed{x} = e\ \mathsf{in}\ (e_1 , \mathcal{E}_1 \langle\!\langle \mathsf{box}\ \boxed{x} \rangle\!\rangle) : B \times C \rrbracket$$

$=\langle$ definition $\rangle$

$$\begin{array}{ll} \mathit{let} & \begin{array}{lll} f & = & \llbracket \Gamma \vdash e : \Box A \rrbracket \\ g & = & \llbracket \Gamma, x : A^s \vdash (e_1 , \mathcal{E}_1 \langle\!\langle \mathsf{box}\ \boxed{x} \rangle\!\rangle) : B \times C \rrbracket \end{array} \\ \mathit{in} & \langle id_\Gamma , f \rangle \,;\, \tau_{\Gamma,\Box A} \,;\, Tg \,;\, \mu_{B \times C} \end{array}$$

$=\langle$ definition $\rangle$

$$\begin{array}{lrcl} & f & = & \llbracket \Gamma \vdash e : \Box A \rrbracket \\ \mathit{let} & g_1 & = & \llbracket \Gamma, x : A^s \vdash e_1 : B \rrbracket \\ & g_2 & = & \llbracket \Gamma, x : A^s \vdash \mathcal{E}_1 \langle\!\langle \mathsf{box}\, \boxed{x} \rangle\!\rangle : C \rrbracket \\ & g & = & \langle g_1 , g_2 \rangle \,;\, \beta_{B,C} \\ \mathit{in} & \multicolumn{3}{l}{\langle id_\Gamma , f \rangle \,;\, \tau_{\Gamma,\Box A} \,;\, Tg \,;\, \mu_{B\times C}} \end{array}$$

$=\langle$ semantic weakening lemma 5.1 $\rangle$

$$\begin{array}{lrcl} & f & = & \llbracket \Gamma \vdash e : \Box A \rrbracket \\ \mathit{let} & g_1 & = & \pi_1 \,;\, \llbracket \Gamma \vdash e_1 : B \rrbracket \\ & g_2 & = & \llbracket \Gamma, x : A^s \vdash \mathcal{E}_1 \langle\!\langle \mathsf{box}\, \boxed{x} \rangle\!\rangle : C \rrbracket \\ & g & = & \langle g_1 , g_2 \rangle \,;\, \beta_{B,C} \\ \mathit{in} & \multicolumn{3}{l}{\langle id_\Gamma , f \rangle \,;\, \tau_{\Gamma,\Box A} \,;\, Tg \,;\, \mu_{B\times C}} \end{array}$$

$=\langle$ simplification $\rangle$

$$\begin{array}{lrcl} & f & = & \llbracket \Gamma \vdash e : \Box A \rrbracket \\ \mathit{let} & g_1 & = & \llbracket \Gamma \vdash e_1 : B \rrbracket \\ & g_2 & = & \llbracket \Gamma, x : A^s \vdash \mathcal{E}_1 \langle\!\langle \mathsf{box}\, \boxed{x} \rangle\!\rangle : C \rrbracket \\ \mathit{in} & \multicolumn{3}{l}{\langle id_\Gamma , f \rangle \,;\, \tau_{\Gamma,\Box A} \,;\, T\langle \pi_1 \,;\, g_1 , g_2 \rangle \,;\, T\beta_{B,C} \,;\, \mu_{B\times C}} \end{array}$$

$=\langle$ simplification $\rangle$

$$\begin{array}{lrcl} & f & = & \llbracket \Gamma \vdash e : \Box A \rrbracket \\ \mathit{let} & g_1 & = & \llbracket \Gamma \vdash e_1 : B \rrbracket \\ & g_2 & = & \llbracket \Gamma, x : A^s \vdash \mathcal{E}_1 \langle\!\langle \mathsf{box}\, \boxed{x} \rangle\!\rangle : C \rrbracket \\ \mathit{in} & \multicolumn{3}{l}{\langle g_1 , \langle id_\Gamma , f \rangle \,;\, \tau_{\Gamma,\Box A} \,;\, Tg_2 \,;\, \mu_C \rangle \,;\, \beta_{B,C}} \end{array}$$

$=\langle$ definition $\rangle$

$$\begin{array}{lrcl} \mathit{let} & g_1 & = & \llbracket \Gamma \vdash e_1 : B \rrbracket \\ & g_2 & = & \llbracket \Gamma \vdash \mathsf{let\ box}\, \boxed{x} = e \,\mathsf{in}\, \mathcal{E}_1 \langle\!\langle \mathsf{box}\, \boxed{x} \rangle\!\rangle : C \rrbracket \\ \mathit{in} & \multicolumn{3}{l}{\langle g_1 , g_2 \rangle \,;\, \beta_{B,C}} \end{array}$$

$=\langle$ induction hypothesis $\rangle$

$$\begin{array}{lrcl} \mathit{let} & g_1 & = & \llbracket \Gamma \vdash e_1 : B \rrbracket \\ & g_2 & = & \llbracket \Gamma \vdash \mathcal{E}_1 \langle\!\langle e \rangle\!\rangle : C \rrbracket \\ \mathit{in} & \multicolumn{3}{l}{\langle g_1 , g_2 \rangle \,;\, \beta_{B,C}} \end{array}$$

$=\langle$ definition $\rangle$

$$\llbracket \Gamma \vdash (e_1 , \mathcal{E}_1 \langle\!\langle e \rangle\!\rangle) : B \times C \rrbracket$$

$\diamond\, \mathcal{E} = (\mathcal{E}_1\,, v)$

$$[\![\Gamma \vdash \text{let box}\,\boxed{x} = e \text{ in } (\mathcal{E}_1 \langle\!\langle \text{box}\,\boxed{x} \rangle\!\rangle\,, v) : C \times B]\!]$$

=⟨ definition ⟩

$$\begin{array}{ll} \textit{let} & \begin{array}{lcl} f & = & [\![\Gamma \vdash e : \Box A]\!] \\ g & = & [\![\Gamma, x : A^s \vdash (\mathcal{E}_1 \langle\!\langle \text{box}\,\boxed{x} \rangle\!\rangle\,, v) : C \times B]\!] \end{array} \\ \textit{in} & \langle id_\Gamma\,, f\rangle \,;\, \tau_{\Gamma,\Box A} \,;\, Tg \,;\, \mu_{C\times B} \end{array}$$

=⟨ definition ⟩

$$\begin{array}{ll} \textit{let} & \begin{array}{lcl} f & = & [\![\Gamma \vdash e : \Box A]\!] \\ g_1 & = & [\![\Gamma, x : A^s \vdash \mathcal{E}_1 \langle\!\langle \text{box}\,\boxed{x} \rangle\!\rangle : C]\!] \\ g_2 & = & [\![\Gamma, x : A^s \vdash v : B]\!] \\ g & = & \langle g_1\,, g_2\rangle \,;\, \beta_{C,B} \end{array} \\ \textit{in} & \langle id_\Gamma\,, f\rangle \,;\, \tau_{\Gamma,\Box A} \,;\, Tg \,;\, \mu_{C\times B} \end{array}$$

=⟨ semantic weakening lemma 5.1 ⟩

$$\begin{array}{ll} \textit{let} & \begin{array}{lcl} f & = & [\![\Gamma \vdash e : \Box A]\!] \\ g_1 & = & [\![\Gamma \vdash \mathcal{E}_1 \langle\!\langle \text{box}\,\boxed{x} \rangle\!\rangle : C]\!] \\ g_2 & = & \pi_1 \,;\, [\![\Gamma, x : A^s \vdash v : B]\!] \\ g & = & \langle g_1\,, g_2\rangle \,;\, \beta_{C,B} \end{array} \\ \textit{in} & \langle id_\Gamma\,, f\rangle \,;\, \tau_{\Gamma,\Box A} \,;\, Tg \,;\, \mu_{C\times B} \end{array}$$

=⟨ simplification ⟩

$$\begin{array}{ll} \textit{let} & \begin{array}{lcl} f & = & [\![\Gamma \vdash e : \Box A]\!] \\ g_1 & = & [\![\Gamma, x : A^s \vdash \mathcal{E}_1 \langle\!\langle \text{box}\,\boxed{x} \rangle\!\rangle : C]\!] \\ g_2 & = & [\![\Gamma \vdash v : B]\!] \end{array} \\ \textit{in} & \langle id_\Gamma\,, f\rangle \,;\, \tau_{\Gamma,\Box A} \,;\, T\langle g_1\,, \pi_1 \,;\, g_2\rangle \,;\, T\beta_{C,B} \,;\, \mu_{C\times B} \end{array}$$

=⟨ simplification ⟩

$$\begin{array}{ll} \textit{let} & \begin{array}{lcl} f & = & [\![\Gamma \vdash e : \Box A]\!] \\ g_1 & = & [\![\Gamma, x : A^s \vdash \mathcal{E}_1 \langle\!\langle \text{box}\,\boxed{x} \rangle\!\rangle : C]\!] \\ g_2 & = & [\![\Gamma \vdash v : B]\!] \end{array} \\ \textit{in} & \langle\langle id_\Gamma\,, f\rangle \,;\, \tau_{\Gamma,\Box A} \,;\, Tg_1 \,;\, \mu_C\,, g_2\rangle \,;\, \beta_{C,B} \end{array}$$

=⟨ definition ⟩

$$\begin{array}{l} \mathit{let} \quad g_1 = [\![\Gamma \vdash \text{let box}\ \boxed{x} = e \text{ in } \mathcal{E}_1 \langle\!\langle \text{box}\ \boxed{x} \rangle\!\rangle : C]\!] \\ \qquad\ g_2 = [\![\Gamma \vdash v : B]\!] \\ \mathit{in} \quad \langle g_1 , g_2 \rangle \,;\, \beta_{C,B} \end{array}$$

$=\langle$ induction hypothesis $\rangle$

$$\begin{array}{l} \mathit{let} \quad g_1 = [\![\Gamma \vdash \mathcal{E}_1 \langle\!\langle e \rangle\!\rangle : C]\!] \\ \qquad\ g_2 = [\![\Gamma \vdash v : B]\!] \\ \mathit{in} \quad \langle g_1 , g_2 \rangle \,;\, \beta_{C,B} \end{array}$$

$=\langle$ definition $\rangle$

$$[\![\Gamma \vdash (\mathcal{E}_1 \langle\!\langle e \rangle\!\rangle , v) : C \times B]\!]$$

$\diamond\ \mathcal{E} = \text{let box}\ \boxed{z} = \mathcal{E}_1 \text{ in } e_1$

$$[\![\Gamma \vdash \text{let box}\ \boxed{x} = e \text{ in } (\text{let box}\ \boxed{z} = \mathcal{E}_1 \langle\!\langle \text{box}\ \boxed{x} \rangle\!\rangle \text{ in } e_1) : B]\!]$$

$=\langle$ definition $\rangle$

$$\begin{array}{l} \mathit{let} \quad f = [\![\Gamma \vdash e : \square A]\!] \\ \qquad\ g = [\![\Gamma, x : A^{s} \vdash \text{let box}\ \boxed{z} = \mathcal{E}_1 \langle\!\langle \text{box}\ \boxed{x} \rangle\!\rangle \text{ in } e_1 : B]\!] \\ \mathit{in} \quad \langle id_\Gamma , f \rangle \,;\, \tau_{\Gamma,\square A} \,;\, Tg \,;\, \mu_B \end{array}$$

$=\langle$ definition $\rangle$

$$\begin{array}{l} \mathit{let} \quad f = [\![\Gamma \vdash e : \square A]\!] \\ \qquad\ g_1 = [\![\Gamma, x : A^{s} \vdash \mathcal{E}_1 \langle\!\langle \text{box}\ \boxed{x} \rangle\!\rangle : \square C]\!] \\ \qquad\ g_2 = [\![\Gamma, x : A^{s}, z : C^{s} \vdash e_1 : B]\!] \\ \qquad\ g = \langle id_{\Gamma\times\square A} , g_1 \rangle \,;\, \tau_{\Gamma\times\square A,\square C} \,;\, Tg_2 \,;\, \mu_B \\ \mathit{in} \quad \langle id_\Gamma , f \rangle \,;\, \tau_{\Gamma,\square A} \,;\, Tg \,;\, \mu_B \end{array}$$

$=\langle$ functoriality of $T$ $\rangle$

$$\begin{array}{l} \mathit{let} \quad f = [\![\Gamma \vdash e : \square A]\!] \\ \qquad\ g_1 = [\![\Gamma, x : A^{s} \vdash \mathcal{E}_1 \langle\!\langle \text{box}\ \boxed{x} \rangle\!\rangle : \square C]\!] \\ \qquad\ g_2 = [\![\Gamma, x : A^{s}, z : C^{s} \vdash e_1 : B]\!] \\ \mathit{in} \quad \langle id_\Gamma , f \rangle \,;\, \tau_{\Gamma,\square A} \,;\, T\langle id_{\Gamma\times\square A} , g_1 \rangle \,;\, T\tau_{\Gamma\times\square A,\square C} \,;\, T^2 g_2 \,;\, T\mu_B \,;\, \mu_B \end{array}$$

$=\langle$ semantic substitution theorem 5.4 and semantic weakening lemma 5.1 $\rangle$

$$\begin{array}{l} \textit{let} \quad f = [\![\Gamma \vdash e : \Box A]\!] \\ \qquad g_1 = [\![\Gamma, x : A^s \vdash \mathcal{E}_1 \langle\!\langle \mathsf{box}\, \boxed{x} \rangle\!\rangle : \Box C]\!] \\ \qquad g_2 = \langle \pi_1 \,;\, \pi_1 \,,\, \pi_2 \rangle \,;\, [\![\Gamma, z : C^s \vdash e_1 : B]\!] \\ \textit{in} \quad \langle id_\Gamma \,,\, f \rangle \,;\, \tau_{\Gamma,\Box A} \,;\, T\langle id_{\Gamma\times\Box A} \,,\, g_1 \rangle \,;\, T\tau_{\Gamma\times\Box A,\Box C} \,;\, T^2 g_2 \,;\, T\mu_B \,;\, \mu_B \end{array}$$

$=\langle$ simplification $\rangle$

$$\begin{array}{l} \textit{let} \quad f = [\![\Gamma \vdash e : \Box A]\!] \\ \qquad g_1 = [\![\Gamma, x : A^s \vdash \mathcal{E}_1 \langle\!\langle \mathsf{box}\, \boxed{x} \rangle\!\rangle : \Box C]\!] \\ \qquad g_2 = [\![\Gamma, z : C^s \vdash e_1 : B]\!] \\ \textit{in} \quad \langle id_\Gamma \,,\, f \rangle \,;\, \tau_{\Gamma,\Box A} \,;\, T\langle id_{\Gamma\times\Box A} \,,\, g_1 \rangle \,;\, T\tau_{\Gamma\times\Box A,\Box C} \\ \qquad ;\, T^2\langle \pi_1 \,;\, \pi_1 \,,\, \pi_2 \rangle \,;\, T^2 g_2 \,;\, T\mu_B \,;\, \mu_B \end{array}$$

$=\langle$ simplification $\rangle$

$$\begin{array}{l} \textit{let} \quad f = [\![\Gamma \vdash e : \Box A]\!] \\ \qquad g_1 = [\![\Gamma, x : A^s \vdash \mathcal{E}_1 \langle\!\langle \mathsf{box}\, \boxed{x} \rangle\!\rangle : \Box C]\!] \\ \qquad g_2 = [\![\Gamma, z : C^s \vdash e_1 : B]\!] \\ \textit{in} \quad \langle id_\Gamma \,,\, \langle id_\Gamma \,,\, f \rangle \,;\, \tau_{\Gamma,\Box A} \,;\, Tg_1 \,;\, \mu_{\Box C} \rangle \,;\, \tau_{\Gamma,\Box C} \,;\, Tg_2 \,;\, \mu_B \end{array}$$

$=\langle$ definition $\rangle$

$$\begin{array}{l} \textit{let} \quad g_1 = [\![\Gamma \vdash \mathsf{let\ box}\, \boxed{x} = e \,\mathsf{in}\, \mathcal{E}_1 \langle\!\langle \mathsf{box}\, \boxed{x} \rangle\!\rangle : \Box C]\!] \\ \qquad g_2 = [\![\Gamma, z : C^s \vdash e_1 : B]\!] \\ \textit{in} \quad \langle id_\Gamma \,,\, g_1 \rangle \,;\, \tau_{\Gamma,\Box C} \,;\, Tg_2 \,;\, \mu_B \end{array}$$

$=\langle$ induction hypothesis $\rangle$

$$\begin{array}{l} \textit{let} \quad g_1 = [\![\Gamma \vdash \mathcal{E}_1 \langle\!\langle e \rangle\!\rangle : \Box C]\!] \\ \qquad g_2 = [\![\Gamma, z : C^s \vdash e_1 : B]\!] \\ \textit{in} \quad \langle id_\Gamma \,,\, g_1 \rangle \,;\, \tau_{\Gamma,\Box C} \,;\, Tg_2 \,;\, \mu_B \end{array}$$

$=\langle$ definition $\rangle$

$$[\![\Gamma \vdash \mathsf{let\ box}\, \boxed{z} = \mathcal{E}_1 \langle\!\langle e \rangle\!\rangle \,\mathsf{in}\, e_1 : B]\!]$$

$\diamond\, \mathcal{E} = \mathsf{let\ box}\, \boxed{z} = v \,\mathsf{in}\, \mathcal{E}_1$

$$[\![\Gamma \vdash \mathsf{let\ box}\, \boxed{x} = e \,\mathsf{in}\, (\mathsf{let\ box}\, \boxed{z} = v \,\mathsf{in}\, \mathcal{E}_1 \langle\!\langle \mathsf{box}\, \boxed{x} \rangle\!\rangle) : B]\!]$$

$$\begin{array}{l} \textit{let} \quad f = [\![\Gamma \vdash e : \Box A]\!] \\ \qquad g = [\![\Gamma, x : A^s \vdash \mathsf{let\ box}\, \boxed{z} = v \,\mathsf{in}\, \mathcal{E}_1 \langle\!\langle \mathsf{box}\, \boxed{x} \rangle\!\rangle : B]\!] \\ \textit{in} \quad \langle id_\Gamma \,,\, f \rangle \,;\, \tau_{\Gamma,\Box A} \,;\, Tg \,;\, \mu_B \end{array}$$

$=\langle$ definition $\rangle$

$$\begin{array}{ll} \mathit{let} & f = [\![\Gamma \vdash e : \Box A]\!] \\ & g_1 = [\![\Gamma, x : A^{\mathsf{s}} \vdash v : \Box C]\!] \\ & g_2 = [\![\Gamma, x : A^{\mathsf{s}}, z : C^{\mathsf{s}} \vdash \mathcal{E}_1 \langle\!\langle \mathsf{box}\ \boxed{x} \rangle\!\rangle : B]\!] \\ & g = \langle id_{\Gamma \times \Box A} , g_1 \rangle ; \tau_{\Gamma \times \Box A, \Box C} ; Tg_2 ; \mu_B \\ \mathit{in} & \langle id_\Gamma , f \rangle ; \tau_{\Gamma, \Box A} ; Tg ; \mu_B \end{array}$$

$=\langle$ functoriality of $T$ $\rangle$

$$\begin{array}{ll} \mathit{let} & f = [\![\Gamma \vdash e : \Box A]\!] \\ & g_1 = [\![\Gamma, x : A^{\mathsf{s}} \vdash v : \Box C]\!] \\ & g_2 = [\![\Gamma, x : A^{\mathsf{s}}, z : C^{\mathsf{s}} \vdash \mathcal{E}_1 \langle\!\langle \mathsf{box}\ \boxed{x} \rangle\!\rangle : B]\!] \\ \mathit{in} & \langle id_\Gamma , f \rangle ; \tau_{\Gamma, \Box A} ; T\langle id_{\Gamma \times \Box A} , g_1 \rangle ; T\tau_{\Gamma \times \Box A, \Box C} ; T^2 g_2 ; T\mu_B ; \mu_B \end{array}$$

$=\langle$ semantic weakening lemma 5.1 $\rangle$

$$\begin{array}{ll} \mathit{let} & f = [\![\Gamma \vdash e : \Box A]\!] \\ & g_1 = \pi_1 ; [\![\Gamma \vdash v : \Box C]\!] \\ & g_2 = [\![\Gamma, x : A^{\mathsf{s}}, z : C^{\mathsf{s}} \vdash \mathcal{E}_1 \langle\!\langle \mathsf{box}\ \boxed{x} \rangle\!\rangle : B]\!] \\ \mathit{in} & \langle id_\Gamma , f \rangle ; \tau_{\Gamma, \Box A} ; T\langle id_{\Gamma \times \Box A} , g_1 \rangle ; T\tau_{\Gamma \times \Box A, \Box C} ; T^2 g_2 ; T\mu_B ; \mu_B \end{array}$$

$=\langle$ simplification $\rangle$

$$\begin{array}{ll} \mathit{let} & f = [\![\Gamma \vdash e : \Box A]\!] \\ & g_1 = [\![\Gamma \vdash v : \Box C]\!] \\ & g_2 = [\![\Gamma, x : A^{\mathsf{s}}, z : C^{\mathsf{s}} \vdash \mathcal{E}_1 \langle\!\langle \mathsf{box}\ \boxed{x} \rangle\!\rangle : B]\!] \\ \mathit{in} & \langle id_\Gamma , f \rangle ; \tau_{\Gamma, \Box A} ; T\langle id_{\Gamma \times \Box A} , \pi_1 ; g_1 \rangle ; T\tau_{\Gamma \times \Box A, \Box C} ; T^2 g_2 ; T\mu_B ; \mu_B \end{array}$$

$=\langle$ semantic substitution theorem 5.4 and semantic weakening lemma 5.1 $\rangle$

$$\begin{array}{ll} \mathit{let} & f = [\![\Gamma \vdash e : \Box A]\!] \\ & g_1 = [\![\Gamma \vdash v : \Box C]\!] \\ & g_2 = \langle \pi_1 ; \pi_1 , \pi_2 , \pi_1 ; \pi_2 \rangle ; [\![\Gamma, z : C^{\mathsf{s}}, x : A^{\mathsf{s}} \vdash \mathcal{E}_1 \langle\!\langle \mathsf{box}\ \boxed{x} \rangle\!\rangle : B]\!] \\ \mathit{in} & \langle id_\Gamma , f \rangle ; \tau_{\Gamma, \Box A} ; T\langle id_{\Gamma \times \Box A} , \pi_1 ; g_1 \rangle ; T\tau_{\Gamma \times \Box A, \Box C} ; T^2 g_2 ; T\mu_B ; \mu_B \end{array}$$

$=\langle$ functoriality of $T$ $\rangle$

$$\begin{array}{ll} \mathit{let} & f = [\![\Gamma \vdash e : \Box A]\!] \\ & g_1 = [\![\Gamma \vdash v : \Box C]\!] \\ & g_2 = [\![\Gamma, z : C^{\mathsf{s}}, x : A^{\mathsf{s}} \vdash \mathcal{E}_1 \langle\!\langle \mathsf{box}\ \boxed{x} \rangle\!\rangle : B]\!] \\ \mathit{in} & \langle id_\Gamma , f \rangle ; \tau_{\Gamma, \Box A} ; T\langle id_{\Gamma \times \Box A} , \pi_1 ; g_1 \rangle ; T\tau_{\Gamma \times \Box A, \Box C} \\ & ; T^2\langle \pi_1 ; \pi_1 , \pi_2 , \pi_1 ; \pi_2 \rangle ; T^2 g_2 ; T\mu_B ; \mu_B \end{array}$$

$=\langle$ semantic weakening lemma 5.1 $\rangle$

$$\begin{array}{ll} \textit{let} & \begin{array}{lcl} f & = & [\![\Gamma, z : C^s \vdash e : \Box A]\!] \\ g_1 & = & [\![\Gamma \vdash v : \Box C]\!] \\ g_2 & = & [\![\Gamma, z : C^s, x : A^s \vdash \mathcal{E}_1 \langle\!\langle \mathsf{box}\ \boxed{x} \rangle\!\rangle : B]\!] \end{array} \\ \textit{in} & \begin{array}{l} \langle id_\Gamma , \pi_1 ; f \rangle ; \tau_{\Gamma,\Box A} ; T\langle id_{\Gamma\times\Box A} , \pi_1 ; g_1 \rangle ; T\tau_{\Gamma\times\Box A,\Box C} \\ ; T^2\langle \pi_1 ; \pi_1 , \pi_2 , \pi_1 ; \pi_2 \rangle ; T^2 g_2 ; T\mu_B ; \mu_B \end{array} \end{array}$$

=⟨ simplification ⟩

$$\begin{array}{ll} \textit{let} & \begin{array}{lcl} f & = & [\![\Gamma, z : C^s \vdash e : \Box A]\!] \\ g_1 & = & [\![\Gamma \vdash v : \Box C]\!] \\ g_2 & = & [\![\Gamma, z : C^s, x : A^s \vdash \mathcal{E}_1 \langle\!\langle \mathsf{box}\ \boxed{x} \rangle\!\rangle : B]\!] \end{array} \\ \textit{in} & \langle id_\Gamma , g_1 \rangle ; \tau_{\Gamma,\Box C} ; T\langle id_{\Gamma\times\Box C} , f \rangle ; T\tau_{\Gamma\times\Box C,\Box A} ; T^2 g_2 ; T\mu_B ; \mu_B \end{array}$$

=⟨ definition ⟩

$$\begin{array}{ll} \textit{let} & \begin{array}{lcl} g_1 & = & [\![\Gamma \vdash v : \Box C]\!] \\ g_2 & = & [\![\Gamma, z : C^s \vdash \mathsf{let\ box}\ \boxed{x} = e\ \mathsf{in}\ \mathcal{E}_1 \langle\!\langle \mathsf{box}\ \boxed{x} \rangle\!\rangle : B]\!] \end{array} \\ \textit{in} & \langle id_\Gamma , g_1 \rangle ; \tau_{\Gamma,\Box C} ; T g_2 ; \mu_B \end{array}$$

=⟨ induction hypothesis ⟩

$$\begin{array}{ll} \textit{let} & \begin{array}{lcl} g_1 & = & [\![\Gamma \vdash v : \Box C]\!] \\ g_2 & = & [\![\Gamma, z : C^s \vdash \mathcal{E}_1 \langle\!\langle e \rangle\!\rangle : B]\!] \end{array} \\ \textit{in} & \langle id_\Gamma , g_1 \rangle ; \tau_{\Gamma,\Box C} ; T g_2 ; \mu_B \end{array}$$

=⟨ definition ⟩

$$[\![\Gamma \vdash \mathsf{let\ box}\ \boxed{z} = v\ \mathsf{in}\ \mathcal{E}_1 \langle\!\langle e \rangle\!\rangle : B]\!]$$

□

## E SUPPLEMENTARY MATERIAL FOR SECTION 7 (EMBEDDING)

Lemma E.1. *For any context* $\Gamma$*, we have* $\underline{\Gamma}^s = \underline{\Gamma}$.

Proof. We do induction on the context $\Gamma$.

(1) $\boxed{\Gamma}$

(2) $\boxed{\Gamma = }$

(3) $\underline{\cdot}^s = {}^s = \; = \underline{\cdot}$ by definition

(4) $\boxed{\Gamma = \Delta, x : A}$

(5) $\underline{\Delta, x : A}^s = (\underline{\Delta}, x : \underline{A}^s)^s$ by definition

(6) $(\underline{\Delta}, x : \underline{A}^s)^s = \underline{\Delta}^s, x : \underline{A}^s$ by definition

(7) $\Delta^s, x : A^s = \Delta, x : A^s$ induction hypothesis

(8) $\Delta, x : A^s = \Delta, x : A^s$

(9) $\Gamma^s = \Gamma$

□

LEMMA E.2. $[e'/x]\, e = [e'/x]\, e$.

PROOF. We proceed by cases on $e$.

(1) $[e'/x]\, e$

(2) $e = ()$

(3) $[e'/x]\ ()$ by definition

(4) $()$ by definition

(5) $[e'/x]\ ()$ by definition

(6) $[e'/x]\ ()$ by definition

(7) $e = x$

(8) $[e'/x]\, x$ by definition

(9) $e'$ by definition

(10) $[e'/x]\, x$ by definition

(11) $[e'/x]\, x$ by definition

(12) $e = y, (y \neq x)$

(13) $[e'/y]\, x$ by definition

(14) $x$ by definition

(15) $x$ by definition

(16) $[e'/y]\, x$ by definition

(17) $[e'/y]\, x$ by definition

(18) $e = \lambda y.\ e_1, (y \neq x)$

(19) $[e'/x]\, \lambda y.\ e_1$ by definition

(20) $\lambda y.\ [e'/x]\ e_1$ by definition

(21) $\lambda z.$ let box $y$ $= z$ in $[e'/x]\ e_1$ by definition

(22) $\lambda z.$ let box $y$ $= [e'/x]\ z$ in $[e'/x]\ e_1$ by definition

(23) $[e'/x]\ \lambda z.$ let box $x$ $= z$ in $e_1$ by definition

(24) $[e'/x]\ \lambda x.\ e_1$ by definition

(25) $e = e_1\ e_2$

(26) $[e'/x]\ e_1\ e_2$ by definition

(27) $[e'/x]\ e_1\ [e'/x]\ e_2$ by definition

(28) $[e'/x]\ e_1$ (box $[e'/x]\ e_2$) by definition

(29) $[e'/x]\ e_1$ ($[e'/x]$ box $e_2$) by definition

(30) $[e'/x]\ e_1$ (box $e_2$) by definition

(31) $[e'/x]\ e_1\ e_2$ by definition

(32) $[e'/x]\ e$

□

Lemma E.3. *If $x : A \in \Gamma$, then $x : A^s \in \Gamma$.*

Proof. We do induction on $x : A \in \Gamma$.

(1) $x : A \in \Gamma$

(2) $\dfrac{}{x : A \in (\Gamma, x : A)}$ ∈-id

(3) $x : A^s \in \Gamma, x : A^s$ ∈-id

(4) $x : A^s \in \Gamma, x : A$ by definition

(5) $\dfrac{x : A \in \Gamma \qquad (x \neq y)}{x : A \in (\Gamma, y : B)}$ ∈-ex

(6) $x : A \in \Gamma$ inversion

(7) $x : A^s \in \Gamma$ induction hypothesis

(8) $x : A^{s} \in \Gamma, y : B^{s}$ ∈-ex

(9) $x : A^{s} \in \Gamma, y : B$ by definition

(10) $x : A^{s} \in \Gamma$

□

**Theorem 7.1** (Preservation of typing). *If* $\Gamma \vdash_{\lambda} e : A$*, then* $\Gamma \vdash e : A$.

**Proof.** We do induction on $\Gamma \vdash_{\lambda} e : A$.

(1) $\Gamma \vdash_{\lambda} e : A$

(2) $\dfrac{}{\Gamma \vdash_{\lambda} () : \mathsf{unit}}$ unitI

(3) $\Gamma \vdash () : \mathsf{unit}$ unitI

(4) $\Gamma \vdash () : \mathsf{unit}$ by definition

(5) $\dfrac{x : A \in \Gamma}{\Gamma \vdash_{\lambda} x : A}$ Var

(6) $x : A \in \Gamma$ inversion

(7) $x : A^{s} \in \Gamma$ lemma E.3

(8) $\Gamma \vdash x : A$ Var

(9) $\Gamma \vdash x : A$ by definition

(10) $\dfrac{\Gamma, x : A \vdash_{\lambda} e : B}{\Gamma \vdash_{\lambda} \lambda x : A.\ e : A \Rightarrow B}$ ⇒I

(11) $\Gamma, x : A \vdash_{\lambda} e : B$ inversion

(12) $\Gamma, x : A \vdash e : B$ induction hypothesis

(13) $\Gamma, x : A^{s} \vdash e : B$ by definition

(14) $\Gamma, z : \Box A \vdash z : \Box A$ Var

(15) $(\Gamma, z : \Box A) \supseteq \Gamma$ ⊇-wk

(16) $(\Gamma, z : \Box A, x : A^{s}) \supseteq (\Gamma, x : A^{s})$ ⊇-cong

(17) $\Gamma, z : \Box A, x : A^{s} \vdash e : B$ lemma 3.1 (16) (13)

(18) $\Gamma, z : \Box A \vdash \mathsf{let\ box}\ \boxed{x} = z\ \mathsf{in}\ e : B$ □E (14) (17)

(19) $\Gamma \vdash \lambda z : \square A.\ \text{let box}\ \boxed{x} = z\ \text{in}\ e : \square A \Rightarrow B$ $\Rightarrow$I

(20) $\Gamma \vdash \lambda x : A.\ e : A \Rightarrow B$ by definition

(21) $$\frac{\Gamma \vdash_\lambda e_1 : A \Rightarrow B \qquad \Gamma \vdash_\lambda e_2 : A}{\Gamma \vdash_\lambda e_1\, e_2 : B}$$ $\Rightarrow$E

(22) $\Gamma \vdash_\lambda e_1 : A \Rightarrow B$ inversion

(23) $\Gamma \vdash_\lambda e_2 : A$ inversion

(24) $\Gamma \vdash e_1 : A \Rightarrow B$ induction hypothesis

(25) $\Gamma \vdash e_1 : \square A \Rightarrow B$ by definition

(26) $\Gamma \vdash e_2 : A$ induction hypothesis

(27) $\Gamma^s \vdash e_2 : A$ lemma E.1

(28) $\Gamma \vdash^s e_2 : A$ CTX-SAFE

(29) $\Gamma \vdash \text{box}\ \boxed{e_2} : \square A$ $\square$I

(30) $\Gamma \vdash e_1\ (\text{box}\ \boxed{e_2}) : B$ $\Rightarrow$E (25) (29)

(31) $\Gamma \vdash e_1\, e_2 : B$ by definition

(32) $\Gamma \vdash e : A$

□

Theorem 7.2 (Preservation of equality). *If* $\Gamma \vdash_\lambda e_1 \approx e_2 : A$, *then* $\Gamma \vdash e_1 \approx e_2 : A$.

Proof. We do induction on $\Gamma \vdash_\lambda e_1 \approx e_2 : A$.

(1) $\Gamma \vdash_\lambda e_1 \approx e_2 : A$

(2) $$\frac{\Gamma, x : A \vdash_\lambda e_1 : B \qquad \Gamma \vdash_\lambda e_2 : A}{\Gamma \vdash_\lambda (\lambda x : A.\ e_1)\ e_2 \approx [e_2/x]e_1 : B}$$ $\Rightarrow \beta$

(3) $\Gamma, x : A \vdash_\lambda e_1 : B$ inversion

(4) $\Gamma, x : A \vdash e_1 : B$ theorem 7.1

(5) $\Gamma, x : A^s \vdash e_1 : B$ by definition

(6) $\Gamma \vdash_\lambda e_2 : A$ inversion

(7) $\Gamma \vdash e_2 : A$ theorem 7.1

(8) $\Gamma^{s} \vdash e_2 : A$ lemma E.1

(9) $\Gamma \vdash \text{let box}\,\boxed{x} = \text{box}\,\boxed{e_2}\ \text{in}\ e \approx [e_2/x]e : B$ $\Box\beta$

(10) $\Gamma \vdash \boxed{\begin{array}{c}(\lambda z : \Box A.\ \text{let box}\,\boxed{x} = z\ \text{in}\ e_1)\ (\text{box}\,\boxed{e_2}) \\ \approx \\ \text{let box}\,\boxed{x} = \text{box}\,\boxed{e_2}\ \text{in}\ e_1\end{array}} : B$ $\Rightarrow\beta$

(11) $\Gamma \vdash \boxed{\begin{array}{c}(\lambda z : \Box A.\ \text{let box}\,\boxed{x} = z\ \text{in}\ e_1)\ (\text{box}\,\boxed{e_2}) \\ \approx \\ [e_1/x]e_2\end{array}} : B$ TRANS

(12) $\Gamma \vdash (\lambda x : A.\ e_1)\ e_2 \approx [e_2/x]e_1 : B$ by definition

(13) $\boxed{\dfrac{\Gamma \vdash_\lambda e : A \Rightarrow B}{\Gamma \vdash_\lambda e \approx \lambda x : A.\ e\ x : A \Rightarrow B}}$ $\Rightarrow\eta$

(14) $\Gamma \vdash_\lambda e : A \Rightarrow B$ inversion

(15) $\Gamma \vdash e : A \Rightarrow B$ theorem 7.1

(16) $\Gamma \vdash e : \Box A \Rightarrow B$ by definition

(17) $\Gamma^{s} \vdash e : \Box A \Rightarrow B$ lemma E.1

(18) $\Gamma \vdash^{s} e : \Box A \Rightarrow B$ CTX-SAFE

(19) $\Gamma \vdash e \approx \lambda z.\ e\ z : \Box A \Rightarrow B$ $\Rightarrow\eta$-SAFE

(20) $\Gamma, z : \Box A \vdash z : \Box A$ VAR

(21) $\Gamma, z : \Box A \vdash e : \Box A \Rightarrow B$ lemma 3.1 (16)

(22) $\Gamma, z : \Box A \vdash e\ z : B$ $\Rightarrow$E

(23) $\Gamma, z : \Box A, x : A^{s} \vdash x : A$ VAR

(24) $\Gamma, z : \Box A, x : A^{s} \vdash \text{box}\,\boxed{x} : \Box A$ $\Box$I

(25) $\Gamma, z : \Box A, x : A^{s} \vdash e\ (\text{box}\,\boxed{x}) : B$ $\Rightarrow$E

(26) $\Gamma, z : \Box A \vdash \text{let box}\,\boxed{x} = z\ \text{in}\ e\ (\text{box}\,\boxed{x}) : B$ $\Box$E

(27) $\Gamma, z : \Box A \vdash \boxed{\begin{array}{c}e\ z \\ \approx \\ \text{let box}\,\boxed{x} = z\ \text{in}\ e\ (\text{box}\,\boxed{x})\end{array}} : B$ $\Box\eta$-impure on $e$ $\mathcal{E}$

(28) $\Gamma \vdash \lambda z.\ e\ z \approx \lambda z.\ \text{let box}\ \boxed{x} = z \text{ in } e\ (\text{box}\ \boxed{x}) : \Box A \Rightarrow B$ λ-CONG

(29) $\Gamma \vdash e \approx \lambda z.\ \text{let box}\ \boxed{x} = z \text{ in } e\ (\text{box}\ \boxed{x}) : \Box A \Rightarrow B$ TRANS

(30) $\Gamma \vdash e \approx \lambda z.\ \text{let box}\ \boxed{x} = z \text{ in } e\ (\text{box}\ \boxed{x}) : \Box A \Rightarrow B$ by definition

(31) $\Gamma \vdash e \approx \lambda x.\ e\ x : A \Rightarrow B$ by definition

(32) $\dfrac{\Gamma \vdash_\lambda e : A}{\Gamma \vdash_\lambda e \approx e : A}$ REFL

(33) $\Gamma \vdash_\lambda e : A$ inversion

(34) $\Gamma \vdash e : A$ theorem 7.1

(35) $\Gamma \vdash e \approx e : A$ REFL

(36) $\dfrac{\Gamma \vdash_\lambda e_1 \approx e_2 : A}{\Gamma \vdash_\lambda e_2 \approx e_1 : A}$ SYM

(37) $\Gamma \vdash_\lambda e_1 \approx e_2 : A$ inversion

(38) $\Gamma \vdash e_1 \approx e_2 : A$ induction hypothesis

(39) $\Gamma \vdash e_2 \approx e_1 : A$ SYM

(40) $\dfrac{\Gamma \vdash_\lambda e_1 \approx e_2 : A \qquad \Gamma \vdash_\lambda e_2 \approx e_3 : A}{\Gamma \vdash_\lambda e_1 \approx e_3 : A}$ TRANS

(41) $\Gamma \vdash_\lambda e_1 \approx e_2 : A$ inversion

(42) $\Gamma \vdash_\lambda e_2 \approx e_3 : A$ inversion

(43) $\Gamma \vdash e_1 \approx e_2 : A$ induction hypothesis

(44) $\Gamma \vdash e_2 \approx e_3 : A$ induction hypothesis

(45) $\Gamma \vdash e_1 \approx e_3 : A$ TRANS

(46) $\dfrac{\Gamma, x : A \vdash_\lambda e_1 \approx e_2 : B}{\Gamma \vdash_\lambda \lambda x : A.\ e_1 \approx \lambda x : A.\ e_2 : A \Rightarrow B}$ λ-CONG

(47) $\Gamma, x : A \vdash_\lambda e_1 \approx e_2 : B$ inversion

(48) $\Gamma, x : A \vdash e_1 \approx e_2 : B$ induction hypothesis

(49) $\Gamma, x : A^{s} \vdash e_1 \approx e_2 : B$ by definition

(50) $\Gamma, z : \Box A, x : A^{s} \vdash e_1 \approx e_2 : B$ lemma 3.1

(51) $\Gamma, z : \Box A \vdash z : \Box A$ VAR

(52) $\Gamma, z : \Box A \vdash z \approx z : \Box A$ REFL

(53) $\Gamma, z : \Box A \vdash \begin{array}{c}(\text{let box } \boxed{x} = z \text{ in } e_1) \\ \approx \\ (\text{let box } \boxed{x} = z \text{ in } e_2)\end{array} : B$ let box-CONG

(54) $\Gamma \vdash \begin{array}{c}(\lambda z.\ \text{let box } \boxed{x} = z \text{ in } e_1) \\ \approx \\ (\lambda z.\ \text{let box } \boxed{x} = z \text{ in } e_2)\end{array} : \Box A \Rightarrow B$ $\lambda$-CONG

(55) $\Gamma \vdash \lambda x.\ e_1 \approx \lambda x.\ e_2 : A \Rightarrow B$ by definition

(56) $\dfrac{\Gamma \vdash_{\lambda} e_1 \approx e_2 : A \Rightarrow B \qquad \Gamma \vdash_{\lambda} e_3 \approx e_4 : A}{\Gamma \vdash_{\lambda} e_1\, e_3 \approx e_2\, e_4 : B}$ APP-CONG

(57) $\Gamma \vdash_{\lambda} e_1 \approx e_2 : A \Rightarrow B$ inversion

(58) $\Gamma \vdash e_1 \approx e_2 : A \Rightarrow B$ induction hypothesis

(59) $\Gamma \vdash e_1 \approx e_2 : \Box A \Rightarrow B$ by definition

(60) $\Gamma \vdash e_3 \approx e_4 : A$ induction hypothesis

(61) $\Gamma^{s} \vdash e_3 \approx e_4 : A$ lemma E.1

(62) $\Gamma \vdash \text{box } \boxed{e_3} \approx \text{box } \boxed{e_4} : \Box A$ box-CONG

(63) $\Gamma \vdash e_1\, (\text{box } \boxed{e_2}) \approx e_3\, (\text{box } \boxed{e_4}) : B$ APP-CONG

(64) $\Gamma \vdash e_1\, e_2 \approx e_3\, e_4 : B$ by definition

(65) $\Gamma \vdash e_1 \approx e_2 : A$

□

We can define a reverse translation which forgets the purity annotations, in figure 21.

We use the notation $\widehat{X}$ to denote the *unembedding* of a syntactic object $X$ from our calculus to STLC. We use $b$ to mean base types, i.e., unit, str and cap.

We prove some properties of the unembedding of an embedded term.

$$
\begin{array}{llcl}
\textsc{Types} & \overline{b} & := & \mathsf{unit} \\
 & \overline{A \Rightarrow B} & := & \overline{A} \Rightarrow \overline{B} \\
 & \overline{\Box A} & := & \overline{A} \\
\\
\textsc{Contexts} & \overline{\cdot} & := & \\
 & \overline{\Gamma, x : A^q} & := & \overline{\Gamma}, x : \overline{A} \\
\\
\textsc{Terms} & \overline{()} & := & () \\
 & \overline{s} & := & () \\
 & \overline{x} & := & x \\
 & \overline{\lambda x : A.\ e} & := & \lambda x : \overline{A}.\ \overline{e} \\
 & \overline{e_1\, e_2} & := & \overline{e_1}\ \overline{e_2} \\
 & \overline{\mathsf{box}\ \boxed{e}} & := & \overline{e} \\
 & \overline{\mathsf{let\ box}\ \boxed{x} = e_1\ \mathsf{in}\ e_2} & := & (\lambda x.\ \overline{e_2})\ \overline{e_1} \\
 & \overline{e_1 . \mathsf{print}(e_2)} & := & ()
\end{array}
$$

Fig. 21. Reverse Translation to STLC

LEMMA E.4. *For any STLC type* $A$, $\underline{\overline{A}} = A$.

PROOF. We do induction on $A$.

| | | |
|---|---|---|
| (1) | $\boxed{A}$ | |
| (2) | $\boxed{\underline{\overline{b}}}$ | |
| (3) | $\overline{b}$ | definition |
| (4) | $b$ | definition |
| (5) | $\boxed{\underline{\overline{A \Rightarrow B}}}$ | |
| (6) | $\overline{\Box \underline{A} \Rightarrow \underline{B}}$ | definition |
| (7) | $\overline{\Box \underline{A}} \Rightarrow \overline{\underline{B}}$ | definition |
| (8) | $\underline{\overline{A}} \Rightarrow \underline{\overline{B}}$ | definition |
| (9) | $A \Rightarrow B$ | induction hypothesis |
| (10) | $\underline{\overline{A}} = A$ | |

□

LEMMA E.5. *For any STLC context* $\Gamma$, $\underline{\overline{\Gamma}} = \Gamma$.

PROOF. We do induction on $\Gamma$.

| | |
|---|---|
| (1) | $\boxed{\Gamma}$ |
| (2) | $\boxed{\underline{\overline{\cdot}}}$ |

□

Lemma E.6. *If* $\Gamma \vdash_\lambda e : A$, *then* $\Gamma \vdash_\lambda \overline{\underline{e}} : A$.

Proof. We do induction on $\Gamma \vdash_\lambda e : A$.

(1) $\Gamma \vdash_\lambda e : A$

(2) $\dfrac{}{\Gamma \vdash_\lambda () : \mathsf{unit}}$ unitI

(3) $\Gamma \vdash_\lambda () : \mathsf{unit}$ unitI

(4) $\dfrac{x : A \in \Gamma}{\Gamma \vdash_\lambda x : A}$ Var

(5) $x : A \in \Gamma$ inversion

(6) $\Gamma \vdash_\lambda x : A$ Var

(7) $\dfrac{\Gamma, x : A \vdash_\lambda e : B}{\Gamma \vdash_\lambda \lambda x : A.\ e : A \Rightarrow B}$ $\Rightarrow$I

(8) $\Gamma, x : A \vdash_\lambda e : B$ inversion

(9) $\Gamma \vdash_\lambda \overline{\underline{\lambda x : A.\ e}} : A \Rightarrow B$

(10) $\dfrac{\Gamma \vdash_\lambda e_1 : A \Rightarrow B \qquad \Gamma \vdash_\lambda e_2 : A}{\Gamma \vdash_\lambda e_1\, e_2 : B}$ $\Rightarrow$E

(11) $\Gamma \vdash_\lambda e_1 : A \Rightarrow B$ inversion

(12) $\Gamma \vdash_\lambda e_2 : A$ inversion

(13) $\Gamma \vdash_\lambda \overline{\underline{e_1\, e_2}} : B$

(14) $\Gamma \vdash_\lambda \overline{\underline{e}} : A$

□

We observe that an embedding followed by an unembedding gives a $\beta\eta$-equal term.

Lemma E.7. *If* $\Gamma \vdash_\lambda e : A$, *then* $\Gamma \vdash_\lambda e \approx \overline{\underline{e}} : A$.

Proof. We do induction on $\Gamma \vdash_\lambda e : A$.

(1) $\Gamma \vdash_\lambda e : A$

(2) $\dfrac{x : A \in \Gamma}{\Gamma \vdash_\lambda x : A}$ Var

| | | |
|---|---|---|
| (3) | $\overline{\underline{x}} = \overline{x} = x$ | definition |
| (4) | $\Gamma \vdash_\lambda \overline{\underline{x}} : A$ | |
| (5) | $\Gamma \vdash_\lambda x \approx \overline{\underline{x}} : A$ | REFL |
| (6) | $\dfrac{\Gamma, x : A \vdash_\lambda e : B}{\Gamma \vdash_\lambda \lambda x : A.\ e : A \Rightarrow B}$ | $\Rightarrow$I |
| (7) | $\Gamma \vdash_\lambda \lambda x : A.\ e \approx \lambda z : A.\ (\lambda x : A.\ e)\ z : B$ | $\Rightarrow \eta$ |
| (8) | $\lambda z : A.\ (\lambda x : A.\ e)\ z = \lambda z : \overline{\underline{A}}.\ (\lambda x : A.\ e)\ z$ | lemma E.4 |
| (9) | $\lambda z : \overline{\underline{A}}.\ (\lambda x : A.\ e)\ z = \lambda z : \overline{\Box \underline{A}}.\ \overline{\text{let box}\ \boxed{x} = z\ \text{in}\ \underline{e}}$ | definition |
| (10) | $\lambda z : \overline{\Box \underline{A}}.\ \overline{\text{let box}\ \boxed{x} = z\ \text{in}\ \underline{e}} = \overline{\lambda z : \Box \underline{A}.\ \text{let box}\ \boxed{x} = z\ \text{in}\ \underline{e}}$ | definition |
| (11) | $\overline{\lambda z : \Box \underline{A}.\ \text{let box}\ \boxed{x} = z\ \text{in}\ \underline{e}} = \overline{\underline{\lambda x : A.\ e}}$ | definition |
| (12) | $\Gamma \vdash_\lambda \lambda x : A.\ e \approx \overline{\underline{\lambda x : A.\ e}} : A \Rightarrow B$ | |
| (13) | $\dfrac{\Gamma \vdash_\lambda e_1 : A \Rightarrow B \qquad \Gamma \vdash_\lambda e_2 : A}{\Gamma \vdash_\lambda e_1\ e_2 : B}$ | $\Rightarrow$E |
| (14) | $\Gamma \vdash_\lambda e_1 : A \Rightarrow B$ | inversion |
| (15) | $\Gamma \vdash_\lambda e_2 : A$ | inversion |
| (16) | $\Gamma \vdash_\lambda e_1 \approx \overline{\underline{e_1}} : A \Rightarrow B$ | induction hypothesis |
| (17) | $\Gamma \vdash_\lambda e_2 \approx \overline{\underline{e_2}} : A$ | induction hypothesis |
| (18) | $\Gamma \vdash_\lambda e_1\ e_2 \approx \overline{\underline{e_1}}\ \overline{\underline{e_2}} : B$ | APP-CONG |
| (19) | $\overline{\underline{e_1}}\ \overline{\underline{e_2}} = \overline{\underline{e_1}}\ \overline{\text{box}\ \boxed{\underline{e_2}}} = \overline{\underline{e_1}\ \text{box}\ \boxed{\underline{e_2}}} = \overline{\underline{e_1\ e_2}}$ | definition |
| (20) | $\Gamma \vdash_\lambda e_1\ e_2 \approx \overline{\underline{e_1\ e_2}} : B$ | |
| (21) | $\Gamma \vdash_\lambda e \approx \overline{\underline{e}} : A$ | |

□

At this point, we could setup a syntactic logical relation to show a conservative extension result. Instead, we will use an abstract trick.

Note that there is a forgetful functor from $\mathcal{C}$ to Set, which forgets the weight assignments. It is easy to see from our definition of $\mathcal{C}$ in section 4 that this functor preserves the cartesian closed structure, and is hence a cartesian closed functor. Forgetting the extra structure of Set, we could instead choose CCC[1], the free cartesian closed category on one generator 1. We consider the forgetful functor $\mathcal{F}$ from $\mathcal{C}$ to CCC[1], which forgets the capability annotations.

$$\begin{aligned}\mathcal{F}(\mathsf{unit}) &:= 1\\ \mathcal{F}(\Sigma^*) &:= 1\\ \mathcal{F}(A \times B) &:= \mathcal{F}(A) \times \mathcal{F}(B)\\ \mathcal{F}(A \Rightarrow B) &:= \mathcal{F}(A) \Rightarrow \mathcal{F}(B)\end{aligned}$$

We note that it maps the monad and comonad to identity.

$$\begin{aligned}\mathcal{F}(\Box A) &= \mathcal{F}(A)\\ \mathcal{F}(TA) &= \mathcal{F}(A)\end{aligned}$$

We observe that the action of this functor $\mathcal{F}$ on embedded terms gives back the original term.

Lemma E.8. *If* $\Gamma \vdash_\lambda e : A$, *then* $\mathcal{F}(\llbracket \Gamma \vdash e : A \rrbracket) = \llbracket \Gamma \vdash_\lambda e : A \rrbracket$.

Proof. We proceed by induction on $\Gamma \vdash_\lambda e : A$.

◇ $\dfrac{x : A \in \Gamma}{\Gamma \vdash_\lambda x : A}$ Var

$\mathcal{F}(\llbracket \Gamma \vdash x : A \rrbracket)$

$=\langle$ definition $\rangle$

$\mathcal{F}(\llbracket x : A \in \Gamma \rrbracket \mathbin{;} \eta_A)$

$=\langle$ functoriality of $\mathcal{F}$ $\rangle$

$\mathcal{F}(\llbracket x : A \in \Gamma \rrbracket) \mathbin{;} \mathcal{F}(\eta_A)$

$=\langle$ definition $\rangle$

$\llbracket x : A \in \Gamma \rrbracket$

$=\langle$ definition $\rangle$

$\llbracket \Gamma \vdash_\lambda e : A \rrbracket$

◇ $\dfrac{\Gamma, x : A \vdash_\lambda e : B}{\Gamma \vdash_\lambda \lambda x : A.\ e : A \Rightarrow B}$ $\Rightarrow$I

$\mathcal{F}(\llbracket \Gamma \vdash \lambda x : A.\ e : A \Rightarrow B \rrbracket)$

$=\langle$ definition $\rangle$

$\mathcal{F}(\llbracket \Gamma \vdash \lambda z : \Box A.\ \mathsf{let\ box}\ \boxed{x} = z\ \mathsf{in}\ e : \Box A \Rightarrow B \rrbracket)$

$=\langle$ definition $\rangle$

$$\mathcal{F}(\text{curry}\,(\llbracket \Gamma, z : \Box A^{\mathsf{i}} \vdash \text{let box}\,\boxed{x} = z \text{ in } e : B \rrbracket) \,;\, \eta_{A \to TB})$$

$=\langle$ functoriality of $\mathcal{F}$ $\rangle$

$$\mathcal{F}(\text{curry}\,(\llbracket \Gamma, z : \Box A^{\mathsf{i}} \vdash \text{let box}\,\boxed{x} = z \text{ in } e : B \rrbracket)) \,;\, \mathcal{F}(\eta_{A \to TB})$$

$=\langle$ definition $\rangle$

$$\begin{array}{ll} \mathit{let} & \begin{array}{lcl} f & = & \llbracket \Gamma, z : \Box A^{\mathsf{i}} \vdash z : \Box A \rrbracket \\ g & = & \llbracket \Gamma, z : \Box A^{\mathsf{i}}, x : A^{\mathsf{s}} \vdash e : B \rrbracket \end{array} \\ \mathit{in} & \mathcal{F}(\text{curry}\,(\langle id_{\Gamma \times \Box A}, f \rangle \,;\, \tau_{\Gamma \times \Box A, \Box A} \,;\, Tg \,;\, \mu_B)) \end{array}$$

$=\langle$ simplification $\rangle$

$$\begin{array}{ll} \mathit{let} & g = \llbracket \Gamma, z : \Box A^{\mathsf{i}}, x : A^{\mathsf{s}} \vdash e : B \rrbracket \\ \mathit{in} & \mathcal{F}(\text{curry}\,(\langle id_{\Gamma \times \Box A}, \pi_2 \,;\, \eta_{\Box A} \rangle \,;\, \tau_{\Gamma \times \Box A, \Box A} \,;\, Tg \,;\, \mu_B)) \end{array}$$

$=\langle$ strength law and monad laws $\rangle$

$$\begin{array}{ll} \mathit{let} & g = \llbracket \Gamma, z : \Box A^{\mathsf{i}}, x : A^{\mathsf{s}} \vdash e : B \rrbracket \\ \mathit{in} & \mathcal{F}(\text{curry}\,(\langle id_{\Gamma \times \Box A}, \pi_2 \rangle \,;\, g)) \end{array}$$

$=\langle$ $\mathcal{F}$ preserves exponentials $\rangle$

$$\text{curry}\,(\mathcal{F}(\llbracket \Gamma, x : A^{\mathsf{s}} \vdash e : B \rrbracket))$$

$=\langle$ definition $\rangle$

$$\text{curry}\,(\mathcal{F}(\llbracket \Gamma, x : A \vdash e : B \rrbracket))$$

$=\langle$ induction hypothesis $\rangle$

$$\text{curry}\,(\llbracket \Gamma, x : A \vdash_{\lambda} e : B \rrbracket)$$

$=\langle$ definition $\rangle$

$$\llbracket \Gamma \vdash_{\lambda} \lambda x : A.\ e : A \Rightarrow B \rrbracket$$

$$\diamond \quad \frac{\Gamma \vdash_{\lambda} e_1 : A \Rightarrow B \qquad \Gamma \vdash_{\lambda} e_2 : A}{\Gamma \vdash_{\lambda} e_1\, e_2 : B} \Rightarrow\!\mathrm{E}$$

$$\mathcal{F}(\llbracket \Gamma \vdash e_1\, e_2 : B \rrbracket)$$

$=\langle$ definition $\rangle$

$$\mathcal{F}(\llbracket \Gamma \vdash e_1 \text{ box}\,\boxed{e_2} : B \rrbracket)$$

$=\langle$ definition $\rangle$

$$\begin{array}{l} \textit{let} \quad f = [\![\Gamma \vdash e_1 : \Box A \Rightarrow B]\!] \\ \phantom{\textit{let}} \quad g = [\![\Gamma \vdash \mathsf{box}\ \boxed{e_2} : \Box A]\!] \\ \textit{in} \quad \mathcal{F}(\langle f, g\rangle \mathbin{;} \beta_{\Box A \to TB, \Box A} \mathbin{;} T\, \mathsf{ev}_{\Box A, TB} \mathbin{;} \mu_B) \end{array}$$

$= \langle$ functoriality of $\mathcal{F}$ $\rangle$

$$\begin{array}{l} \textit{let} \quad f = [\![\Gamma \vdash e_1 : \Box A \Rightarrow B]\!] \\ \phantom{\textit{let}} \quad g = [\![\Gamma \vdash \mathsf{box}\ \boxed{e_2} : \Box A]\!] \\ \textit{in} \quad \mathcal{F}(\langle f, g\rangle) \mathbin{;} \mathcal{F}(\beta_{\Box A \to TB, \Box A}) \mathbin{;} \mathcal{F}(T\, \mathsf{ev}_{\Box A, TB}) \mathbin{;} \mathcal{F}(\mu_B) \end{array}$$

$= \langle$ action of $\mathcal{F}$ $\rangle$

$$\begin{array}{l} \textit{let} \quad f = \mathcal{F}([\![\Gamma \vdash e_1 : A \Rightarrow B]\!]) \\ \phantom{\textit{let}} \quad g = \mathcal{F}([\![\Gamma \vdash e_2 : A]\!]) \\ \textit{in} \quad \langle f, g\rangle \mathbin{;} \mathsf{ev}_{A,B} \end{array}$$

$= \langle$ induction hypothesis $\rangle$

$$\begin{array}{l} \textit{let} \quad f = [\![\Gamma \vdash_\lambda e_1 : A \Rightarrow B]\!] \\ \phantom{\textit{let}} \quad g = [\![\Gamma \vdash_\lambda e_2 : A]\!] \\ \textit{in} \quad \langle f, g\rangle \mathbin{;} \mathsf{ev}_{A,B} \end{array}$$

$= \langle$ definition $\rangle$

$$[\![\Gamma \vdash_\lambda e_1\, e_2 : B]\!]$$

□

Theorem 7.3 (Conservative Extension). *If $\Gamma \vdash_\lambda e_1 : A, \Gamma \vdash_\lambda e_2 : A$, and $\Gamma \vdash e_1 \approx e_2 : A$, then $\Gamma \vdash_\lambda e_1 \approx e_2 : A$.*

Proof.

(1) $\Gamma \vdash_\lambda e_1 : A, \Gamma \vdash_\lambda e_2 : A$

(2) $\Gamma \vdash e_1 \approx e_2 : A$

(3) $[\![\Gamma \vdash e_1 : A]\!] = [\![\Gamma \vdash e_2 : A]\!]$ — soundness of $\approx$ theorem 6.1

(4) $\mathcal{F}([\![\Gamma \vdash e_1 : A]\!]) = \mathcal{F}([\![\Gamma \vdash e_2 : A]\!])$ — congruence

(5) $[\![\Gamma \vdash_\lambda e_1 : A]\!] = [\![\Gamma \vdash_\lambda e_2 : A]\!]$ — lemma E.8

(6) $\Gamma \vdash_\lambda e_1 \approx e_2 : A$ — completeness of STLC

□